\documentclass[11pt,a4paper]{article}
\pdfoutput=1
\usepackage{jheppub}
\usepackage{amsfonts,amssymb,amsmath}
\usepackage{epsfig, graphicx,yfonts}

\newcommand{\be}{\begin{equation}}
\newcommand{\ee}{\end{equation}}
\newcommand{\bea}{\begin{eqnarray}}
\newcommand{\eea}{\end{eqnarray}}
\newcommand{\eqn}[1]{(\ref{#1})}

\newcommand{\sac}{\, , \qquad}
\newcommand{\mt}[1]{\textrm{\tiny #1}}
\newcommand{\jem}{J^\mt{EM}}
\newcommand{\cf}{{\cal F}}
\newcommand{\cb}{{\cal B}}
\newcommand{\ch}{{\cal H}}

\newcommand{\vev}[1]{\langle #1\rangle}
\newcommand{\wn}{{\textswab{w}}}
\newcommand{\qn}{{\textswab{q}}}
\def\nc {N_\mt{c}}
\def\nf {N_\mt{f}}
\def\uh {u_\mt{H}}
\def\gym {g_\mt{YM}}

\title{More on thermal probes of a  \\ strongly coupled anisotropic plasma}
\author[a]{Viktor Jahnke,}
\author[b]{Andr\'es Luna,}
\author[b]{Leonardo Pati\~no,}
\author[a]{and Diego Trancanelli} 
\affiliation[a]{Instituto de F\'isica, Universidade de S\~ao Paulo, 05314-970 S\~ao Paulo, Brazil} 
\affiliation[b]{Departamento de F\'isica, Facultad de Ciencias, Universidad Nacional Aut\'onoma de M\'exico, A.P. 50-542, M\'exico D.F. 04510, M\'exico}

\abstract{
We extend the analysis of {\tt 1211.2199}, where the photon production rate of an anisotropic strongly coupled plasma with $\nf\ll \nc$ massless quarks was considered. We allow here for non-vanishing quark masses and study how these affect the spectral densities and conductivities. We also compute another important probe of the plasma, the dilepton production rate. We consider generic angles between the anisotropic direction and the photon and dilepton wave vectors, as well as arbitrary quark masses and arbitrary values of the anisotropy parameter. Generically, the anisotropy increases the production rate of both photons and dileptons, compared with an isotropic plasma at the same temperature.
}
  

\emailAdd{viktor@if.usp.br} 
\emailAdd{andresluna@ciencias.unam.mx} 
\emailAdd{leopj@ciencias.unam.mx} 
\emailAdd{dtrancan@fma.if.usp.br} 
  
\begin{document}  

\maketitle
\setlength{\parskip}{3pt}


\section{Introduction}

The quark-gluon plasma (QGP) produced in the relativistic scattering of heavy ions at RHIC \cite{rhic,rhic2} and LHC \cite{lhc} seemingly behaves as a strongly coupled fluid \cite{fluid,fluid2}, thus rendering a perturbative approach in small $\alpha_\mt{S}$ problematic. In the absence of other reliable computational tools,\footnote{The lattice approach to QCD is the preferred choice for studying thermodynamics and equations of state, but  is not particularly well suited to compute dynamical quantities such as transport coefficients.} it is then interesting to apply the AdS/CFT correspondence \cite{duality,duality2,duality3} to such a system and try to extract qualitative feature of its strongly coupled dynamics and, possibly, model independent quantities, such as the shear viscosity to entropy density ratio~\cite{Kovtun:2004de}.   

Besides being strongly coupled, the QGP also exhibits another important characteristic which is worth modeling in a holographic setup, namely the presence of a spatial anisotropy in the initial stages of the evolution. Such an anisotropy has recently been described at strong coupling via a type IIB supergravity black brane solution with an anisotropic horizon \cite{Mateos:2011ix,Mateos:2011tv}.\footnote{This geometry is the finite temperature generalization of the geometry found in \cite{ALT}.} Understanding how this anisotropy affects various physical observables has recently received some attention. Quantities that have been studied include the shear viscosity to entropy density ratio \cite{rebhan_viscosity,mamo}, the drag force experienced by a heavy quark \cite{Chernicoff:2012iq,giataganas}, the energy lost by a quark rotating in the transverse plane \cite{fadafan}, the stopping distance of a light probe \cite{stopping}, the jet quenching parameter of the medium \cite{giataganas,Rebhan:2012bw,jet}, the potential between a quark and antiquark pair, both static \cite{giataganas,Rebhan:2012bw,Chernicoff:2012bu,indians} and in a plasma wind \cite{Chernicoff:2012bu}, including its imaginary part \cite{Fadafan:2013bva}, Langevin diffusion and Brownian motion \cite{langevin,langevin2}, chiral symmetry breaking \cite{Ali-Akbari:2013txa}, and the production of thermal photons \cite{photon,Wu:2013qja,Arciniega:2013dqa}.\footnote{For a review of many of these studies, see \cite{Giataganas:2013lga}.}

This last observable is particularly interesting since it furnishes valuable data about the conditions of the in-medium location of production of the photons. This is because, given the limited spatial extend of the plasma and the weakness of the electromagnetic interaction, photons produced in the plasma escape from it virtually unperturbed. Some of the holographic studies of this quantity include \cite{CaronHuot:2006te,Parnachev:2006ev,Mateos:2007yp,Atmaja:2008mt,Bu:2012zza,Jo:2010sg,corr1,corr2,corr25,corr26,corr3,corr31,corr34,corr35}.\footnote{At weak coupling, this has been studied in the presence of anisotropy in, for example, \cite{weak1}.} Here we extend the analysis started in \cite{photon} in two different directions. 

First,  we consider non-equatorial embeddings of the flavor D7-branes introduced in \cite{photon}, corresponding to quarks with non-vanishing masses, thus making our analysis closer to the real-world system. We allow for arbitrary values of the anisotropy parameter $a/T$ and for arbitrary angles between the photon wave vectors and the anisotropic direction, or beam direction. We also study the DC conductivity as a function of the quark mass. We find that, in general, an anisotropic plasma glows brighter than its isotropic counterpart at the same temperature. This holds for all values of the quark masses and for all angles between the anisotropic direction and the photon wave vector. This same computation for a specific value of the anisotropy and for wave vectors either parallel or perpendicular to the anisotropic direction has already been performed in \cite{Wu:2013qja}, where a strong, external magnetic field was also included. 

As a second extension of \cite{photon}, we study thermal production, via virtual photon decay, of lepton/antilepton pairs ({\it dileptons}) in the same background. This quantity is also of phenomenological interest and is obtained by considering time-like momenta for the emitted particles, which can be massive. Compared to the photon production calculation, there is now an extra parameter, namely the magnitude of the spatial momentum. We find that the dilepton production rate is generically larger than the corresponding rate of an isotropic plasma at the same temperature, except for a small range of anisotropies, if the quark mass and the frequency are sufficiently large. These quantities are generically monotonically dependent (either increasing or decreasing) on the angle between the momentum and the anisotropic direction.

The paper is organized as follows. In Sec.~\ref{sec1} we review how to compute the production rate of photons and dileptons in an anisotropic plasma first in the gauge theory side and then via holography using the anisotropic background of \cite{Mateos:2011ix,Mateos:2011tv}. In Sec.~\ref{sec2} we present our results for the spectral densities, conductivities, and total production rates for photons in a plasma with massive quarks. In Sec.~\ref{sec3} we do the same for dileptons, which is essentially the extension of the previous computation to the case in which the emitted particles have a time-like momentum, rather than a light-like one. We discuss our results in Sec.~\ref{sec5}. We relegate to two appendices some technical details of the computation.


\section{Photon and dilepton production in an anisotropic plasma}
\label{sec1}

Here we briefly recall the basic setup of \cite{photon}. The gauge theory we shall consider is obtained via an isotropy-breaking deformation of four-dimensional ${\cal N}=4$ super Yang-Mills (SYM) with gauge group $SU(\nc)$, at large $\nc$ and large 't~Hooft coupling $\lambda=\gym^2\nc$. The deformation consists in including in the action a theta-term which depends linearly on one of the spatial directions, say $z$, \cite{ALT}
\be
S_{SU(\nc)}=S_{{\cal N}=4}+\frac{1}{8\pi^2}\int \theta(z)\, \mathrm{Tr}F\wedge F\,,
\qquad 
\theta(z)\propto z \,,
\label{andef}
\ee
where the proportionality constant in $\theta(z)$ has dimensions of energy and will be related to the parameter $a$ that we shall introduce in the next subsection. The rotational $SO(3)$ symmetry in the space directions is broken by the new term down to $SO(2)$ in the $xy$-plane. For this reason we shall call the $z$-direction the longitudinal (or anisotropic) direction, while $x$ and $y$ will be the transverse directions. This theory has matter fields in the adjoint representation of the gauge group. We can also introduce $\nf$ flavors of scalars $\Phi^a$ and fermions $\Psi^a$ in the fundamental representation, with the index $a=1,\ldots, \nf$. With an abuse of language, we will refer to these fundamental fields indistinctly as `quarks'. 

To study photon production we turn on a dynamical photon by including a $U(1)$ kinetic term in the action (\ref{andef}) and a coupling to the fields that we want to be charged under this Abelian symmetry. In order to realize a situation as similar to QCD as possible, we require that only the fundamental fields be charged, while the adjoint fields are to remain neutral.  We do not know the gravitational dual of the full $SU(\nc)\times U(1)$ theory, but fortunately this will not be necessary for our purposes. It was in fact shown in \cite{CaronHuot:2006te} that to compute the two-point correlation function of the electromagnetic current to leading order in the electromagnetic coupling $\alpha_\mt{EM}$, it is enough to consider the $SU(\nc)$ theory only, whose dual is known from \cite{Mateos:2011ix,Mateos:2011tv}. Our computation will then be to leading order in $\alpha_\mt{EM}$, since the coupling of the photons to the surrounding medium is small, but fully non-perturbative in the 't~Hooft coupling $\lambda$ of the $SU(\nc)$ theory. 

In general, photon production in differential form is given by the expression \cite{lebellac,CaronHuot:2006te,Mateos:2007yp}
\bea
\frac{d\Gamma_\gamma}{d\vec k} = \frac{e^2}{(2\pi)^3 2|\vec k|}\Phi(k)\eta^{\mu\nu}\, \chi_{\mu\nu}(k)\Big|_{k^0=|\vec k|}\,,
\label{difftr}
\eea
with $\eta^{\mu\nu}$ the Minkowski metric (our convention is $(-+++)$), $k^\mu=(k^0,\vec k)$ the photon null momentum and $\Phi(k)$ the distribution function, which for thermal equilibrium, as in our case, reduces to the Bose-Einstein distribution $n_B(k^0)=1/(e^{k^0/T}-1)$. The spectral density is $\chi_{\mu\nu}(k)=-2 \mbox{ Im } G^\mt{R}_{\mu\nu}(k)$, with
\bea
G^\mt{R}_{\mu\nu}(k) = -i \int d^4x \, e^{-i k\cdot x}\, \Theta(t) \vev{[\jem_\mu(x),\jem_\nu(0)]}
\eea
the retarded correlator of two electro-magnetic currents $\jem_\mu$.

If the theory also includes fermions bearing electric charge $e_\ell$, these can be produced in particle/antiparticle pairs (called {\it dileptons} in the following) via virtual photon decay processes. The spectral density above can then be used to compute the dilepton production rate by means of the expression \cite{lebellac}
\bea
\frac{d\Gamma_{\ell\bar{\ell}}}{d k}& =& \frac{e^2{e_\ell}^2}{(2\pi)^4 6\pi | k|^5}\Phi(k)\,\Theta(k_0)\Theta(-k^2-4{m_\ell}^2)(-k^2-4{m_\ell}^2)^{1/2}(-k^2+2{m_\ell}^2)\eta^{\mu\nu}\, \chi_{\mu\nu}(k)\,,\cr &&
\label{difflep}
\eea
where $m_\ell$ is the mass of the lepton/antilepton and the spectral function is now evaluated on the time-like four-momentum $k^\mu$ of the virtual photon.

A consequence of the Ward identity $k^\mu\chi_{\mu\nu}=0$ for null $k^\mu$ is that, for the photon production rate, only the transverse spectral functions contribute. A simple way to extract this contribution is by not taking the whole trace as in (\ref{difftr}), but by summing over the projections into the polarization vectors for the photon that are mutually orthogonal and orthogonal to $\vec k$:
\bea
\frac{d\Gamma_\gamma}{d\vec k} = \frac{e^2}{(2\pi)^3 2|\vec k|}\Phi(k)\sum_{s=1,2} \epsilon^\mu_{(s)}(\vec k)\,  \epsilon^\nu_{(s)}(\vec k)\, \chi_{\mu\nu}(k)\Big|_{k^0=|\vec k|}\,.
\label{diff}
\eea
Each term of the sum stands for the number of photons emitted with polarization vector $\vec\epsilon_{(s)}$.

\begin{figure}
    \begin{center}
        \includegraphics[width=0.65\textwidth]{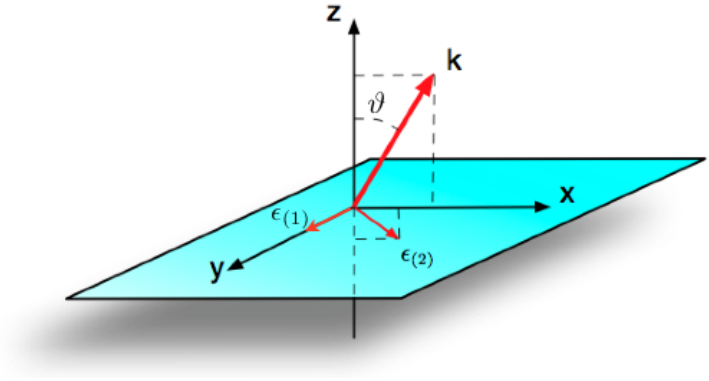}
        \caption{Momentum and polarization vectors. Because of the rotational symmetry in the $xy$-plane, the momentum can be chosen to be contained in the $xz$-plane, forming an angle $\vartheta$ with the $z$-direction. $\vec \epsilon_{(1)}$ is oriented along the $y$-direction and $\vec\epsilon_{(2)}$ is contained in the $xz$-plane, orthogonally to $\vec k$.}
        \label{momentum}
    \end{center}
\end{figure}

Given the $SO(2)$ symmetry in the $xy$-plane, we can choose without  loss of generality $\vec k$ to lie in the $xz$-plane, forming an angle $\vartheta$ with the $z$-direction -- see Fig.~\ref{momentum}. 
Specifically, we set
\be
\vec k = q (\sin \vartheta, 0, \cos \vartheta) \,.
\ee
In the photon production computation it will be $q=k_0$, while in the dilepton production computation $q$ will be an independent parameter. This means that we can choose the polarization vectors as 
\be
\vec\epsilon_{(1)}=(0,1,0)\sac \vec\epsilon_{(2)} = (\cos \vartheta, 0, - \sin \vartheta) \,.
\ee
Production of photons with polarization $\vec\epsilon_{(1)}$ is then proportional to $\chi_{yy} \sim \mbox{Im}\, \langle \jem_y \jem_y \rangle$, whereas for those with polarization $\vec\epsilon_{(2)}$ it is proportional to\footnote{Note that $\chi_{xz}=\chi_{zx}$; see e.g.~\cite{CaronHuot:2006te}.} 
\be
\epsilon^\mu_{(2)}\,  \epsilon^\nu_{(2)}\, \chi_{\mu\nu} = 
\cos^2 \vartheta \, \chi_{xx} + \sin^2 \vartheta \, \chi_{zz} 
- 2 \cos \vartheta \sin \vartheta  \, \chi_{xz} \,. 
\label{combination}
\ee
For the dilepton production, on the other hand, we will just compute the trace of the spectral density, as it appears in (\ref{difflep}). We see then that we need to compute the different correlators $\chi_{\mu\nu}$ of the current for both null and time-like momenta, and plug them in the production densities described above. In the following section we will see how these correlators can be obtained from gravity.


\subsection{Gravity set-up}

The dual gravitational background for the theory (\ref{andef}) at finite temperature is the type IIB supergravity geometry found in \cite{Mateos:2011ix,Mateos:2011tv}, whose string frame metric reads
\bea
ds^2=\frac{L^2}{u^2}\left(-{\cal B} {\cal F} \, dt^2+dx^2+dy^2+{\cal H} dz^2 +\frac{du^2}{{\cal F}}\right)+L^2\, e^{\frac{1}{2}\phi}d\Omega_5^2\,,
\label{metric}
\eea
with ${\cal H}=e^{-\phi}$ and $\Omega_5$ the volume form of a round 5-sphere. The gauge theory coordinates are $(t,x,y,z)$ while $u$ is the AdS radial coordinate, with the black hole horizon lying at $u=\uh$ (where $\cf$ vanishes) and the boundary at $u=0$. As mentioned already, we refer to the $z$-direction as the longitudinal direction and to $x$ and $y$ as the transverse directions. $L$ is set to unity in the following. Besides the metric and the dilaton $\phi$, the forms
\bea
F_5 =4 \left(\Omega_5 + \star \Omega_5\right)\,,\qquad F_1=a \, dz
\eea
are also turned on, with $a$ being a parameter with units of energy that controls the degree of anisotropy of the system. The potential for the 1-form is a linear axion, $\chi=a\, z$. This acts as an isotropy-breaking external source that forces the system into an anisotropic equilibrium state.

\begin{figure}
\begin{center}
\includegraphics[scale=.65]{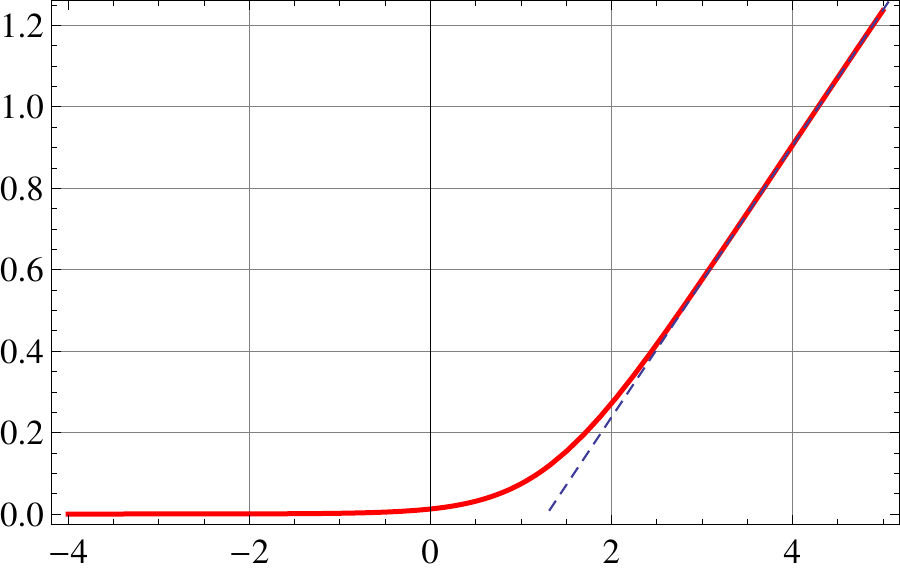}
\put(-40,-10){\small $\log (a/T)$}
\put(-185,50){\rotatebox{90}{\small $\log (s/ s_\mt{iso})$}}
\caption{\small Log-log plot of the entropy density as a function of $a/T$. The dashed blue line is a straight line with slope 1/3.}
\label{scalings}
\end{center} 
\end{figure}

The functions ${\cal B}, {\cal F}$, and $\phi$ depend solely on $u$. They are known analytically in limiting regimes of low and high temperature, and numerically in intermediate regimes \cite{Mateos:2011tv}. For $u\to 0$ (independently of the value of $a$) they asymptote to the $AdS_5\times S^5$ metric, $\cf=\cb=\ch=1$ and $\phi=0$, while for $a=0$ they reduce to the black D3-brane solution
\be
\cb=\ch=1\,, \qquad \phi=\chi=0\,, \qquad \cf=1-\frac{u^4}{\uh^4}\,,
\label{isometric}
\ee
which has temperature  and entropy density given by \cite{peet}
\be
T_\mt{iso}=\frac{1}{\pi\uh}\,, \qquad s_\mt{iso}= \frac{\pi^2}{2} \nc^2 T^3 \,.
\label{siso}
\ee
The temperature and entropy density of the anisotropic geometry are given by \cite{Mateos:2011tv}
\be
T=\frac{e^{-\frac{1}{2}\phi_\mt{H}}\sqrt{{\cal B}_\mt{H}}(16+a^2 \uh^2 e^{\frac{7}{2}\phi_\mt{H}})}{16 \pi \uh}\,,
\qquad
s=\frac{\nc^2}{2\pi\uh^3}e^{-\frac{5}{4}\phi_\mt{H}}\,,
\label{temperature}
\ee
where $\phi_\mt{H}\equiv\phi(u=\uh)$ and $\cb_\mt{H}\equiv\cb(u=\uh)$. As depicted in Fig.~\ref{scalings}, the entropy density of the system interpolates smoothly between the isotropic scaling above for small $a/T$ and the scaling \cite{Mateos:2011tv,ALT}
\be
s \simeq 3.21\, \nc^2 T^3 \left(\frac{a}{T}\right)^\frac{1}{3} \,,
\label{saniso}
\ee
for large $a/T$, the transition between the two behaviors taking place at approximately $a/T\simeq 3.7$. The space can then be interpreted as a domain-wall-like solution interpolating between an AdS geometry in the UV and a Lifshitz-like geometry in the IR, with the radial position at which the transition takes place being set by the anisotropic scale $a$: when $T\gg a$ the horizon lies in the asymptotic AdS region with scaling (\ref{siso}), whereas for $T\ll a$ it lies in the anisotropic region with scaling (\ref{saniso}).

\begin{figure}[tb]
\begin{center}
\includegraphics[scale=0.65]{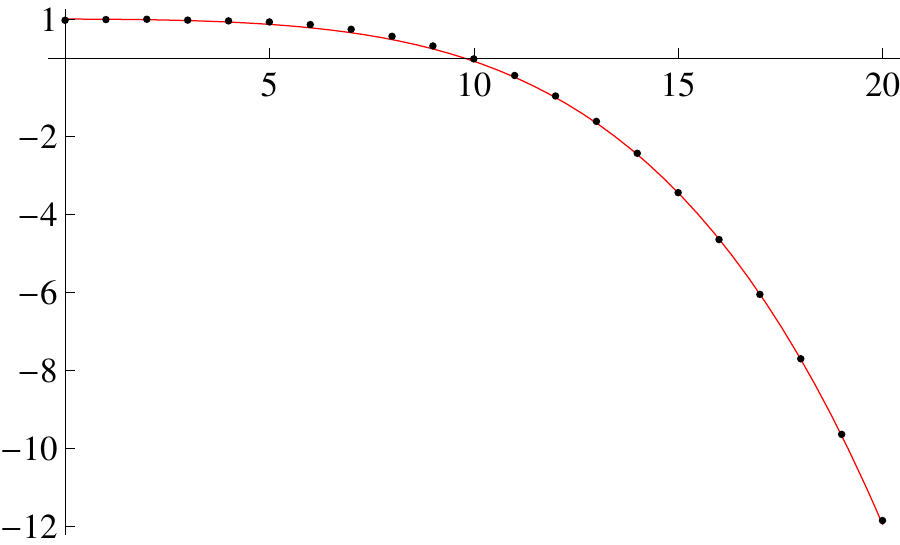}
 \begin{picture}(0,0)
   \put(0,75){$a/T$}
      \put(-185,50){\rotatebox{0}{$\alpha$}}
 \end{picture}
\caption{\small  Ratio \eqn{ratio} as a function of $a/T$. The blue dots are the actual values of the ratio, and the red curve is the fit \eqn{fit}. 
\label{fitplot}
}
 \end{center} 
 \end{figure}
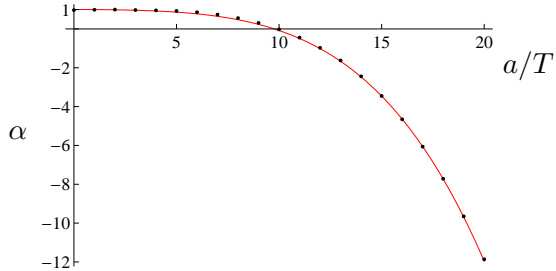

It might be useful to compare the anisotropy introduced in this setup with the anisotropy of other holographic models, or even weak coupling computations. To do this one could consider the following ratio \cite{Chernicoff:2012bu}
\be
\alpha = \frac{4 E + P_\perp -P_\mt{L}}{3 T s} \,,
\label{ratio}
\ee
where $E$  is the energy density and $P_\perp, P_\mt{L}$ are the transverse and longitudinal pressures, respectively. These quantities are presented in great detail in \cite{Mateos:2011tv}. For the isotropic ${\cal N}=4$ super Yang-Mills plasma $\alpha=1$, whereas for $0< a/T \lesssim 20$ the ratio is well approximated by the expression
\be
\alpha \simeq 1 - 0.0036 \left( \frac{a}{T}\right)^2 - 0.000072 \left( \frac{a}{T}\right)^4
\,, \label{fit}
\ee
as shown in Fig.~\ref{fitplot}.

A feature of the anisotropic geometry of \cite{Mateos:2011ix,Mateos:2011tv} is the presence of a conformal anomaly that appears during the renormalization of the theory, introducing a reference scale $\mu$. This anomaly implies that some physical quantities (such as, for example, the energy density and pressures) do not depend only on the ratio $a/T$, but on two independent dimensionless ratios that can be built out of $a$, $T$, and $\mu$. Fortunately, as we shall see in the following, all the quantities computed in this paper are not affected by this anomaly and will be independent of $\mu$.\footnote{The same happens for the quantity $\alpha$ introduced in (\ref{ratio}), which does not depend on $a$ and $T$ separately, but only on the combination $a/T$.} 


The introduction of $N_\mt{f}$ flavors of quarks is achieved by placing $N_\mt{f}$ probe D7-branes in the background (\ref{metric}). To keep the system in a deconfined phase we will work with `black-hole embeddings' \cite{thermobrane} for the D7 branes. The complete system can then be thought of as a D3/D7 system with two different kinds of D7-branes, one kind sourcing the anisotropy \cite{Mateos:2011tv,ALT}%
\footnote{These $N_\mt{D7}$ branes are smeared homogeneously along the $z$-direction and can be thought of as giving rise to a density $n_\mt{D7}=N_\mt{D7}/L_z$ of extended charges, with $L_z$ being the (infinite) length of the $z$-direction. This charge density is related to the anisotropy parameter $a$ through $a=\gym^2 n_\mt{D7}/4\pi$ \cite{Mateos:2011tv}.
\label{footnote}} 
and the other kind sourcing flavor \cite{flavor1,flavor2}; see \cite{photon}. As argued in \cite{CaronHuot:2006te}, at leading order in $\alpha_\mt{EM}$ it suffices to evaluate the correlators needed for (\ref{difftr}) and (\ref{difflep}) in the $SU(\nc)$ gauge theory with no dynamical photons. At strong 't Hooft coupling and large $\nc$, these correlators can be calculated holographically, as we explain now. 

Let $A_m$ $(m=0, \ldots, 7)$ be the gauge field associated to the overall $U(1)\subset U(\nf)$ gauge symmetry on the D7-branes. Upon dimensional reduction on the $3$-sphere wrapped by the flavor D7-branes, $A_m$ gives rise to a massless gauge field  $(A_\mu, A_u)$, three massless scalars, and a tower of massive Kaluza-Klein (KK) modes. All these fields propagate on the five non-compact dimensions of the D7-branes. We will work in the gauge $A_u=0$,\footnote{This gauge choice will be immaterial in the following, since we shall switch to gauge invariant quantities, but it has the advantage of simplifying our formulas.} and we will consistently set the scalars and the higher KK modes to zero, since these are not of interest here. According to the prescription of \cite{duality2,duality3}, correlation functions of $\jem_\mu$ can be calculated by varying the string partition function with respect to the value of $A_\mu$ at the boundary of the spacetime \eqn{metric}. 

We now proceed to write down the action for the D7-branes. It is easy to realize that there is no Wess-Zumino coupling of the branes to the background $F_5$, because of the particular brane orientation that has been chosen, nor a coupling to the background axion, which would be quartic in the $U(1)$ field strength $F=dA$ \cite{photon}.
This means that the Dirac-Born-Infeld (DBI) action is all we need to consider:
\bea
S&=&-\nf \, T_\mt{D7} \int_\mt{D7} \hskip -.2cm d^8\sigma \, e^{-\phi} \sqrt{-\det\left(g+2\pi \ell_s^2  F\right)}\,,
\label{DBI}
\eea
where $g$ is the induced metric on the D7-branes and $T_\mt{D7} = 1/(2\pi \ell_s)^7 g_s \ell_s$ is the D7-brane tension. To obtain the equations of motion for $A_\mu$, it suffices to expand the action above and use the quadratic part only:
\be
S = - \nf T_\mt{D7} \int_\mt{D7} d^{8}\sigma \, e^{-\phi} \sqrt{-\det g} 
 \, \frac{(2\pi \ell_s^2)^2}{4} F^2  \,,
\label{sdq}
\ee
where $F^2 = F_{mn} F^{mn}$. The embedding of the branes inside the $S^5$ of the geometry can be parametrized by the polar angle $\xi$ of the $S^5$ with $\cos\xi\equiv \psi(u)$. The induced metric on the branes is then given by
\bea
ds^2_\mt{D7}&=& \frac{1}{u^2}\left(-\cf \cb\,   dt^2+dx^2+dy^2+\ch\,  dz^2\right)+
\frac{1-\psi^2+u^2 \cf e^{\frac{1}{2}\phi} \psi'^2}{u^2 \cf(1-\psi^2)}du^2 
\cr && 
+e^{\frac{1}{2}\phi}(1-\psi^2)d\Omega_3^2\,.
\label{indmetric}
\eea
After the dimensional reduction on the three-sphere, the action reduces to
\be
S=-K_\mt{D7}\int dt\, d^{3}\vec{x}du\, M\, F^{mn}F_{mn}
\label{sq5}
\ee
where
\bea
M&=&\frac{e^{-\frac{3}{4}\phi}\sqrt{\mathcal{B}}}{u^{5}}\left(1-\psi^{2}\right)\sqrt{1-\psi^{2}+u^{2}\mathcal{F}e^{\frac{\phi}{2}}\psi'^{2}},
\cr 
K_\mt{D7}&=&2 \pi^4  N_\mt{f}T_\mt{D7} \ell_s^4=\frac{1}{16\pi^2}\nc\nf\,,
\eea
and $F_m$ is restricted to the components  $m=(\mu,u)$.

As argued in \cite{thermobrane,Mateos:2007yp}, in order to calculate the photon emission rate, we may consistently proceed by finding the embedding of the D7-branes that extremizes (\ref{DBI}) in the absence of the gauge field, and then solving for the gauge field perturbations propagating on that embedding considered as a fixed background. By checking that the gauge field obtained in this way does not grow beyond the perturbation limit, we can ensure that no modes of the metric or of the background fields will be excited when following this procedure. We set to zero the components of the gauge field on the three-sphere wrapped by the D7-branes and Fourier decompose the remaining as 
\be
A_\mu(t, \vec x, u) = \int \frac{d k^0 d \vec k}{(2\pi)^{4}} \, 
e^{-i k^0 t + i \vec k \cdot \vec x} \, A_\mu (k^0, \vec k, u) \,,\qquad \vec k=(k_x,0,k_z)=q(\sin \vartheta, 0, \cos \vartheta)\,.
\label{fourier}
\ee
This is possible because the state we consider, although anisotropic, is translationally invariant along the gauge theory directions \cite{Mateos:2011tv}. As mentioned above, in the photon production computation it will be $q=k_0$, while in the dilepton production computation $q$ will be an independent parameter.

Doing so, the equations for the gauge field deriving from (\ref{sq5}) split into the following decoupled equation for $A_y$ (primes denote derivatives with respect to $u$)
\bea
\left(M g^{uu}g^{yy}A'_y\right)'-M g^{yy}\left(g^{tt}k_0^2+g^{xx}k_x^2+g^{zz}k_z^2\right)A_y=0\,,
\label{eomy1}
\eea
together with a coupled system of three equations for the  remaining components $A_{t,x,z}$:
\begin{eqnarray}
&& \hskip -.7cm  
(M g^{uu}g^{tt}A'_t)'-M g^{tt}\left[ g^{xx}k_x(k_xA_t-k_0A_x)+g^{zz}k_z(k_zA_t-k_0A_z)\right] = 0\,, 
\label{eom1} \\
&& \hskip -.7cm 
(M g^{uu}g^{xx}A'_x)'-M g^{xx}\left[ g^{tt}k_0(k_0A_x-k_xA_t)+g^{zz}k_z(k_zA_x-k_xA_z)\right] = 0\,, 
\label{eom2} \\
&& \hskip -.7cm 
(M g^{uu}g^{zz}A'_z)'-M g^{zz}\left[ g^{tt}k_0(k_0A_z-k_zA_t)+g^{xx}k_x(k_xA_z-k_zA_x)\right] = 0\, . 
\label{eom3}
\end{eqnarray}
The inverse metric can be read off directly from (\ref{indmetric}). Equations (\ref{eomy1})-(\ref{eom3}) constitute the set of equations that we shall solve in the next sections, with the appropriate boundary conditions, to obtain the correlation functions of the electromagnetic currents $\jem_\mu$.


\subsection{Quark masses}

\begin{figure}[h!]
    \begin{center}
        \includegraphics[width=0.45\textwidth]{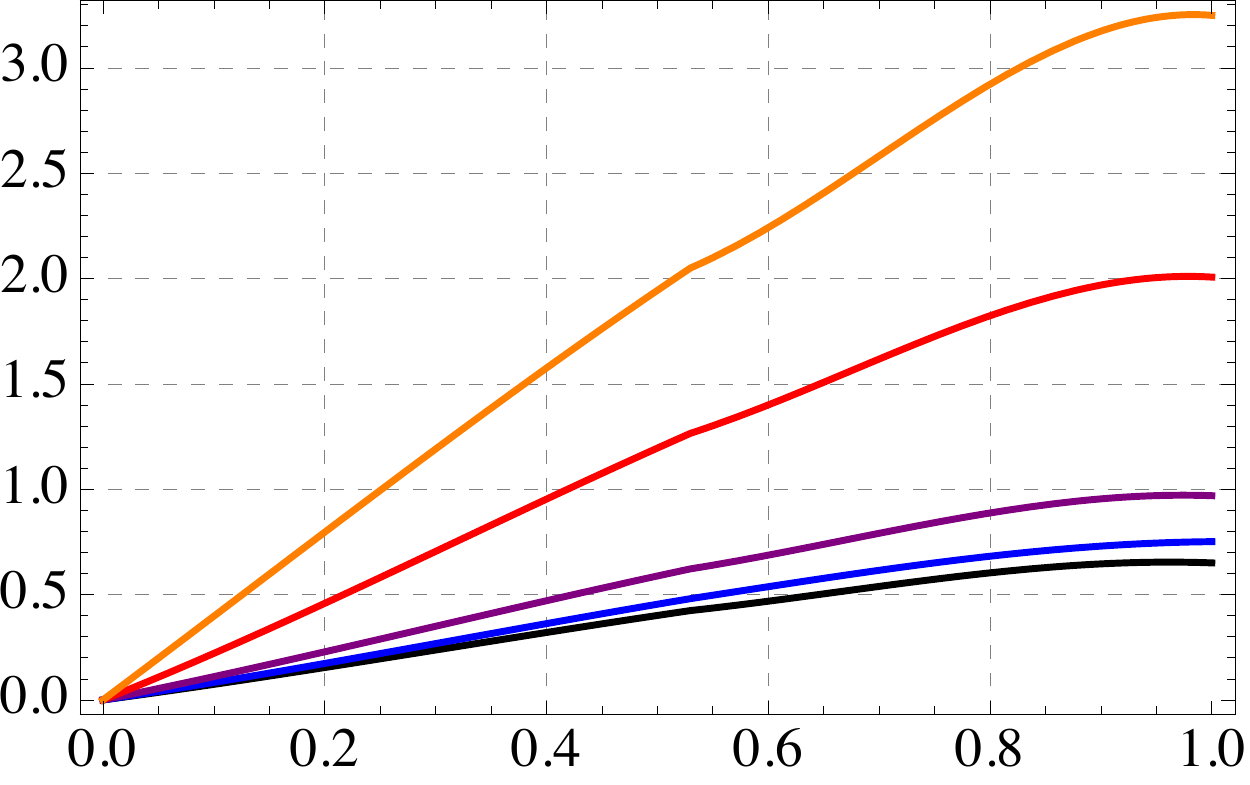}
         \put(-250,100){$M_\mt{q}/\sqrt{\lambda} T$}
         \put(10,5){$\psi_\mt{H}$}
        \caption{The curves correspond, from bottom to top, to $a/T=0,\, 4.41,\, 12.2,\, 86,\, 249$. }
        \label{plotpsi}
    \end{center}
\end{figure}

Given that both $M$ and $g^{uu}$ depend on $\psi$, we need to know this embedding function of the D7-branes to solve (\ref{eomy1})-(\ref{eom3}). The action (\ref{sq5}) for the D7-branes in the absence of the gauge field, takes the form
\begin{equation}
S_\psi=-K_\mt{D7}\int dt\, d\vec{x}\,du\, M_{0}\left(1-\psi^{2}\right)\sqrt{1-\psi^{2}+u^{2}\mathcal{F}e^{\frac{\phi}{2}}\psi'^{2}}\label{eq:accion embedding},
\end{equation}
where
\begin{equation}
M_{0}=\frac{e^{-\frac{3}{4}\phi}\sqrt{\mathcal{B}}}{u^{5}}\,.
\end{equation}
By varying $S_\psi$ with respect to $\psi(u)$ one obtains the equation for the D7-branes embedding 
\begin{equation}
\left(\frac{M_{0}\left(1-\psi^{2}\right)u^{2}\mathcal{F}e^{\frac{\phi}{2}}\psi'}{\sqrt{1-\psi^{2}+u^{2}\mathcal{F}e^{\frac{\phi}{2}}\psi'^{2}}}\right)'+M_{0}\frac{3\psi\left(1-\psi^{2}\right)+2u^{2}\mathcal{F}e^{\frac{\phi}{2}}\psi\psi'^{2}}{\sqrt{1-\psi^{2}+u^{2}\mathcal{F}e^{\frac{\phi}{2}}\psi'^{2}}}=0\,.\label{psiEOM}
\end{equation}
This equation can be solved near the boundary $u = 0$ using the near-boundary expansions of the metric \cite{Mateos:2011tv}
\bea
\mathcal{F}&=&1+\frac{11}{24}a^{2}u^{2}+\mathcal{F}_{4}u^{4}+\frac{7}{12}a^{4}u^{4}\mathrm{log}\, u+O\left(u^{6}\right)\label{eq:expfondo0}\,,\cr
\mathcal{B}&=&1-\frac{11}{24}a^{2}u^{2}+\mathcal{B}_{4}u^{4}-\frac{7}{12}a^{4}u^{4}\mathrm{log}\, u+O\left(u^{6}\right)\,,\cr
\phi&=&-\frac{a^{2}}{2}u^{2}+\left(\frac{1152\mathcal{B}_{4}+121a^{4}}{4032}\right)u^{4}-\frac{a^{4}}{6}u^{4}\mathrm{log}u+O\left(u^{6}\right)\,,
\label{expansions}
\eea
where $\cf_4$ and $\cb_4$ are integration constants which are undetermined by the boundary equations of motion, but that can be read off from the numerics \cite{Mateos:2011tv}. The result for the near-boundary expansion of $\psi(u)$ is
\be
\psi=\psi_{1}u+\left(\psi_{3}+\frac{5}{24}a^{2}\psi_{1}\mathrm{log}u\right)u^{3}+O\left(u^{5}\right)\,,
\label{psibdry}
\ee
where $\psi_1$ and $\psi_3$ are related to the quark mass and condensate, respectively. To solve (\ref{psiEOM}), we follow \cite{Mateos:2007yp} and specify the boundary conditions at the horizon as $\psi(\uh)=\psi_\mt{H}$ and $\psi'(\uh)=0$. We determine $\psi_1$ and $\psi_3$ by fitting the numerical solution near the boundary. 
The relation between $\psi_1$ and the quark mass is given by \cite{Mateos:2007yp}
\be
M_\mt{q}=\sqrt{\lambda} \, T\,  \uh \frac{\psi_1}{\sqrt{2}}\,,
\ee
and the explicit dependence of  the dimensionless ratio $M_\mt{q}/\sqrt{\lambda} T$ for given $\psi_\mt{H}$ and $a/T$ is detailed in Fig.~\ref{plotpsi}.

Note that in the isotropic case  the maximum value of $\psi_\mt{H}$ corresponding to a stable embedding of the D7-branes was $\psi_\mt{H}=0.941$ \cite{thermobrane}. In the presence of anisotropy this will presumably change and some of the higher values of $\psi_\mt{H}$ might correspond to metastable or unstable embeddings.\footnote{All the values of $\psi_\mt{H}$ we have considered result, however, in numerically stable evaluations.} To settle this issue one should analyze phase transitions between black-hole and Minkowski embeddings in the presence of anisotropy, which is something that goes beyond the scope of the present paper.
		

\section{Photon production with massive quarks from holography}
\label{sec2}

In this section we extend the massless quark analysis of \cite{photon} to non-vanishing quark masses and we refer the reader to that reference for more details. The motivation for this extension is to bring our analysis closer to the real-world system studied in the RHIC and LHC experiments.

To compute the various correlation functions,\footnote{The original references for this prescription include \cite{recipe,PSS,PSS2,KS}.} we start by writing the boundary action as%
\be
S_{\epsilon}=-2K_\mt{D7}\int dt\,d\vec{x}\left[Mg^{uu}\left(g^{tt}A_{t}A_{t}'+g^{xx}A_{x}A_{x}'+g^{yy}A_{y}A_{y}'+g^{zz}A_{z}A_{z}'\right)\right]_{u=\epsilon}\label{eq:accion bdry}\,,
\ee
where the limit $\epsilon \to 0$ is intended.

\subsection{Spectral density for the polarization $\epsilon_{(1)}$}

As in the massless case, the spectral density $\chi_{yy}\equiv\chi_{(1)}$ is the easiest to compute, since $A_y$ does not couple to any other mode. The calculation is very similar to the one in \cite{photon}, the only difference being that now the induced metric has a non-trivial brane embedding, $\psi(u)\neq 0$, contained in the new expression for $M$ and $g^{uu}$.  

\begin{figure}[h!]
    \begin{center}
        \includegraphics[width=0.5\textwidth]{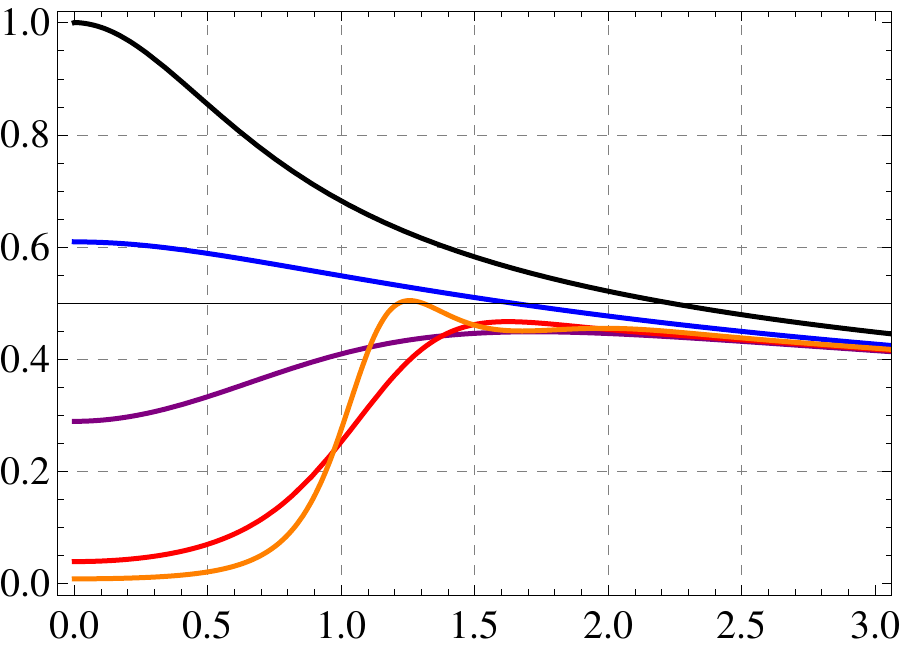}
         \put(-240,70){\rotatebox{90}{$\chi_\mt{iso}/8\tilde{\cal N}_\mt{D7}\,\wn$}}
         \put(15,0){$\wn$}
        \caption{The isotropic correlator $\chi_\mt{iso}$ for $T=0.33$ and, from top to bottom on the left side of the plot, $\psi_\mt{H}=$0 (black), 0.53 (blue), 0.75 (purple), 0.941(red), 0.98 (orange). Here $\tilde{\cal N}_\mt{D7}=4 K_\mt{D7}/\uh^2$ and $\wn=k_0/2\pi T$ is the dimensionless frequency. This color code will be respected throughout this section.}
        \label{plotiso1}
    \end{center}
\end{figure}

Before proceeding further, we recall the isotropic result of \cite{Mateos:2007yp}, since ultimately we want to understand whether the presence of an anisotropy increases or decreases the isotropic photon production and conductivity. Unfortunately, it does not seem possible to obtain analytical results for the spectral density when the quark mass is not zero. One then needs to resort to numerics. In order to compare an anisotropic plasma with the isotropic one, we need that both be at the same temperature. We fix the temperature in the isotropic case by adjusting the position of the black hole horizon, since $T_\mt{iso}=1/\pi \uh$. We then obtain isotropic plots corresponding to the particular temperatures used in the anisotropic geometry. More specifically, we are using $T=0.33,\, 0.36,\, 0.48,\, 0.58$ which are the temperatures for the geometries with $a/T=4.41,\, 12.2,\, 86,$ and 249, respectively, that we consider below. The results for $T=0.33$ are plotted in Fig.~\ref{plotiso1}, for the various masses of interest.

In principle we could also compare the anisotropic plasma with an isotropic plasma at the same entropy density but different temperature. We have checked that the quantities studied in this paper do not depend strongly on whether the comparison is made at the same temperature or at the same entropy density, unlike what happened for other observables, as the ones studied in \cite{Chernicoff:2012iq,jet,Chernicoff:2012bu}. For this reason we do not include here plots with curves normalized with an isotropic plasma with the same entropy density.

\begin{figure}
\begin{center}
\begin{tabular}{cc}
\setlength{\unitlength}{1cm}
\hspace{-0.9cm}
\includegraphics[width=7cm]{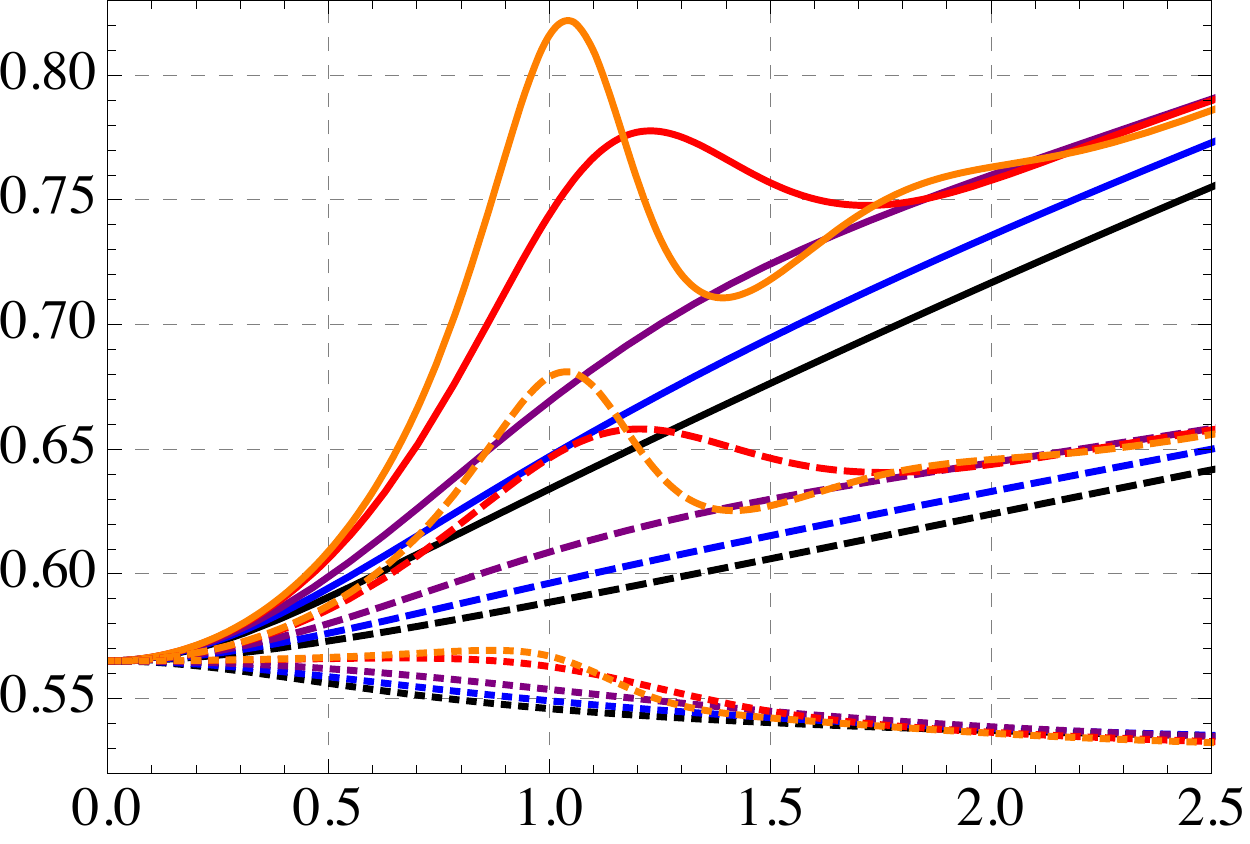} 
\qquad\qquad & 
\includegraphics[width=7cm]{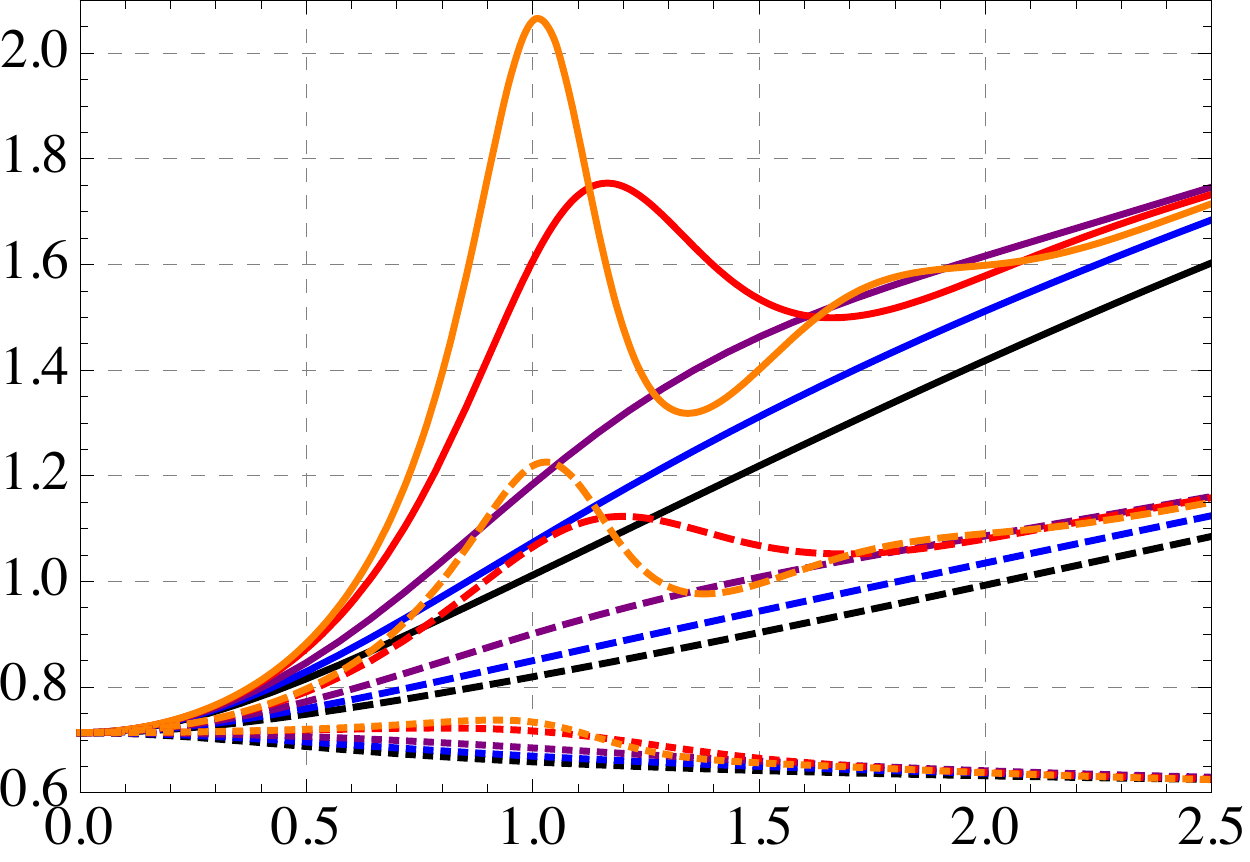}
\qquad
  \put(-455,40){\rotatebox{90}{$\chi_{(1)}/\chi_\mt{iso}(T)$}}
         \put(-250,-10){$\wn$}
         \put(-215,40){\rotatebox{90}{$\chi_{(1)}/\chi_\mt{iso}(T)$}}
         \put(-17,-10){$\wn$}
\\
(a) & (b)\\
& \\
\hspace{-0.9cm}
\includegraphics[width=7cm]{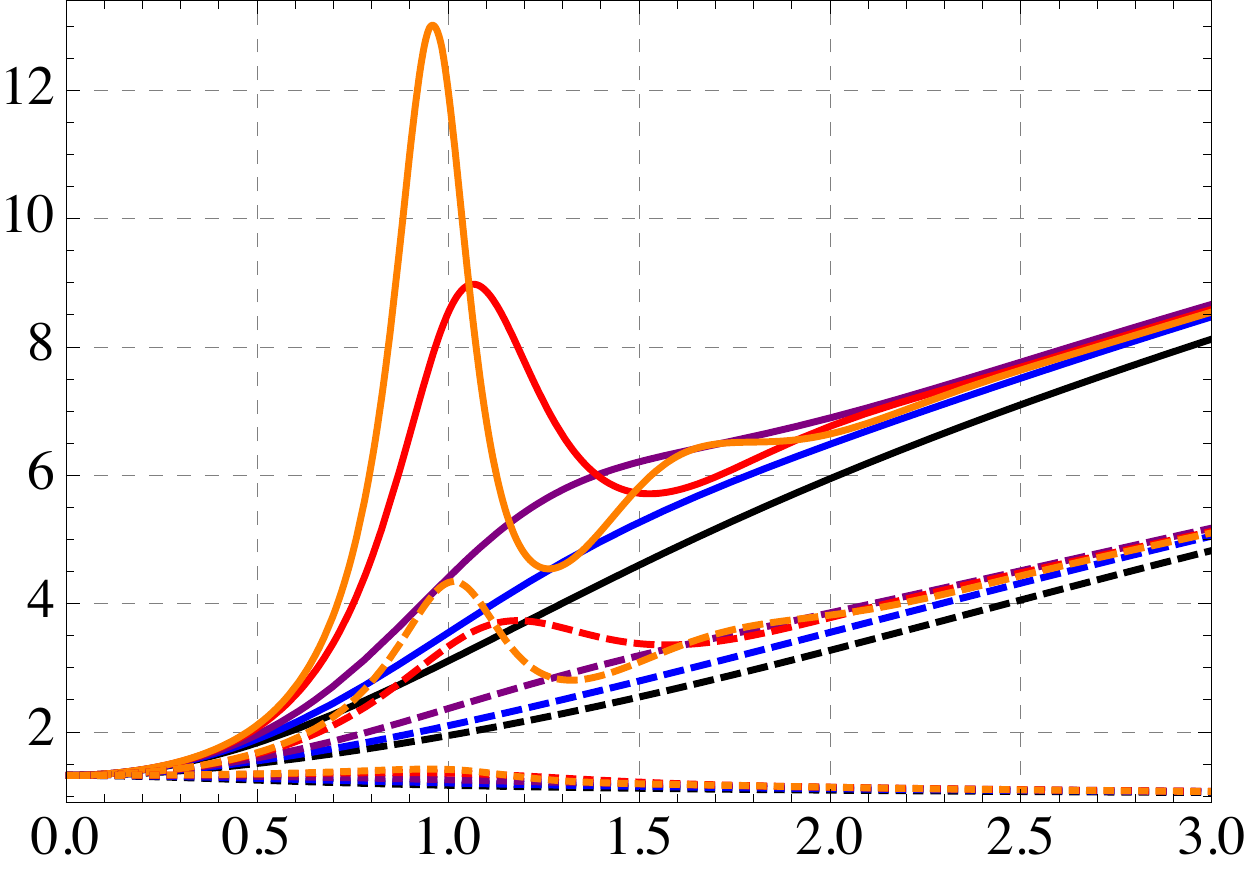} 
\qquad\qquad & 
\includegraphics[width=7cm]{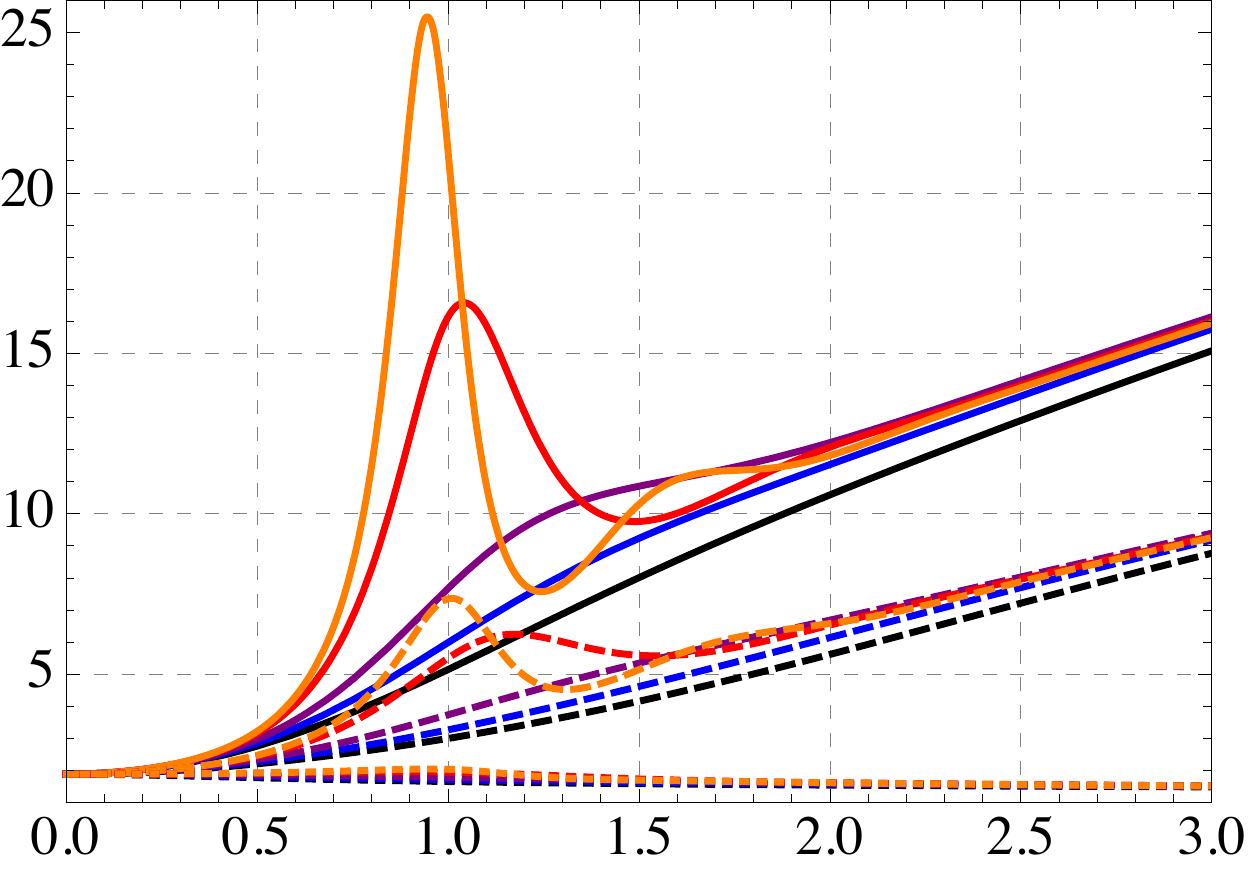}
\qquad
 \put(-455,40){\rotatebox{90}{$\chi_{(1)}/\chi_\mt{iso}(T)$}}
         \put(-250,-10){$\wn$}
         \put(-215,40){\rotatebox{90}{$\chi_{(1)}/\chi_\mt{iso}(T)$}}
         \put(-17,-10){$\wn$}
         \\
(c)& (d) 
\end{tabular}
\end{center}
\caption{\small Plots of the spectral density $\chi_{(1)}$ corresponding to the polarization $\epsilon_{(1)}$, normalized with respect to the isotropic result at fixed temperature $\chi_\mt{iso}(T)$. The curves correspond from top to bottom to the angles $\vartheta=0$ (solid), $ \pi/4$ (dashed), $ \pi/2$ (dotted). Within each group of curves the values of the mass are given, from bottom to top on the right side (black to orange), by $\psi_\mt{H}=0,\, 0.53,\, 0.75,\, 0.941,\, 0.98$. The four plots correspond to the cases $a/T=4.41$ (a), $12.2$ (b), $86$ (c), $249$ (d).}
\label{cyyplotT}
\end{figure}

The correlation function is given by
\begin{equation}
G_{yy}^\mt{R}=-\frac{4K_\mt{D7}}{\left|A_{y}\left(k_{0},0\right)\right|^{2}}\lim_{u\rightarrow0}Q\left(u\right)A_{y}^{*}\left(k_{0},u\right)A_{y}'\left(k_{0},u\right),
\end{equation}
where
\begin{equation}
Q\left(u\right)\equiv  Mg^{uu}g^{yy}.
\label{Q}
\end{equation}
The spectral density then reads
\begin{equation}
\chi_{(1)}=\frac{\nc\nf}{2\pi^{2}\left|A_{y}\left(k_{0},0\right)\right|^{2}}\mbox{Im} \lim_{u\rightarrow0}\, Q\left(u\right)A_{y}^{*}\left(k_{0},u\right)A_{y}'\left(k_{0},u\right)\,,
\label{Xyyphot}
\end{equation}
and is plotted in Fig.~\ref{cyyplotT}. The curves are normalized with the results for an isotropic plasma at the same temperature.

\begin{figure}
\begin{center}
\vskip.5cm
\includegraphics[width=7cm]{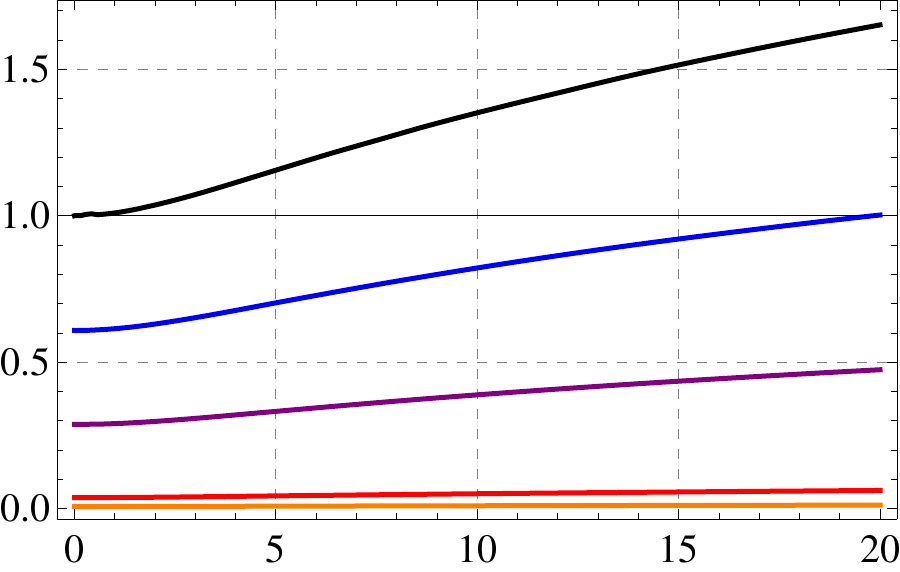} 
          \put(-225,70){$\tilde\sigma_{(1)}$}
         \put(15,5){$a/T$}
\end{center}
 \caption{\small Plot of the conductivity $\tilde\sigma_{(1)}$ corresponding to the polarization $\epsilon_{(1)}$ as a function of $a/T$ for, from top to bottom, $\psi_\mt{H}=0,\, 0.53,\, 0.75,\, 0.941,\, 0.98$.
 }
\label{conductivity}
\end{figure}

The zero-frequency limit of the spectral density gives the electric DC conductivity. For photons with polarization $\epsilon_{(1)}$ this would be the conductivity along the transverse $y$-direction. The quantity
\bea
\sigma_{(1)}(T)=\lim_{k_0\to 0}\frac{\chi_{(1)}}{\chi_{(1),\mt{iso}}(T)}=
\lim_{k_0\to 0}2\frac{\chi_{(1)}}{\chi_{\mt{iso}}(T)}
\label{y_conductivity}
\eea
is mass independent, and therefore given by Fig.~8 of  \cite{photon}. In Fig.~\ref{conductivity} we plot the conductivity
\be
\tilde{\sigma}_{(1)}=2\lim_{\wn\to 0}\,\frac{\chi_{(1)}}{8\tilde{\cal N}_\mt{D7}\wn}
\ee
not normalized with the isotropic result, for the various values of the quark mass. Here $\tilde{\cal N}_\mt{D7}=\nc\nf T^2/4$ and $\wn=k_0/2\pi T$.
We observe that the conductivity decreases as the quark mass increases.

It is worth pointing out that, as it will be discussed in Appendix \ref{concur}, the imaginary part of (\ref{Xyyphot}) is independent of $u$. Numerical accuracy can then be improved by evaluating this quantity at the horizon instead of at the boundary, since we know the analytic values for the metric functions and the ingoing fields at $\uh$.


\subsection{Spectral density for the polarization $\epsilon_{(2)}$}
\label{eps2}

We now move on to compute $\chi_{(2)}$, the correlator corresponding to $\epsilon_{(2)}$. To obtain this, it is easier to work in terms of the gauge invariant fields $E_{i}\equiv\partial_{i}A_{t}-\partial_{t}A_{i}$. Equations (\ref{eomy1})-(\ref{eom3}) can be rewritten in terms of $E_{i}$ with the aid of the constraint 
\begin{equation}
-g^{tt}k_{0}A_{t}'+g^{xx}k_{x}A_{x}'+g^{zz}k_{z}A_{z}'=0\,,
\end{equation}
resulting in
\begin{equation}
E_{x}''+\left[\left(\mathrm{log}Mg^{uu}g^{xx}\right)'+\left(\mathrm{log}\frac{g^{xx}}{g^{tt}}\right)'\frac{k_{x}^{2}}{u^{2}\overline{k}^{2}}g^{xx}\right]E_{x}'+\frac{u^{2}\overline{k}^{2}}{g^{uu}}E_{x}+\left(\mathrm{log}\frac{g^{xx}}{g^{tt}}\right)'\frac{k_{z}k_{x}}{u^{2}\overline{k}^{2}}g^{zz}E_{z}'=0\label{eq:Ex}\,,
\end{equation}
\begin{equation}
E_{z}''+\left[\left(\mathrm{log}Mg^{uu}g^{zz}\right)'+\left(\mathrm{log}\frac{g^{zz}}{g^{tt}}\right)'\frac{k_{z}^{2}}{u^{2}\overline{k}^{2}}g^{zz}\right]E_{z}'+\frac{u^{2}\overline{k}^{2}}{g^{uu}}E_{z}+\left(\mathrm{log}\frac{g^{zz}}{g^{tt}}\right)'\frac{k_{z}k_{x}}{u^{2}\overline{k}^{2}}g^{xx}E_{x}'=0\label{eq:Ez}\,,
\end{equation}
where $u^{2}\overline{k}^{2}\equiv-g^{tt}k_{0}^{2}-g^{xx}k_{x}^{2}-g^{zz}k_{z}^{2}$.
The action (\ref{eq:accion bdry}) can also be written in terms of these
fields as
\begin{multline}
S_{\epsilon}=-2K_\mt{D7}\int dt\,d\vec{x}\,\frac{Mg^{uu}}{-k_{0}^{2}u^{2}\overline{k}^{2}}\left[\left(-g^{tt}k_{0}^{2}-g^{zz}k_{z}^{2}\right)g^{xx}E_{x}E_{x}'+u^{2}\overline{k}^{2}g^{yy}E_{y}E_{y}'+\right.\\
+g^{xx}g^{zz}k_{x}k_{z}\left(E_{x}E_{z}\right)'+\left(-g^{tt}k_{0}^{2}-g^{xx}k_{x}^{2}\right)g^{zz}E_{z}E_{z}'\Biggr]_{u=\epsilon}\label{eq:boundarycampos}\,.
\end{multline}

Since we need to take the limit $\epsilon\rightarrow0$, before proceeding any further, we need to verify that the correlators will remain finite in this limit. To this end we use the near-boundary expansion of the metric (\ref{expansions}) and of the embedding $\psi$ (\ref{psibdry}) to solve the equations (\ref{eq:Ex}) and (\ref{eq:Ez}) perturbatively. We find%
\begin{equation}
E_{x}=E_{x}^{(0)}+E_{x}^{(2)}\cos\vartheta \,u^{2}-\frac{1}{24}\left(\frac{3}{4}E_{x}^{(0)}k_{0}^{2}\cos\vartheta+\left(5-\frac{24\psi_{1}^{2}}{a^{2}}\right)E_{x}^{(2)}\right)\cos\vartheta \, a^{2}\, u^{4}+O\left(u^{6}\right)\,,
\end{equation}
\begin{equation}
E_{z}=E_{z}^{(0)}+E_{x}^{(2)}\sin\vartheta\,  u^{2}+E_{z}^{(4)}u^{4}-\frac{a^{2}k_{0}^{2}\cos\vartheta}{16}\left(E_{z}^{(0)}\cos\theta+E_{x}^{(0)}\sin\vartheta\right)u^{4}\log u+O\left(u^{6}\right)\,.
\end{equation}
Using these expressions, we can rewrite (\ref{eq:boundarycampos}) as
\be
S_{\epsilon}=-2K_\mt{D7}\int dt\, d\vec{x}\left[\mathcal{L}_{1}+\mathcal{L}_{2}+\mathcal{L}_{3}+\mathcal{L}_\mt{m}+\ldots
+O\left(u^{2}\right)\right]_{u=\epsilon}\,,
\ee
where
\bea
\mathcal{L}_{1}&=&-\frac{3}{4}\sin^{2}\vartheta \,E_{x}^{(0)2}-\frac{1}{4}\cos^{2}\vartheta \,E_{z}^{(0)2}-\sin\vartheta\cos\vartheta \,E_{x}^{(0)}E_{z}^{(0)}\,,\cr 
\mathcal{L}_{2}&=&\frac{1}{3k_{0}^{2}}\left[\frac{1+5\cos2\vartheta}{\cos\vartheta}\,E_{x}^{(0)}E_{x}^{(2)}+\frac{48}{a^{2}}\tan\vartheta \,E_{x}^{(0)}E_{z}^{(4)}-10\sin\vartheta \,E_{z}^{(0)}E_{x}^{(2)}+\frac{48}{a^{2}}E_{z}^{(0)}E_{z}^{(4)}\right]\,,\cr
\mathcal{L}_{3}&=&-\left(E_{x}^{(0)}\sin\vartheta+E_{z}^{(0)}\cos\vartheta\right)^{2}\log u\,,\cr
\mathcal{L}_\mt{m}&=&\frac{16\psi_{1}^{2}}{a^{2}k_{0}^{2}}\tan\vartheta\left(E_{x}^{(0)}E_{x}^{(2)}\sin\vartheta+E_{z}^{(0)}E_{x}^{(2)}\cos\vartheta\right)\,,
\eea
and the ellipsis stands for the terms in the $y$-components that have been already dealt with. Notice that $\mathfrak{\mathcal{L}}_{1}$, $\mathfrak{\mathcal{L}}_{2}$, $\mathfrak{\mathcal{L}}_{3}$ are the same as in the $\psi=0$ case of  \cite{photon}.

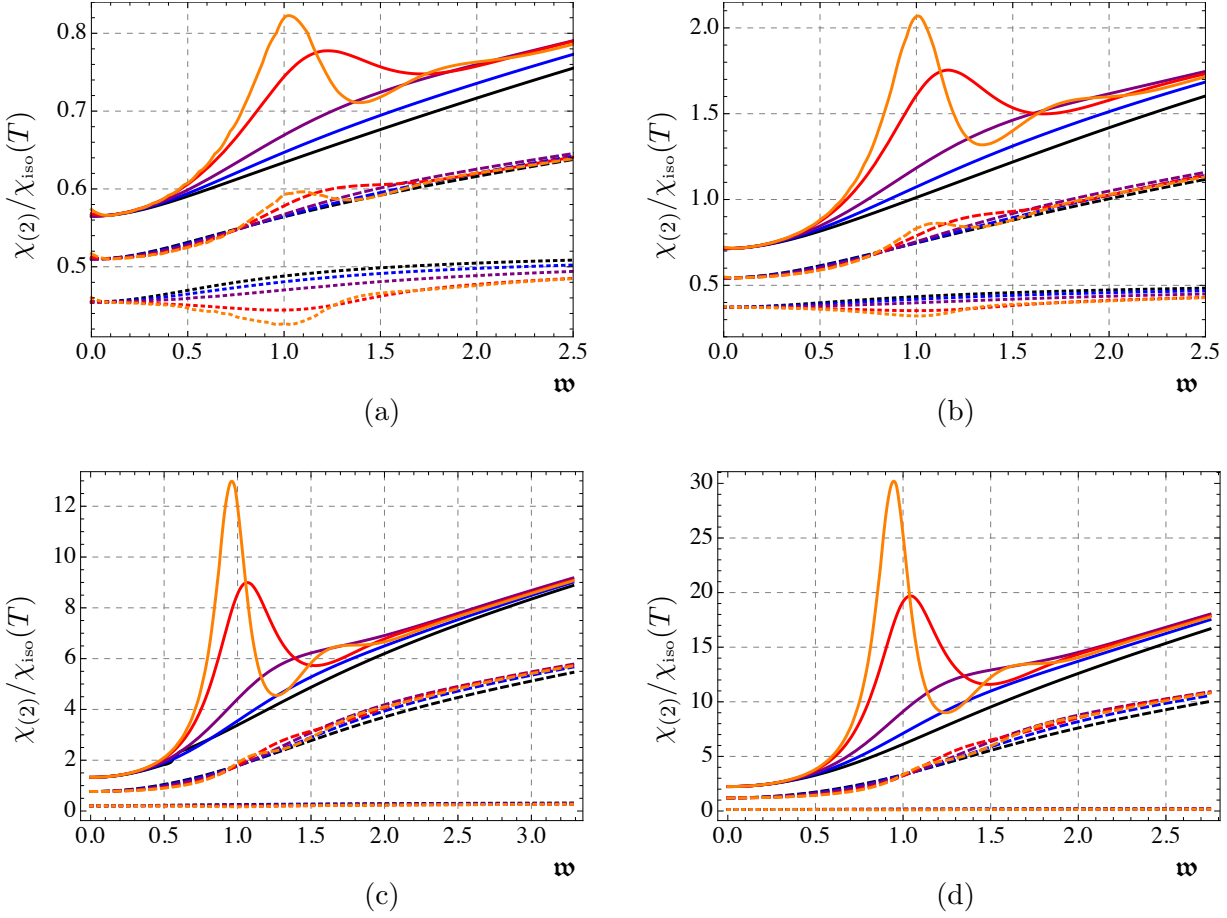
\begin{figure}
\begin{center}
\begin{tabular}{cc}
\setlength{\unitlength}{1cm}
\hspace{-0.9cm}
\includegraphics[width=7cm]{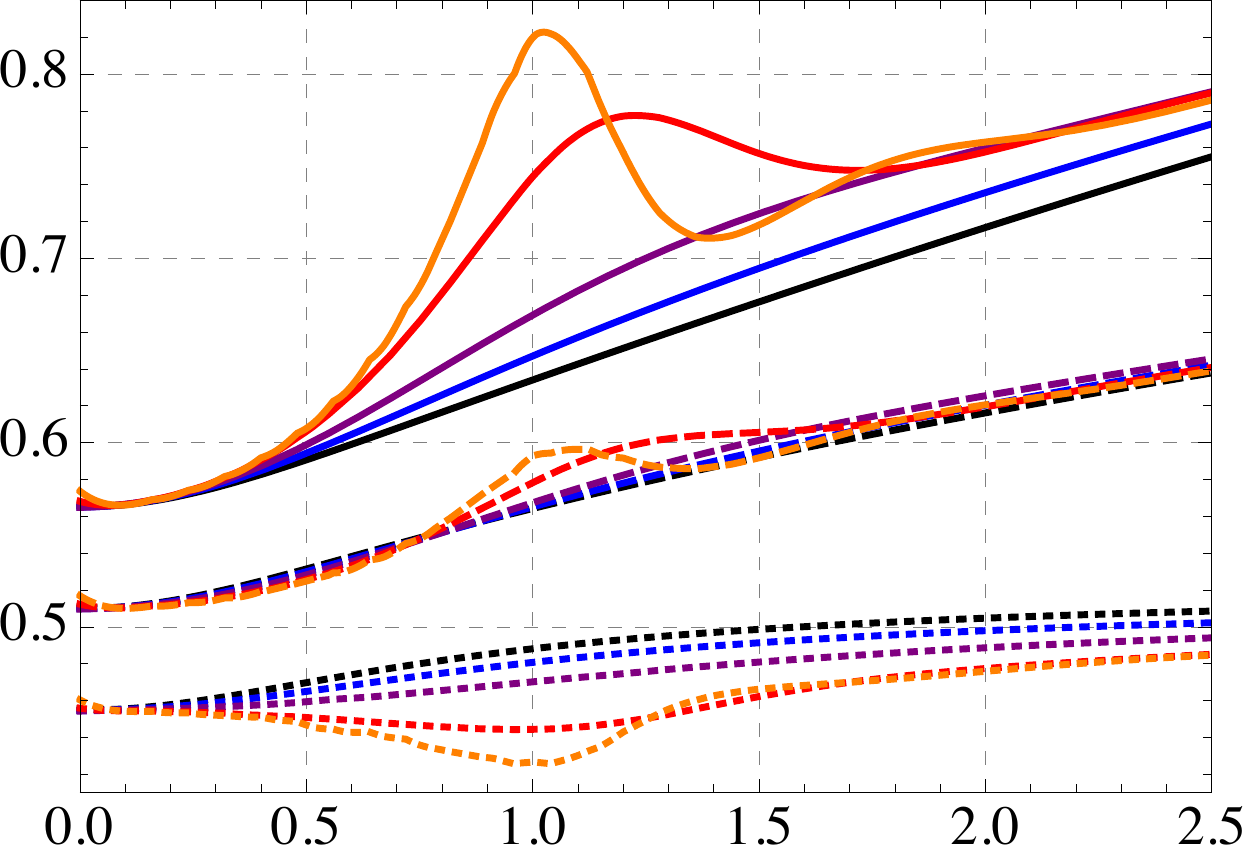} 
\qquad\qquad & 
\includegraphics[width=7cm]{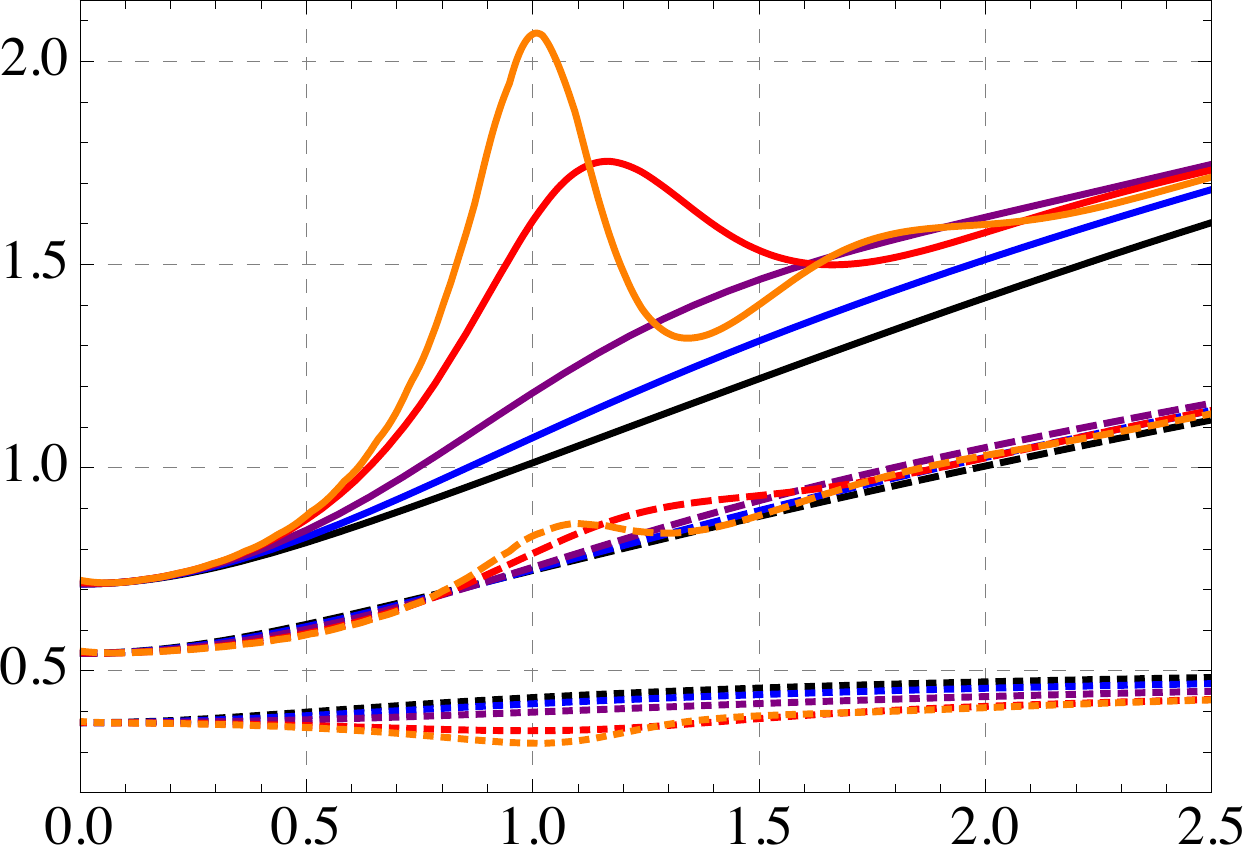}
\qquad
  \put(-455,40){\rotatebox{90}{$\chi_{(2)}/\chi_\mt{iso}(T)$}}
         \put(-250,-10){$\wn$}
         \put(-215,40){\rotatebox{90}{$\chi_{(2)}/\chi_\mt{iso}(T)$}}
         \put(-17,-10){$\wn$}
\\
(a) & (b)\\
& \\
\hspace{-0.9cm}
\includegraphics[width=7cm]{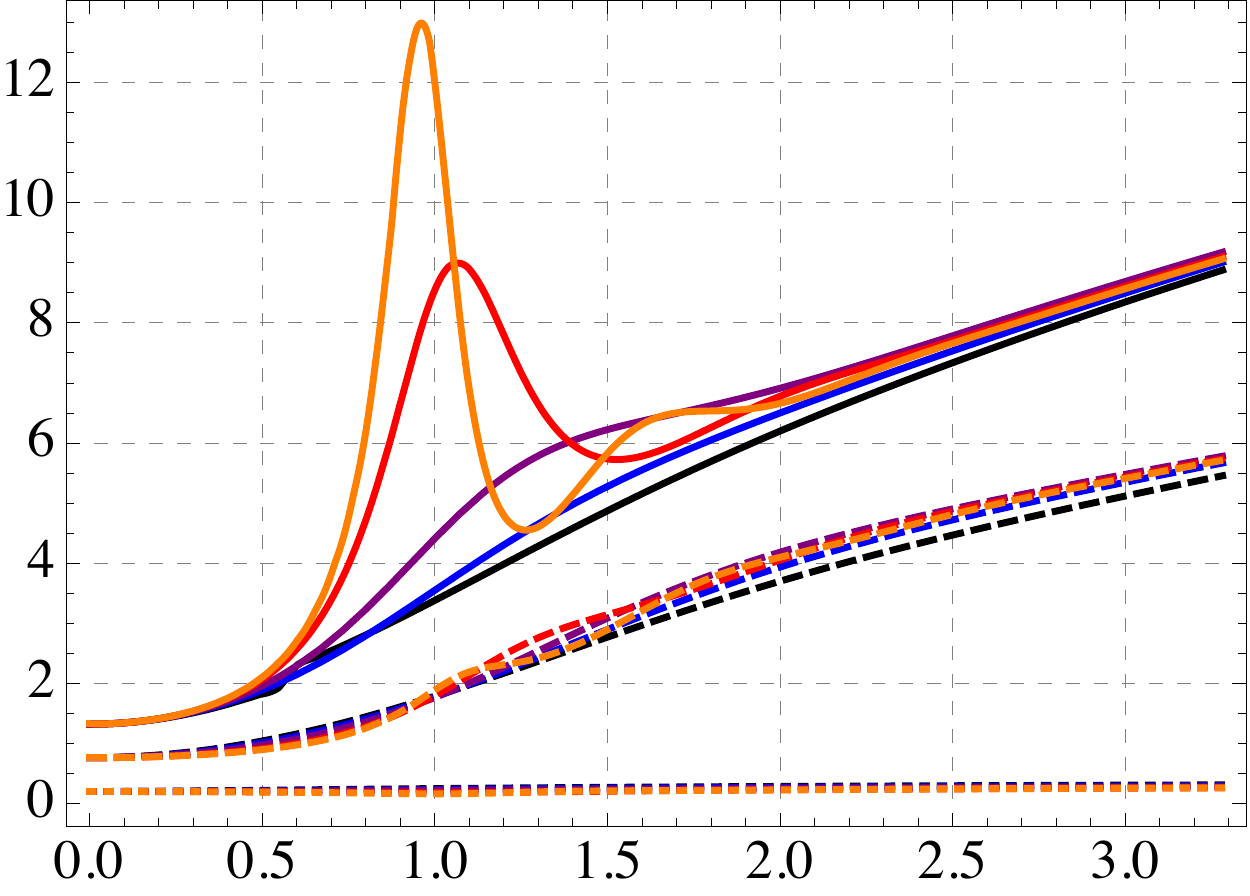} 
\qquad\qquad & 
\includegraphics[width=7cm]{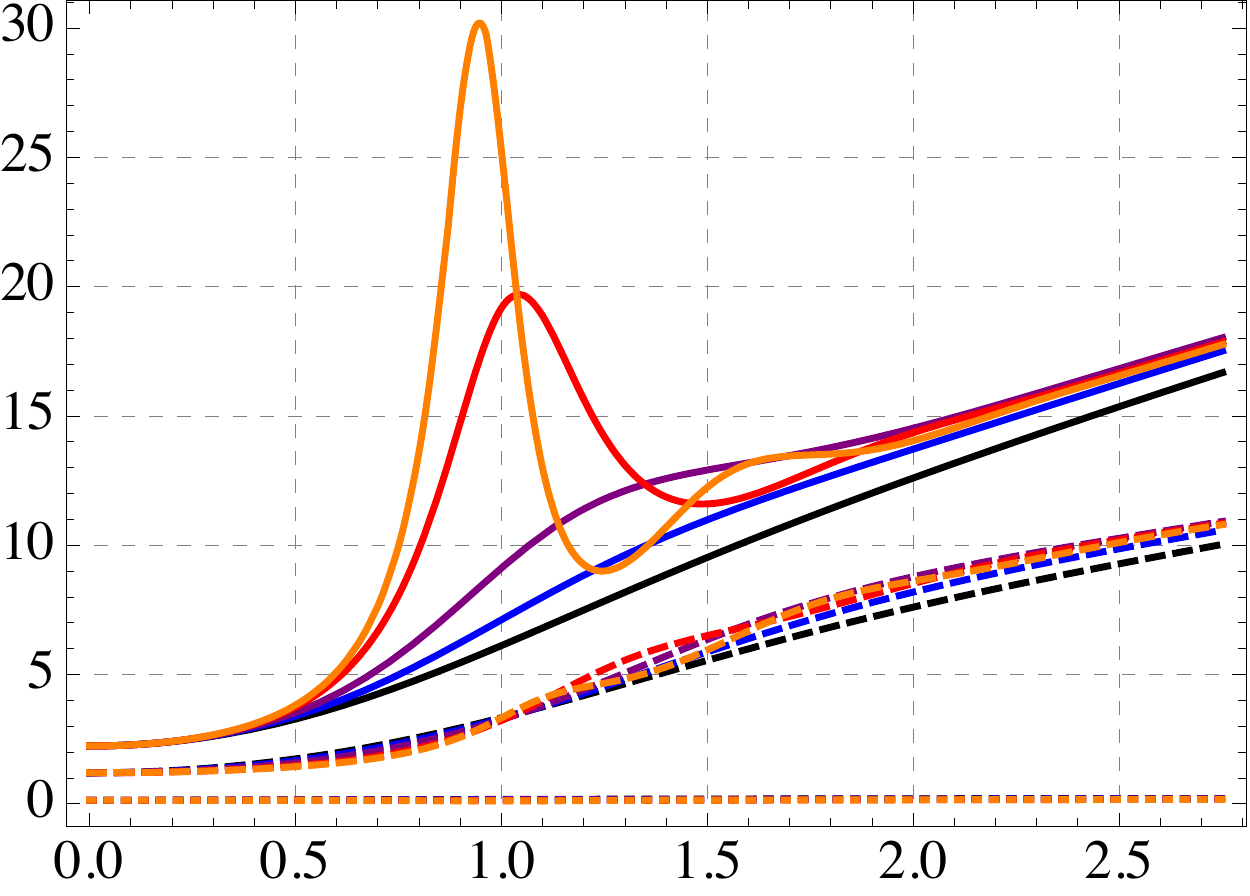}
\qquad
 \put(-455,40){\rotatebox{90}{$\chi_{(2)}/\chi_\mt{iso}(T)$}}
         \put(-250,-10){$\wn$}
         \put(-215,40){\rotatebox{90}{$\chi_{(2)}/\chi_\mt{iso}(T)$}}
         \put(-17,-10){$\wn$}
         \\
(c)& (d) 
\end{tabular}
\end{center}
\caption{\small Plots of the spectral density $\chi_{(2)}$ corresponding to the polarization $\epsilon_{(2)}$, normalized with respect to the isotropic result at fixed temperature $\chi_\mt{iso}(T)$. The curves correspond from top to bottom to the angles $\vartheta=0$ (solid), $\pi/4$ (dashed),  $\pi/2$ (dotted).  Within each group of curves the values of the mass are given, from bottom to top on the right side (black to orange), by $\psi_H=0,\, 0.53,\, 0.75,\, 0.941,\, 0.98$. The four plots correspond to the cases $a/T=4.41$ (a), $12.2$ (b), $86$ (c), $249$ (d).}
\label{mixedT}
\end{figure}

The contribution of $\mathfrak{\mathcal{L}}_\mt{m}$ to the production of photons with polarization $\epsilon_{(2)}$ is proportional to
\be
\cos^{2}\vartheta\frac{\delta^{2}\mathcal{L}_\mt{m}}{\delta E_{x}^{(0)2}}+\sin^{2}\vartheta\frac{\delta^{2}\mathcal{L}_\mt{m}}{\delta E_{z}^{(0)2}}-2\sin\vartheta \cos\vartheta\frac{\delta^{2}\mathcal{L}_\mt{m}}{\delta E_{z}^{(0)}\delta E_{x}^{(0)}}=0\,,
\ee
and therefore vanishes identically, and so does the divergent term ${\cal L}_3$, as shown in \cite{photon}. We obtain then the simple result
\be
\chi_{(2)}\equiv\epsilon_{(2)}^{\mu}\epsilon_{(2)}^{\nu}\chi_{\mu\nu}=16K_\mt{D7}\mathrm{Im}\left[\cos\vartheta\frac{\delta E_{x}^{(2)}}{\delta E_{x}^{(0)}}-\sin\vartheta\frac{\delta E_{x}^{(2)}}{\delta E_{z}^{(0)}}\right]\,.
\label{chi2simplified}
\ee
We can now proceed as in \cite{photon} to determine how $E_{x}^{(2)}$ varies with respect of $E_{x}^{(0)}$ and $E_{z}^{(0)}$. Alternatively, we will explain in Appendix \ref{concur} how to apply the technology developed in \cite{Kaminski:2009dh} to obtain $\chi_{(2)}$ using the values of the fields at the horizon. As a check of our results, we have verified that we obtain the same results using both methods. We display the results in Fig.~\ref{mixedT} for various values of the anisotropy, of the angles, and of the quark masses.

\begin{figure}
\begin{center}
\includegraphics[width=7cm]{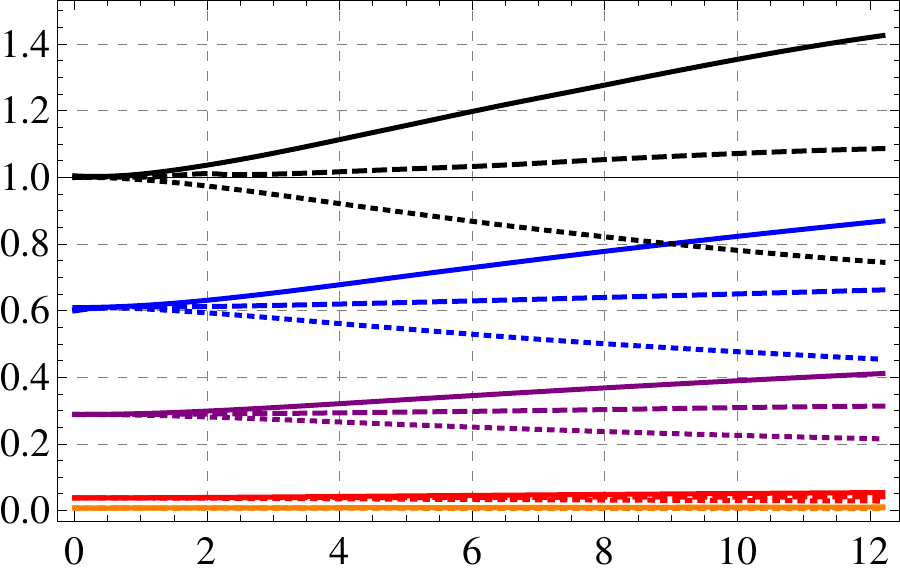} 
          \put(-230,110){$\tilde\sigma_{(2)}$}
         \put(10,10){$a/T$}
\end{center}
 \caption{\small Plot of the conductivity $\tilde{\sigma}_{(2)}$ corresponding to the polarization $\epsilon_{(2)}$ as a function of $a/T$. The groups of curves correspond from top to bottom to $\psi_\mt{H}=0,\, 0.53,\, 0.75,\, 0.941$. Inside each group we plot the angles $\vartheta=0$ (solid), $ \pi/4$ (dashed),  and $\pi/2$ (dotted).}
\label{mixed_conductivity_aoverT}
\end{figure}

\begin{figure}
    \begin{center}
    \vskip .5cm
        \includegraphics[width=7cm]{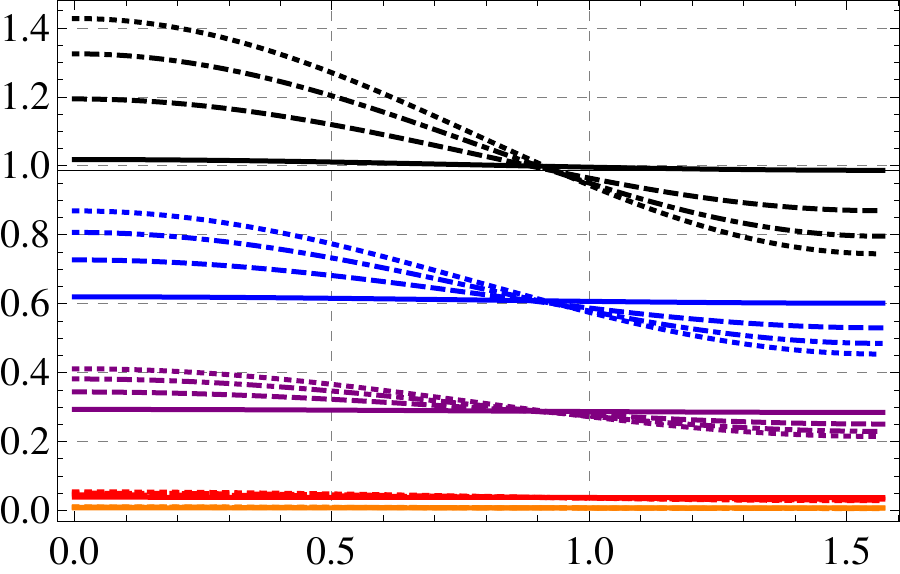}
         \put(-230,110){$\tilde\sigma_{(2)}$}
         \put(10,10){$\vartheta$}
        \caption{\small Plot of the conductivity $\tilde\sigma_{(2)}$ corresponding to the polarization $\epsilon_{(2)}$ as a function of the angle $\vartheta$. The groups of curves correspond, from top to bottom, to $\psi_\mt{H}=0,\, 0.53,\, 0.75,\, 0.941,\, 0.98$. Within each group we have, from bottom to top on the left side of the graph, $a/T=$1.38 (solid), 5.9 (dashed), 9.25 (dot-dashed), and 12.2 (dotted).}
        \label{mixed-conductivity}
    \end{center}
\end{figure}

For photons with polarization along $\epsilon_{(2)}$, the conductivity
\bea
\sigma_{(2)}(T)=\lim_{k_0\to 0}\frac{\chi_{(2)}}{\chi_{(2),\mt{iso}}(T)}=\lim_{k_0\to 0}2\frac{\chi_{(2)}}{\chi_\mt{iso}(T)}
\eea
depends not only on the anisotropy and quark mass, as was the case for the polarization along the $y$-direction, but also on the angle $\vartheta$.  If we normalize with respect to the isotropic case, the conductivity does not depend on the quark masses and is therefore identical to the one depicted in Figs.~11 and 12 of \cite{photon}.  
We can then define unnormalized conductivities, as done above for $\tilde \sigma_{(1)}$,
\be
\tilde{\sigma}_{(2)}=2\lim_{\wn\to 0}\,\frac{\chi_{(2)}}{8\tilde{\cal N}_\mt{D7}\wn}\,,
\ee
which do depend on the masses and are reported in Figs.~\ref{mixed_conductivity_aoverT} (as a function of $a/T$ for fixed $\vartheta$) and \ref{mixed-conductivity} (as a function of $\vartheta$ for fixed $a/T$).


\subsection{Total photon production rate}

\begin{figure}
\begin{center}
\begin{tabular}{cc}
\setlength{\unitlength}{1cm}
\hspace{-0.9cm}
\includegraphics[width=7cm]{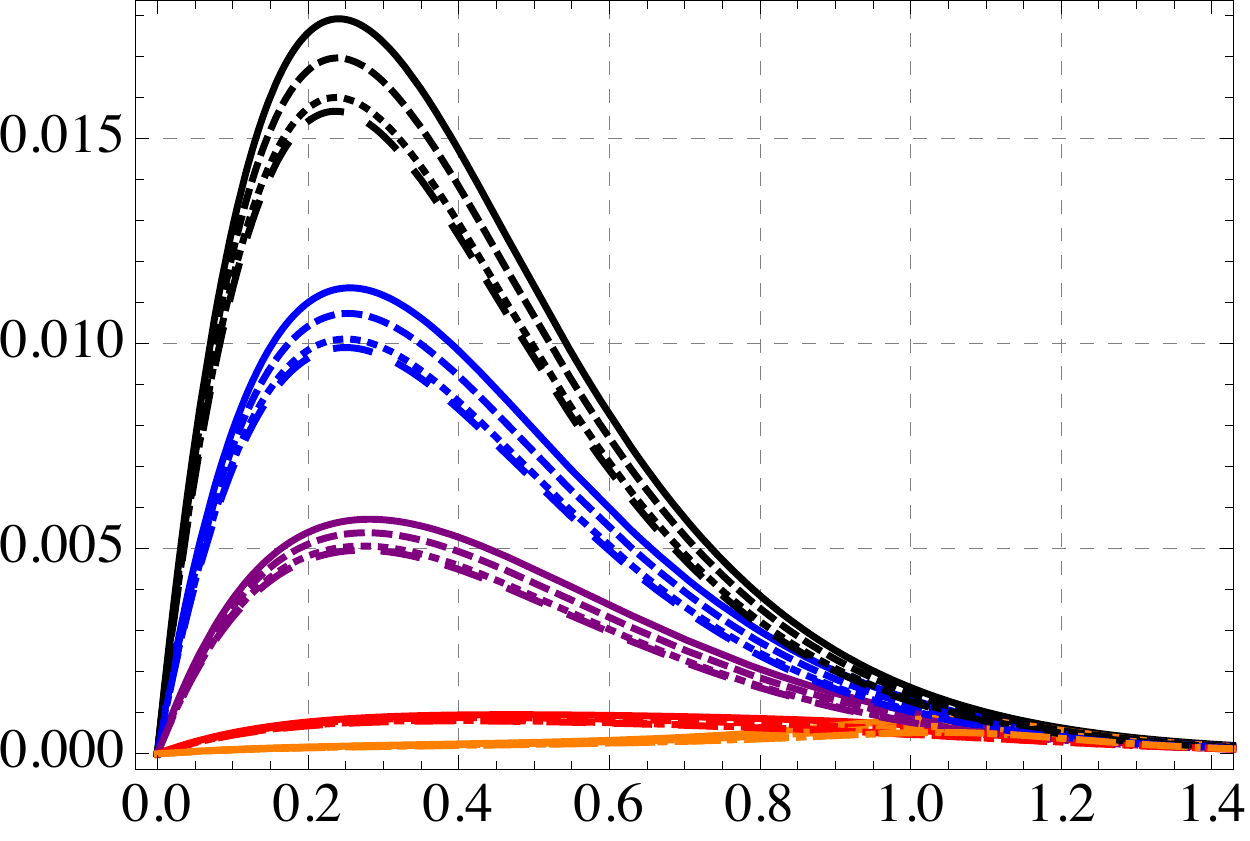} 
\qquad\qquad & 
\includegraphics[width=7cm]{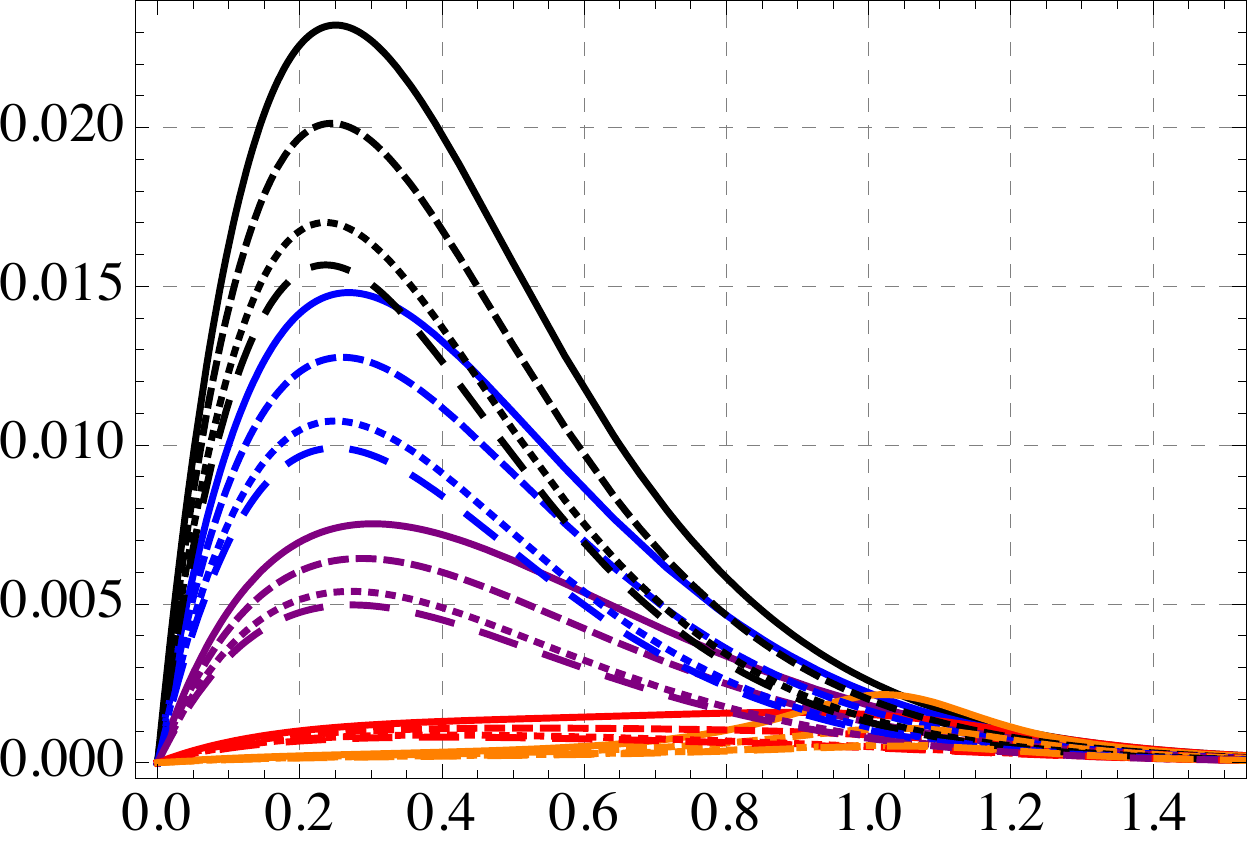}
\qquad
  \put(-460,20){\rotatebox{90}{$\frac{-1}{2\alpha_\mt{EM}\nc\nf T^3}\frac{d\Gamma_\gamma}{d\cos\vartheta\, dk^0}$}}
         \put(-250,-10){$\wn$}
         \put(-220,20){\rotatebox{90}{$\frac{-1}{2\alpha_\mt{EM}\nc\nf T^3}\frac{d\Gamma_\gamma}{d\cos\vartheta\, dk^0}$}}
         \put(-17,-10){$\wn$}
\\
(a) & (b)\\
& \\
\hspace{-0.9cm}
\includegraphics[width=7cm]{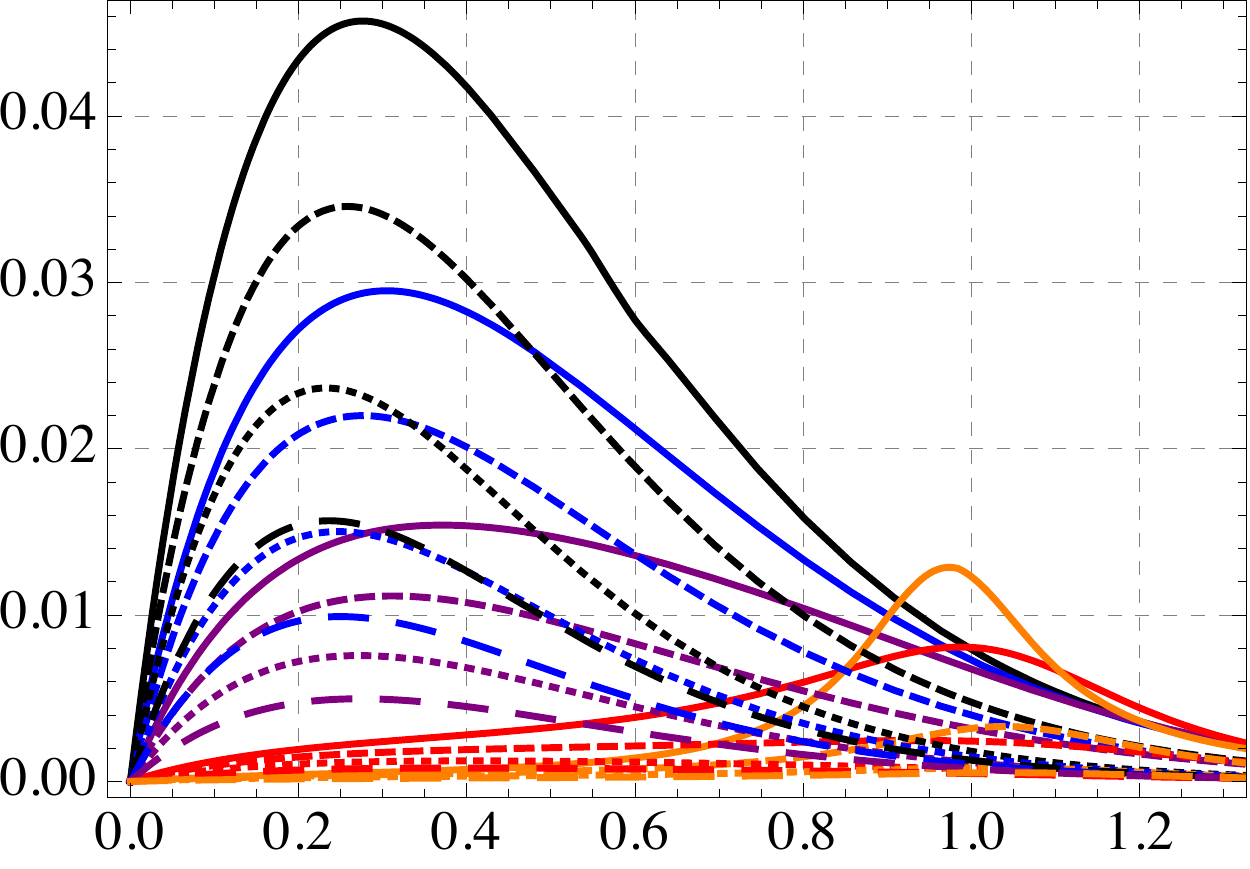} 
\qquad\qquad & 
\includegraphics[width=7cm]{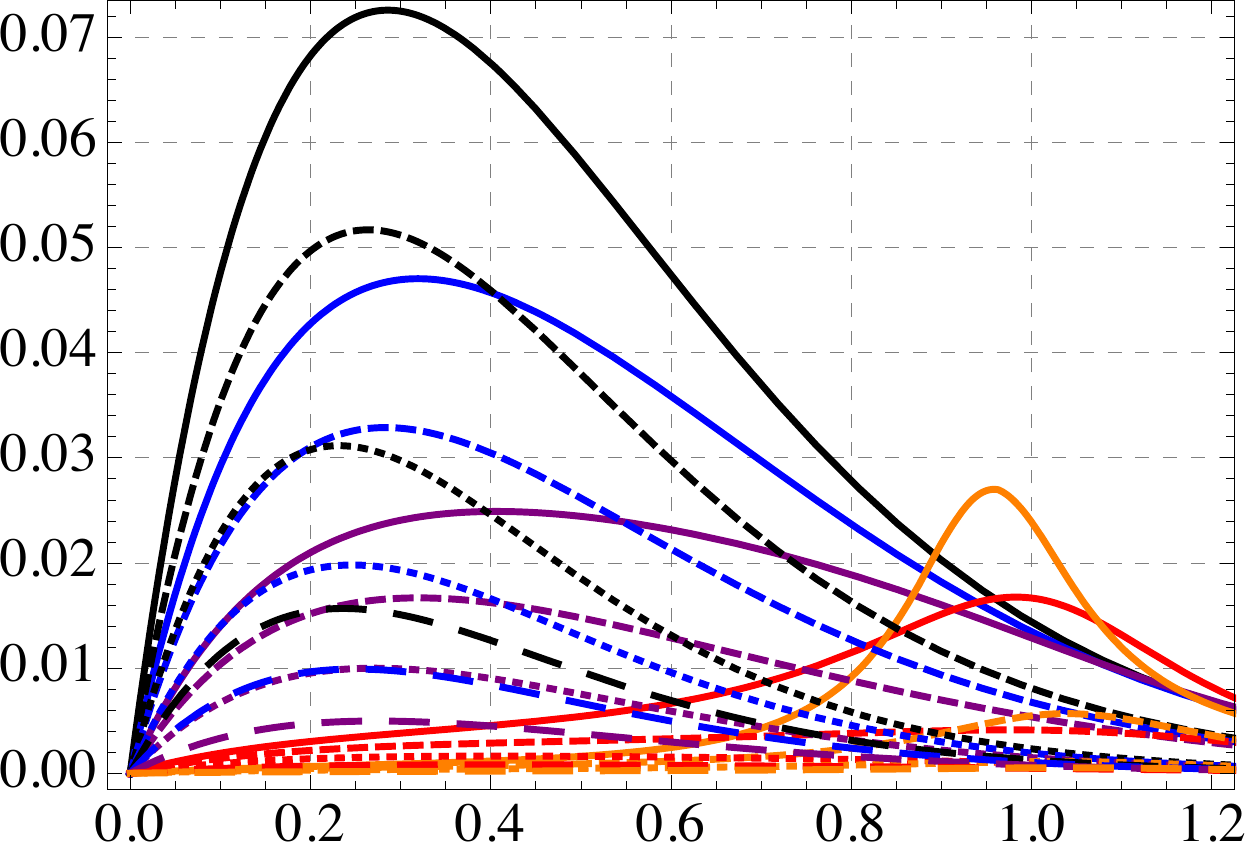}
\qquad
 \put(-460,20){\rotatebox{90}{$\frac{-1}{2\alpha_\mt{EM}\nc\nf T^3}\frac{d\Gamma_\gamma}{d\cos\vartheta\, dk^0}$}}
         \put(-250,-10){$\wn$}
         \put(-220,20){\rotatebox{90}{$\frac{-1}{2\alpha_\mt{EM}\nc\nf T^3}\frac{d\Gamma_\gamma}{d\cos\vartheta\, dk^0}$}}
         \put(-17,-10){$\wn$}
         \\
(c)& (d) 
\end{tabular}
\end{center}
\caption{\small Plots of the total production rate. The groups of curves correspond from top to bottom to $\psi_\mt{H}=0,\, 0.53,\, 0.75,\, 0.941,\, 0.98$. Within each group we plot the angles $\vartheta=0$ (solid), $\pi/4,$ (dashed), and  $\pi/2$ (dotted). The four plots correspond to the cases $a/T=4.41$ (a), $12.2$ (b), $86$ (c), $249$ (d). The temperatures in the four cases are, respectively, $T=0.33,\, 0.36,\, 0.49,\, 0.58$. The isotropic results at the same temperatures and masses are the coarsely dashed curves.}
\label{totalrate}
\end{figure}

We have now all the ingredients to calculate the total emission rate (\ref{diff}).  We convert this quantity to the emission rate per unit photon energy in a infinitesimal angle around $\vartheta$. Using that the photon momentum is light-like, we have 
\be
\frac{-1}{2\alpha_{\mt{EM}}\nc\nf T^3}\frac{d \Gamma_\gamma}{d\cos \vartheta\,dk^0}=\frac{\wn}{2N_c N_f T^2}\frac{1}{e^{2 \pi \wn}-1}\left(\chi_{(1)} + \chi_{(2)}\right)\,,
\ee
which is plotted in Fig.~\ref{totalrate} for different values of $a/T$, $\vartheta$ and $\psi_\mt{H}$. The isotropic result at the same temperature cannot be calculated analytically, since we only have a numerical solution for $\psi$. So, we calculated this quantity numerically and the results are shown in the figures as coarsely dashed curves. We observe that, even in the massive quark case, the anisotropic plasma emits more photons, in total, than the corresponding isotropic plasma at the same temperature.


\section{Dilepton production from holography}
\label{sec3}

The same electromagnetic current considered above, if evaluated for time-like momenta, allows to compute the rate of emission of lepton/antilepton pairs, which are produced via decay of virtual photons. From (\ref{difflep}) we see that the total dilepton production rate is proportional to the trace of the spectral density
\begin{equation}
\eta^{\mu \nu}\chi_{\mu \nu} = -\chi_{tt}+\chi_{x x}+\chi_{y y}+\chi_{z z}\,.
\end{equation}
Since the equation of motion for the $y$-component of the gauge field is decoupled from the equations for the other components, we can split the trace as
\begin{equation}
\eta^{\mu \nu}\chi_{\mu \nu} = \chi_{(1)}+\chi_{(2)}\,,
\end{equation}
where
\bea
\chi_{(1)} \equiv \chi_{y y}\,, \qquad
\chi_{(2)} \equiv  -\chi_{tt}+\chi_{x x}+\chi_{z z}\,.
\eea
The spectral densities are calculated again as
$
\chi_{\mu \nu} = - 2\,\mathrm{Im}\, G_{\mu \nu}^\mt{R}\,.
$
We will obtain the retarded Green functions $G_{\mu \nu}^\mt{R}$ by varying the boundary action (\ref{eq:boundarycampos}) with respect to the values of the gauge fields at the boundary  $A_{\mu}^{(0)}$:
\begin{equation}
\nonumber
G_{\mu \nu}^\mt{R} = \frac{\delta^2 S_{\epsilon}}{\delta A^{\mu(0)}\,\delta A^{\nu(0)}}.
\end{equation}
As in the previous section, it will prove convenient to work with the gauge invariant quantities $E_i = \partial_i A_t - \partial_t A_i$ instead of the gauge fields. 
The Green functions can be obtained as
\begin{align}
&G_{ii}^\mt{R} =-k_0^2\,\frac{\delta^2 S_{\epsilon}}{\delta E_i^{(0)\,2}}\,, \qquad i=x,y,z\,,\\
&G_{tt}^\mt{R} =-k_x^2\,\frac{\delta^2 S_{\epsilon}}{\delta E_x^{(0)\,2}}
-k_z^2\,\frac{\delta^2 S_{\epsilon}}{\delta E_z^{(0)\,2}}-2 k_x\,k_z\,\frac{\delta^2 S_{\epsilon}}{\delta E_x^{(0)} \delta E_z^{(0)}},\label{g00}
\end{align}
where $E_i^{(0)}$ are the values of the gauge invariant fields evaluated at the boundary. In writing (\ref{g00}) we have already used the fact that $k_y$ is zero and that the equation for $E_y$ decouples from the rest. We arrive at
\begin{align}
&\chi_{(1)} =  2\,\mathrm{Im}\left[k_0^2\,\frac{\delta^2 S_{\epsilon}}{\delta E_y^{(0)\,2}}\right]\label{chi1chi2prima}\,,\\
&\chi_{(2)} =  -2\,\mathrm{Im}\left[(k_x^2-k_0^2)\,\frac{\delta^2 S_{\epsilon}}{\delta E_x^{(0)2}}+(k_z^2-k_0^2)\,\frac{\delta^2 S_{\epsilon}}{\delta E_z^{(0)2}}+2\,k_x\,k_z\,\frac{\delta^2 S_{\epsilon}}{\delta E_x^{(0)} \delta E_z^{(0)}}\right]\,.
\label{chi1chi2}
\end{align}
In terms of the spectral densities the latter equation is
\bea
\chi_{(2)} = \left(1-\frac{k_x^2}{k_0^2}\right) \chi_{xx}+\left(1-\frac{k_z^2}{k_0^2}\right) \chi_{zz}-2\frac{k_xk_z}{k_0^2} \chi_{xz}\, .
\label{chi1chi2spectral}
\eea
When light-like momentum is considered, the previous calculation coincides, as it should, with the one for the photon production. 
For dilepton production, the spatial part of the momentum will be given by $\vec{k}=q(\mathrm{sin}\,\vartheta,0,\mathrm{cos}\,\vartheta)$ for $q< k_0$, and the equation (\ref{chi1chi2}) will read 
\begin{align}
\chi_{(2)} =  -2\,\mathrm{Im}\left[(q^2\mathrm{sin}^2\vartheta-k_0^2)\,\frac{\delta^2 S_{\epsilon}}{\delta E_x^{(0)2}}+(q^2\mathrm{cos}^2\vartheta-k_0^2)\,\frac{\delta^2 S_{\epsilon}}{\delta E_z^{(0)2}}+2\,q^2\mathrm{sin}\,\vartheta\,\mathrm{cos}\,\vartheta\,\frac{\delta^2 S_{\epsilon}}{\delta E_x^{(0)} \delta E_z^{(0)}}\right]\,.
\label{chi1chi2dil}
\end{align}

As a warm up, we will begin by performing the calculation in the isotropic limit. This will be used to normalize the results for the anisotropic plasma.


\subsection{Isotropic limit}

In the isotropic limit (\ref{isometric}) we can use the $SO(3)$ symmetry to set $\vartheta=\pi/2$, fixing the spatial component of $k$ in the $x$-direction. We have $k_x=q$, $k_y=k_z=0$ which simplifies the equations above  to
\bea
\chi_{(1)\mt{iso}} = \chi_{yy,\mt{iso}}\label{chi1chi2primaspectraliso}\,,\qquad 
\chi_{(2)\mt{iso}} = \left(1-\frac{q^2}{k_0^2}\right) \chi_{xx, \mt{iso}}+\chi_{zz, \mt{iso}}.
\label{chi1chi2spectraliso}
\eea

We will compute $\chi_{yy}$ repeating the same steps used in the photon production for polarization $\epsilon_{(1)}$. This spectral density  reads
\begin{eqnarray}
\frac{\chi_{(1)\mt{iso}}}{8\tilde{\mathcal{N}}_{D7}\,\textswab{w}}=\frac{1}{2\pi Tk_0\left| A_{y,\mt{iso}}(k,0)\right|^2}\mbox{Im}\,  \lim_{u\to \uh}Q(u)A_{y,\mt{iso}}^{\prime}(k,u)A_{y,\mt{iso}}^*(k,u)
\end{eqnarray}
where $Q(u)$ was defined in (\ref{Q}) and 
$A_{y,\mt{iso}}$ solves equation (\ref{eomy1}), in the isotropic limit (\ref{isometric}) but with $q\neq k_0$.

\begin{figure}
\begin{center}
\begin{tabular}{cc}
\setlength{\unitlength}{1cm}
\hspace{-0.9cm}
\includegraphics[width=6.5cm]{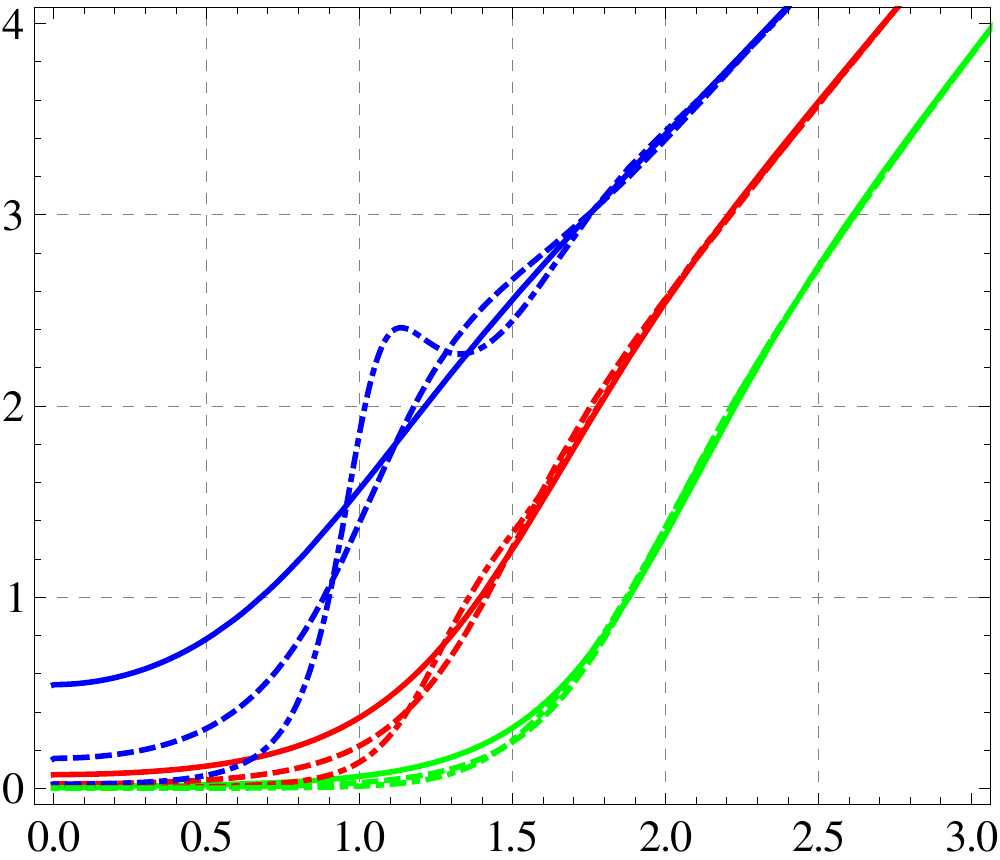} 
\qquad\qquad & 
\includegraphics[width=6.5cm]{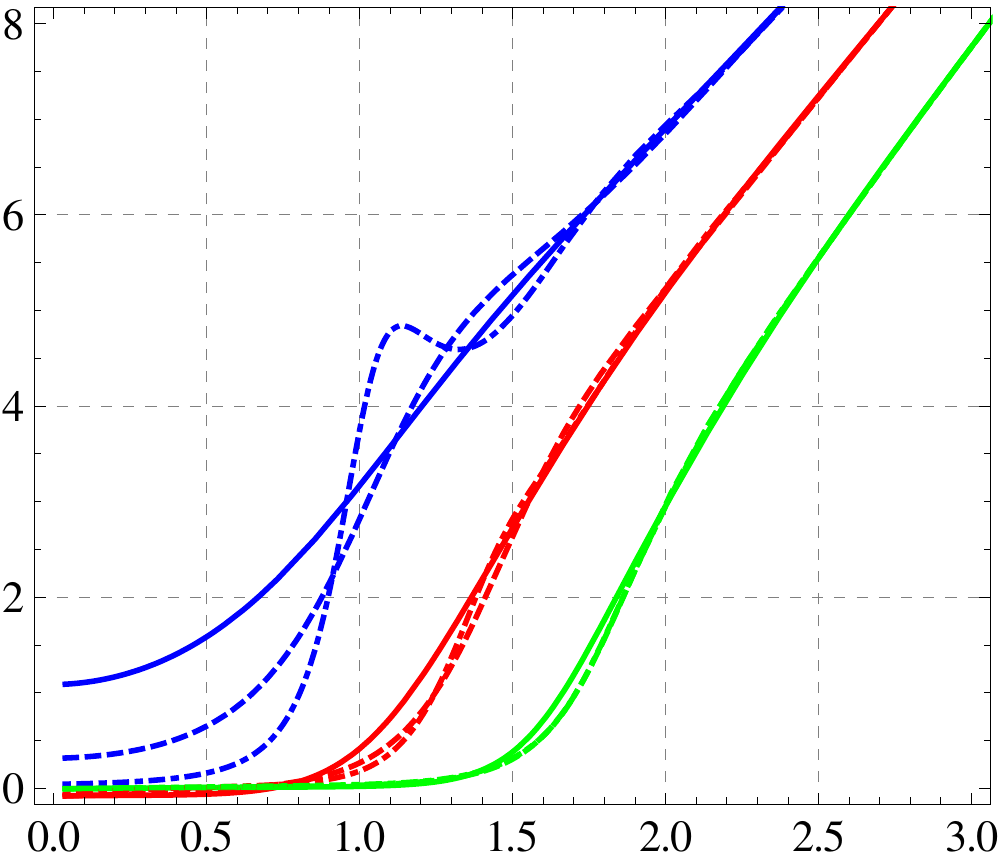}
\qquad
  \put(-430,50){\rotatebox{90}{$\chi_{(1)\mt{iso}}/8\tilde{\mathcal{N}}_\mt{D7}\,\textswab{w}$}}
         \put(-240,-10){$\wn$}
         \put(-205,50){\rotatebox{90}{$\chi_{(2)\mt{iso}}/8\tilde{\mathcal{N}}_\mt{D7}\,\textswab{w}$}}
         \put(-17,-10){$\wn$}
\end{tabular}
\end{center}
\caption{\small Plots of the spectral densities $\chi_{(1)\mt{iso}}$ and $\chi_{(2)\mt{iso}}$. Here we have fixed $T=0.33$. Curves of different colors and traits denote different values of $\qn$ ($\qn=$0 (blue), 1 (red), 1.5 (green)) and of $\psi_\mt{H}$ ($\psi_\mt{H}=$0 (solid), 0.75 (dashed), 0.941 (dot-dashed)). The curves for $\qn=0$ are identical in the two plots, up to an overall factor of 2, as it should be, considering that (\ref{chixxiso}) and (\ref{chizziso}) coincide in this case.}
\label{plotiso}
\end{figure}

To compute $\chi_{xx,\mt{iso}}$ and $\chi_{zz,\mt{iso}}$, we make two observations. First, for $\vartheta =\pi /2$, equations (\ref{eq:Ex}) and (\ref{eq:Ez}) decouple from each other. Second, the action (\ref{eq:boundarycampos}) will have no mixed terms, so we can vary the action with respect to $E_{x,\mt{iso}}$ and $E_{z,\mt{iso}}$ in a similar fashion to what has been done for $A_{y,\mt{iso}}$, and get 
\begin{align}
&\frac{\chi_{xx,\mt{iso}}}{8\tilde{\mathcal{N}}_{D7}\,\textswab{w}}=\frac{k_0}{2\pi T\left| E_{x,\mt{iso}}(k,0)\right|^2}\mbox{Im}\,  \lim_{u\to \uh}Q_x(u)E_{x,\mt{iso}}^{\prime}(k,u)E_{x,\mt{iso}}^*(k,u)\label{chixxiso}\,,\\
&\frac{\chi_{zz,\mt{iso}}}{8\tilde{\mathcal{N}}_{D7}\,\textswab{w}}=\frac{k_0}{2\pi T\left| E_{z,\mt{iso}}(k,0)\right|^2}\mbox{Im}\,  \lim_{u\to \uh}Q_z(u)E_{z,\mt{iso}}^{\prime}(k,u)E_{z,\mt{iso}}^*(k,u)\,,
\label{chizziso}
\end{align}
where $Q_x(u)=\frac{Mg^{uu}}{u^{2}\overline{k}^{2}}g^{tt}g^{xx}$ and $Q_z(u)=\frac{Mg^{uu}}{-k_{0}^{2}u^{2}\overline{k}^{2}}\left(-g^{tt}k_{0}^{2}-g^{xx}q^{2}\right)g^{zz}$ are, respectively, the coefficients multiplying the $E_xE_{x}^\prime$ and $E_zE_{z}^\prime$ terms in the boundary action (\ref{eq:boundarycampos}). $E_{x,\mt{iso}}$ and $E_{z,\mt{iso}}$ are solutions to (\ref{eq:Ex}) and (\ref{eq:Ez}) in the isotropic limit (\ref{isometric}). These quantities reduce to the expressions in \cite{CaronHuot:2006te} and are plotted as a function of $\wn\equiv k_0/2\pi T$ for various values of $\qn\equiv q/2\pi T$ in Fig.~\ref{plotiso}.


\subsection{Dilepton spectral density $\chi_{(1)}$}

\begin{figure}
\begin{center}
\begin{tabular}{ccc}
\setlength{\unitlength}{1cm}
\hspace{-0.9cm}
\includegraphics[width=5cm]{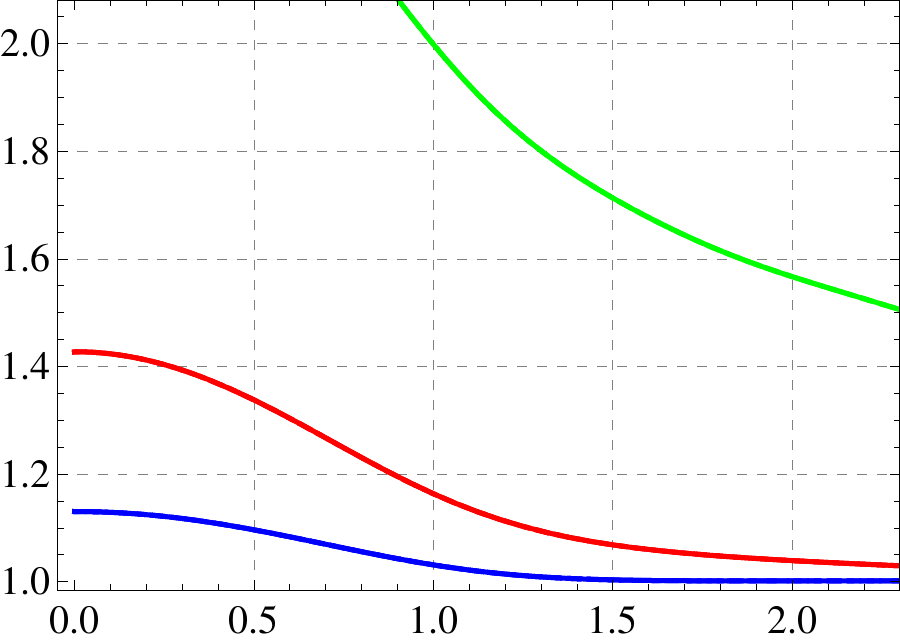} 
& 
\qquad \includegraphics[width=5cm]{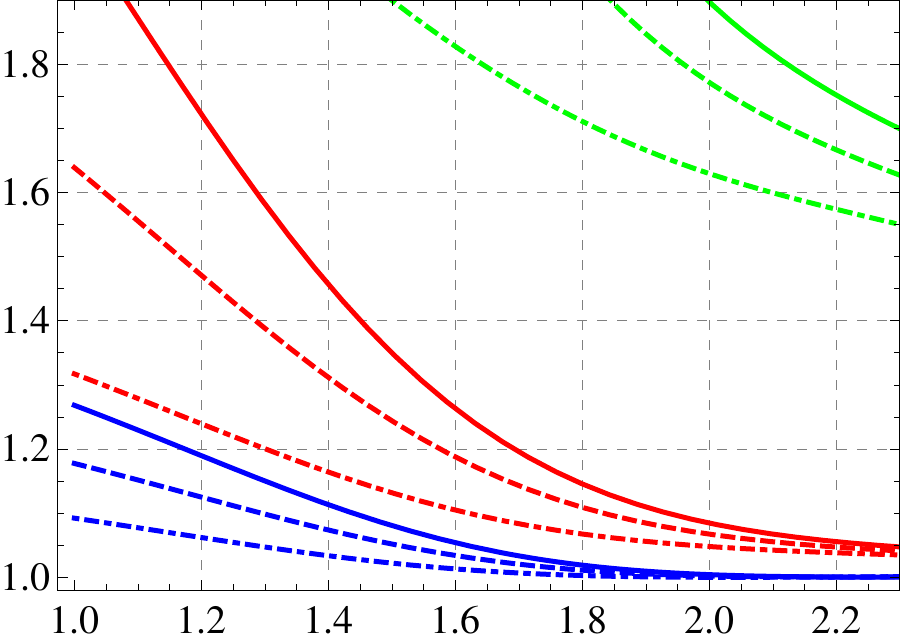}
& 
\qquad \includegraphics[width=5cm]{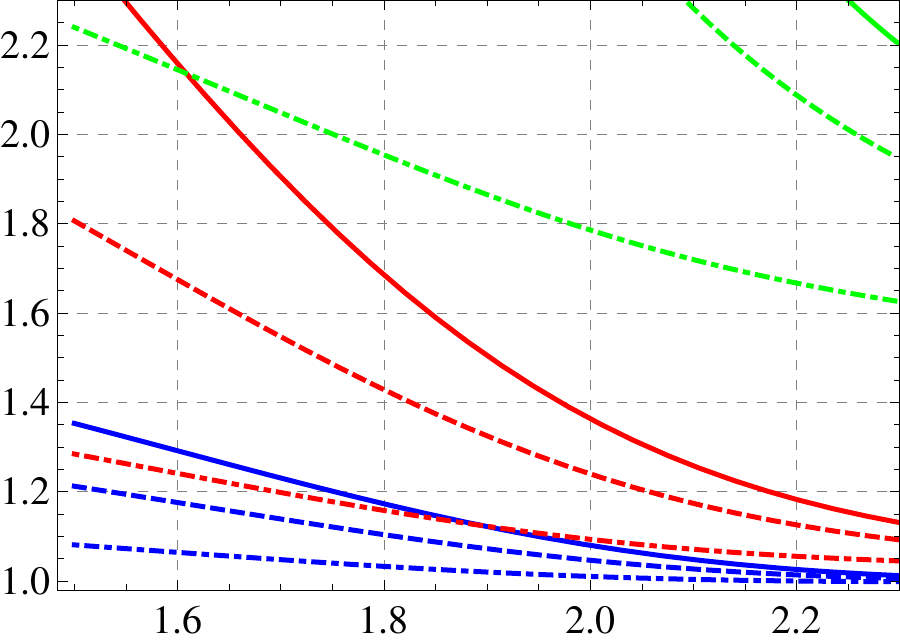}
\qquad
  \put(-510,25){\rotatebox{90}{$\chi_{(1)}/\chi_{(1)\mt{iso}}(T)$}}
         \put(-370,-10){$\textswab{w}$}
         \put(-333,25){\rotatebox{90}{$\chi_{(1)}/\chi_{(1)\mt{iso}}(T)$}}
         \put(-195,-10){$\textswab{w}$}
          \put(-160,25){\rotatebox{90}{$\chi_{(1)}/\chi_{(1)\mt{iso}}(T)$}}
         \put(-15,-10){$\textswab{w}$}
\\
(a) & \qquad(b) & \qquad(c)\\
& \\
\hspace{-0.9cm}
\includegraphics[width=5cm]{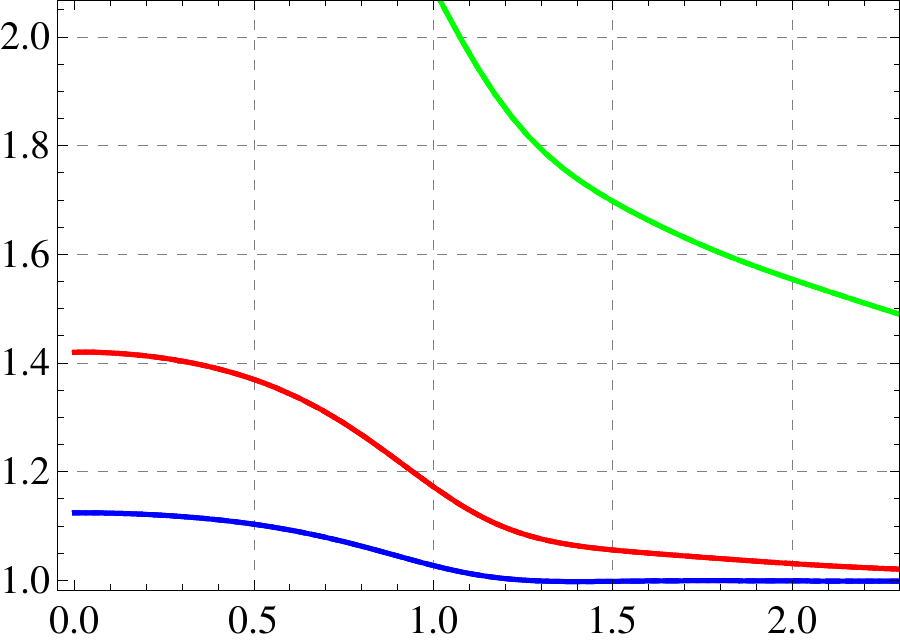} 
& 
\qquad \includegraphics[width=5cm]{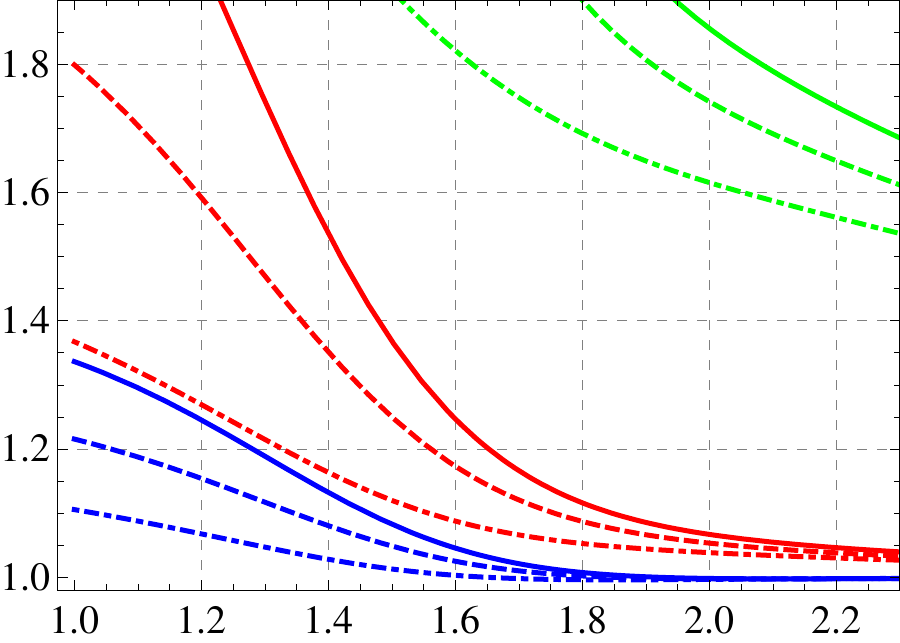}
& 
\qquad \includegraphics[width=5cm]{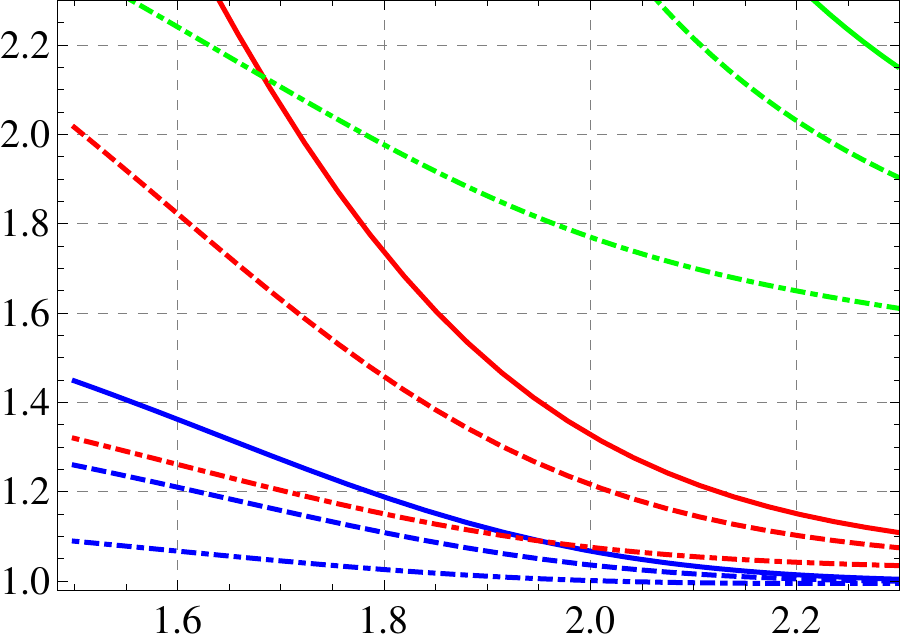}
\qquad
  \put(-510,25){\rotatebox{90}{$\chi_{(1)}/\chi_{(1)\mt{iso}}(T)$}}
         \put(-370,-10){$\textswab{w}$}
         \put(-333,25){\rotatebox{90}{$\chi_{(1)}/\chi_{(1)\mt{iso}}(T)$}}
         \put(-195,-10){$\textswab{w}$}
          \put(-160,25){\rotatebox{90}{$\chi_{(1)}/\chi_{(1)\mt{iso}}(T)$}}
         \put(-15,-10){$\textswab{w}$}
         \\
(d) & \qquad(e) & \qquad(f) \\
& \\
\hspace{-0.9cm}
\includegraphics[width=5cm]{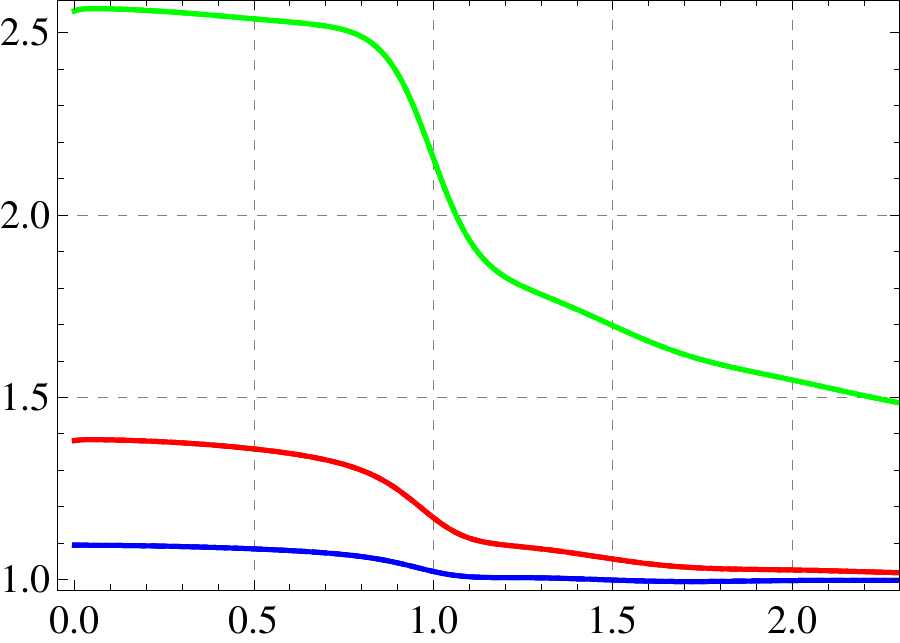} 
& 
\qquad \includegraphics[width=5cm]{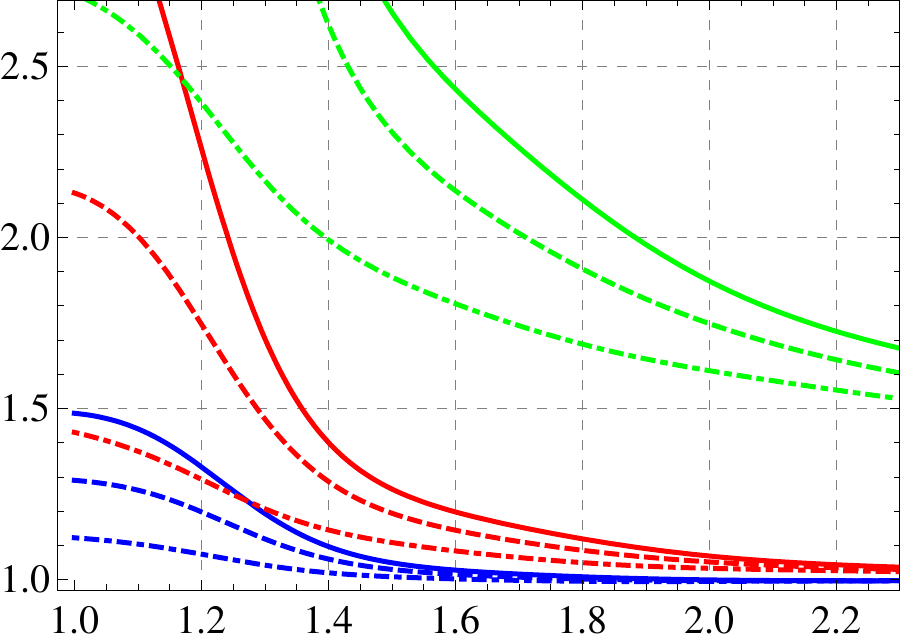}
& 
\qquad \includegraphics[width=5cm]{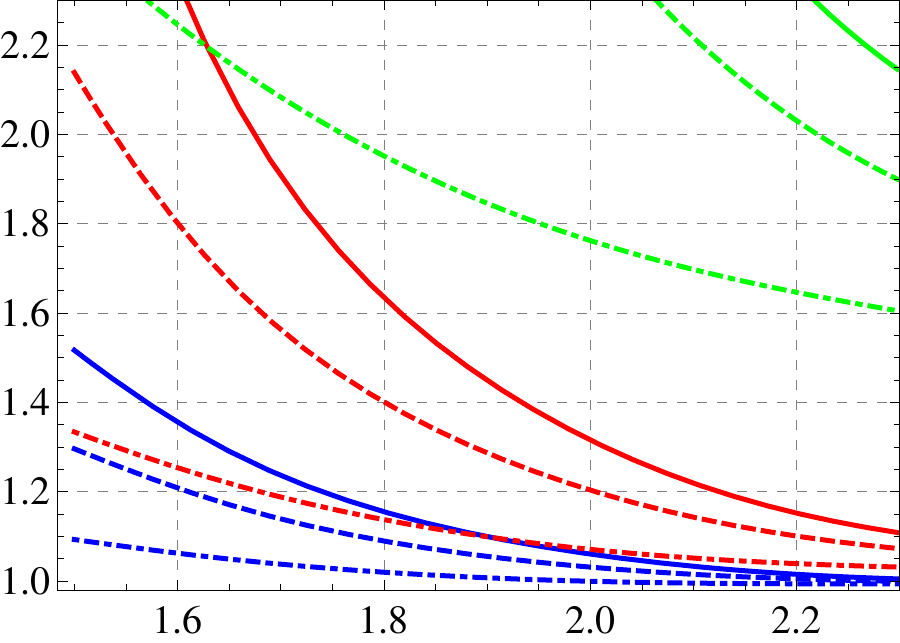}
\qquad
  \put(-510,25){\rotatebox{90}{$\chi_{(1)}/\chi_{(1)\mt{iso}}(T)$}}
         \put(-370,-10){$\textswab{w}$}
         \put(-333,25){\rotatebox{90}{$\chi_{(1)}/\chi_{(1)\mt{iso}}(T)$}}
         \put(-195,-10){$\textswab{w}$}
          \put(-160,25){\rotatebox{90}{$\chi_{(1)}/\chi_{(1)\mt{iso}}(T)$}}
         \put(-15,-10){$\textswab{w}$}
         \\
(g) & \qquad(h) & \qquad(i) \\
\end{tabular}
\end{center}
\caption{\small Plots of the spectral density $\chi_{(1)}$ normalized with respect to the isotropic result at fixed temperature $\chi_{(1)\mt{iso}}(T)$. Curves of different colors denote different values of $a/T$ as follows  $a/T=$4.41 (blue), 12.2 (red), 86 (green). The angles are $\vartheta=0$ (solid), $\pi/4$ (dashed), $\pi/2$ (dot-dashed). Columns correspond to different values of $\textswab{q}$: from left to right it is $\textswab{q}=0,1,1.5$. Rows correspond to different values of the quark mass: from top to bottom it is $\psi_\mt{H}=0,0.75,0.941$. Then, for instance, (h) corresponds to $\textswab{q}=1$, $\psi_\mt{H}=0.941$. }
\label{c1w}
\end{figure}

\begin{figure}
\begin{center}
\begin{tabular}{ccc}
\setlength{\unitlength}{1cm}
\hspace{-0.9cm}
\includegraphics[width=5cm]{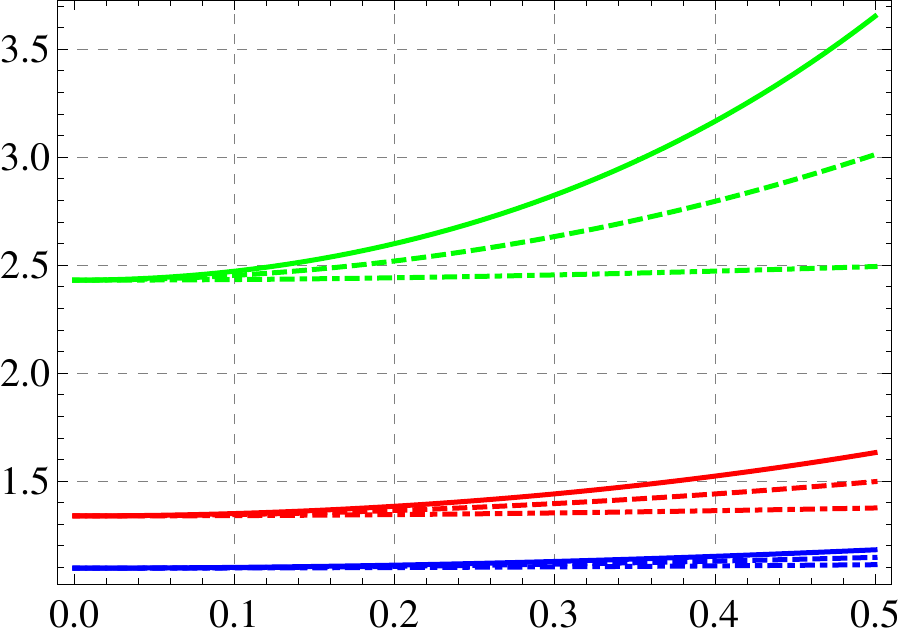} 
& 
\qquad \includegraphics[width=5cm]{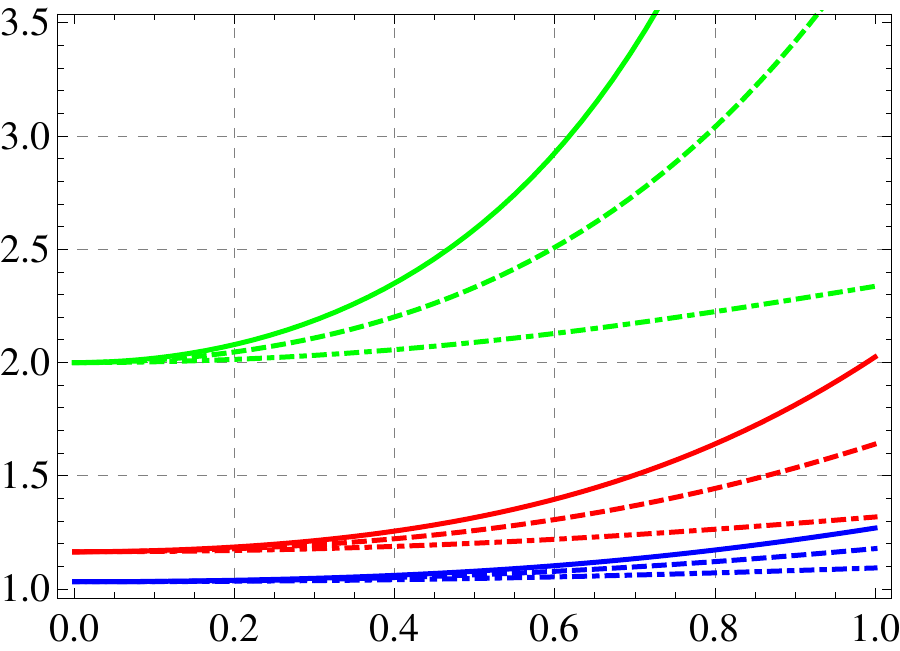}
& 
\qquad \includegraphics[width=5cm]{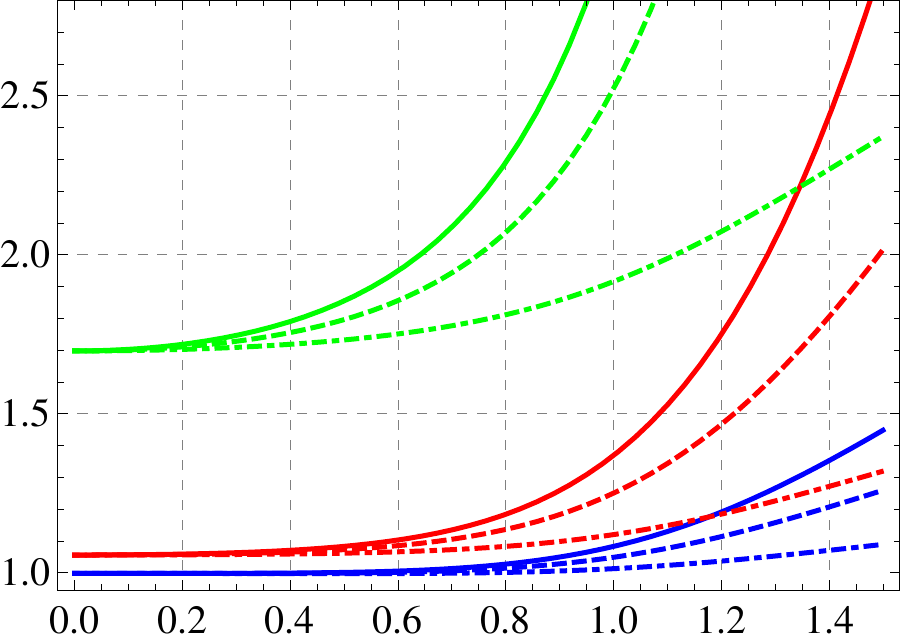}
\qquad
  \put(-510,25){\rotatebox{90}{$\chi_{(1)}/\chi_{(1)\mt{iso}}(T)$}}
         \put(-370,-10){$\textswab{q}$}
         \put(-333,25){\rotatebox{90}{$\chi_{(1)}/\chi_{(1)\mt{iso}}(T)$}}
         \put(-195,-10){$\textswab{q}$}
          \put(-160,25){\rotatebox{90}{$\chi_{(1)}/\chi_{(1)\mt{iso}}(T)$}}
         \put(-15,-10){$\textswab{q}$}
\\
(a) & \qquad(b) & \qquad(c)\\
& \\
\hspace{-0.9cm}
\includegraphics[width=5cm]{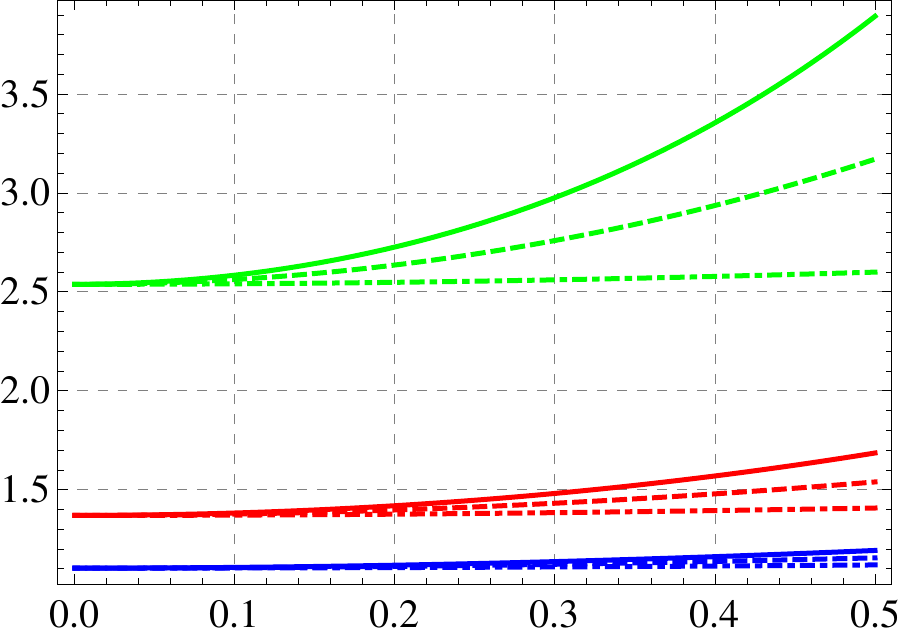} 
& 
\qquad \includegraphics[width=5cm]{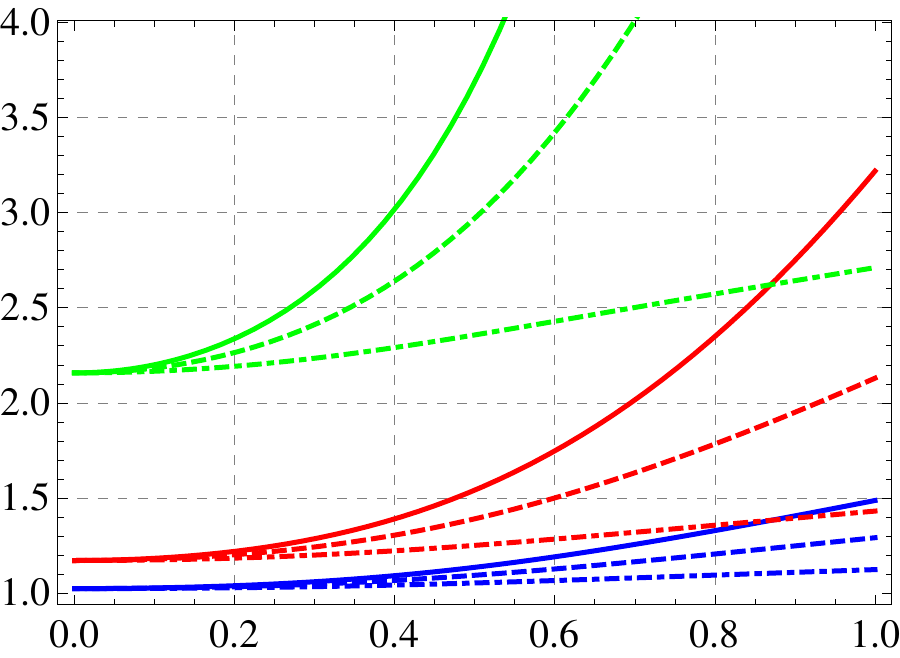}
& 
\qquad \includegraphics[width=5cm]{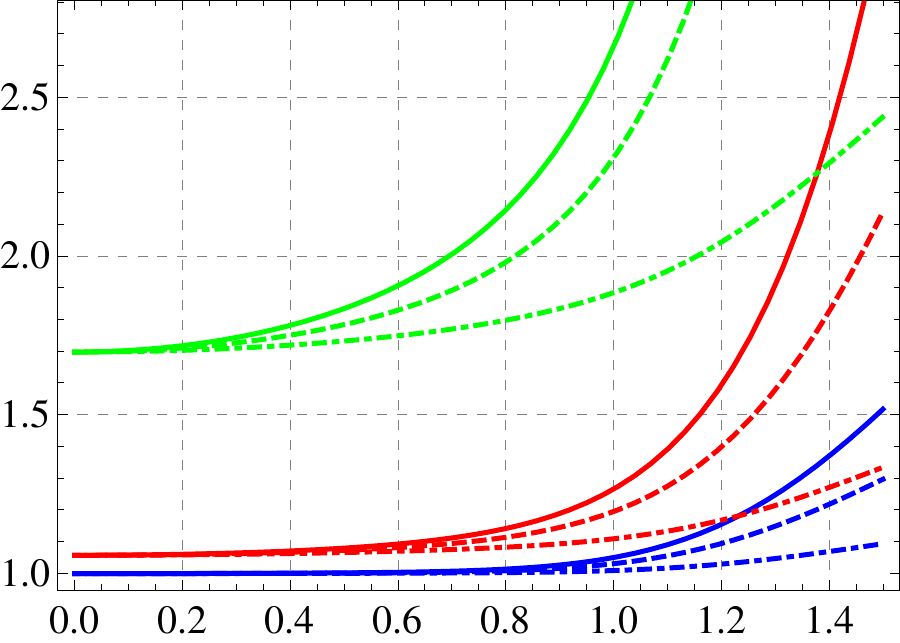}
\qquad
  \put(-510,25){\rotatebox{90}{$\chi_{(1)}/\chi_{(1)\mt{iso}}(T)$}}
         \put(-370,-10){$\textswab{q}$}
         \put(-333,25){\rotatebox{90}{$\chi_{(1)}/\chi_{(1)\mt{iso}}(T)$}}
         \put(-195,-10){$\textswab{q}$}
          \put(-160,25){\rotatebox{90}{$\chi_{(1)}/\chi_{(1)\mt{iso}}(T)$}}
         \put(-15,-10){$\textswab{q}$}
         \\
(d) & \qquad(e) & \qquad(f) \\
& \\
\hspace{-0.9cm}
\includegraphics[width=5cm]{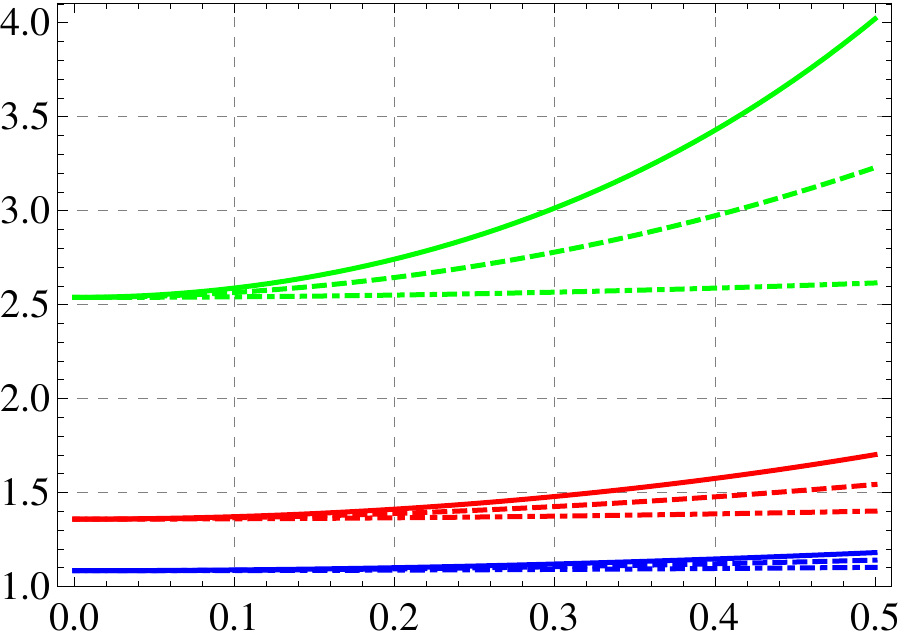} 
& 
\qquad \includegraphics[width=5cm]{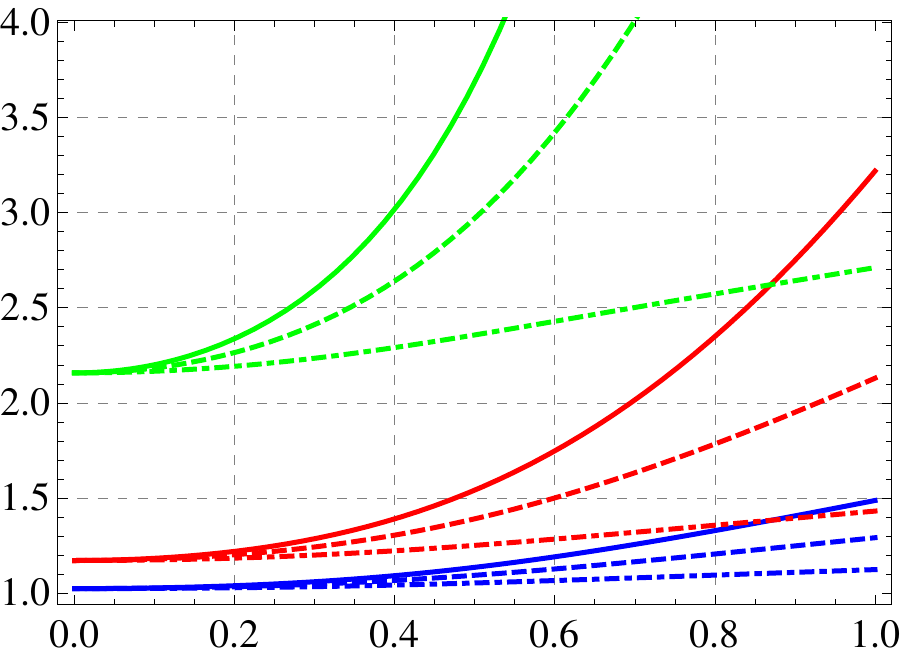}
& 
\qquad \includegraphics[width=5cm]{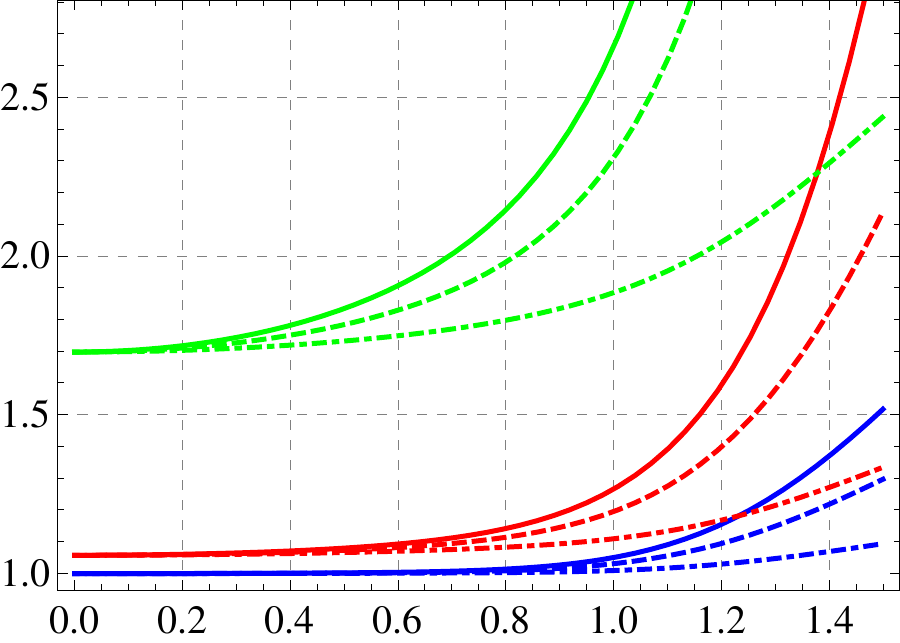}
\qquad
  \put(-510,25){\rotatebox{90}{$\chi_{(1)}/\chi_{(1)\mt{iso}}(T)$}}
         \put(-370,-10){$\textswab{q}$}
         \put(-333,25){\rotatebox{90}{$\chi_{(1)}/\chi_{(1)\mt{iso}}(T)$}}
         \put(-195,-10){$\textswab{q}$}
          \put(-160,25){\rotatebox{90}{$\chi_{(1)}/\chi_{(1)\mt{iso}}(T)$}}
         \put(-15,-10){$\textswab{q}$}
         \\
(g) & \qquad(h) & \qquad(i) \\
\end{tabular}
\end{center}
\caption{\small Plots of the spectral density $\chi_{(1)}$ normalized with respect to the isotropic result at fixed temperature $\chi_{(1)\mt{iso}}(T)$. Curves of different colors denote different values of $a/T$ as follows  $a/T=$4.41 (blue), 12.2 (red), 86 (green). The angles are $\vartheta=0$ (solid), $\pi/4$ (dashed), $\pi/2$ (dash-dotted). Columns correspond to different values of $\textswab{w}$: from left to right it is $\textswab{w}=1/2,1,1.5$. Rows correspond to different values of the quark mass: from top to bottom it is $\psi_\mt{H}=0,0.75,0.941$. Then, for instance, (f) corresponds to $\textswab{w}=3/2$, $\psi_\mt{H}=0.75$. }
\label{c1q}
\end{figure}

\begin{figure}
\begin{center}
\begin{tabular}{ccc}
\setlength{\unitlength}{1cm}
\hspace{-0.9cm}
\includegraphics[width=5cm]{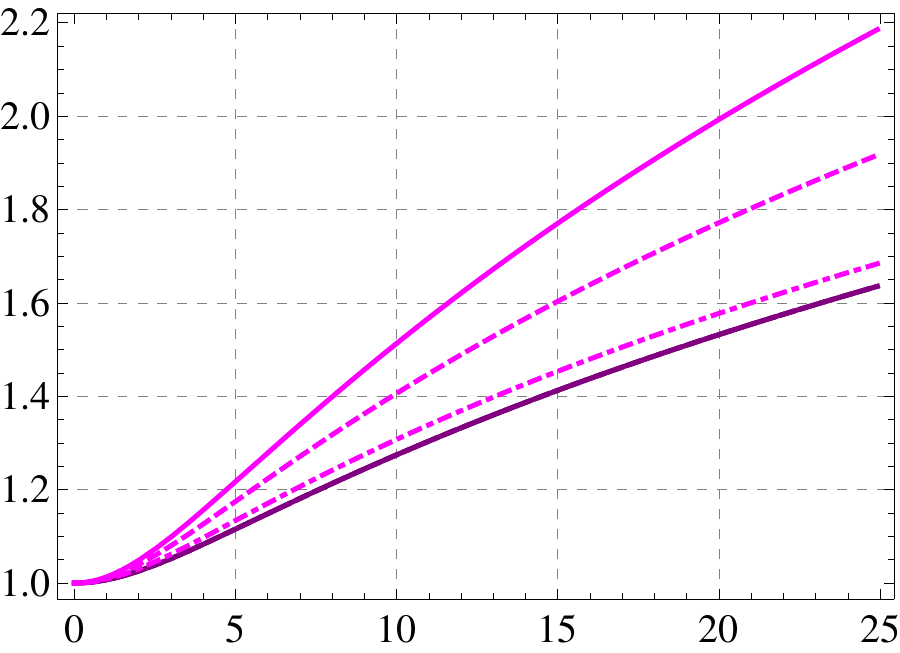} 
& 
\qquad \includegraphics[width=5cm]{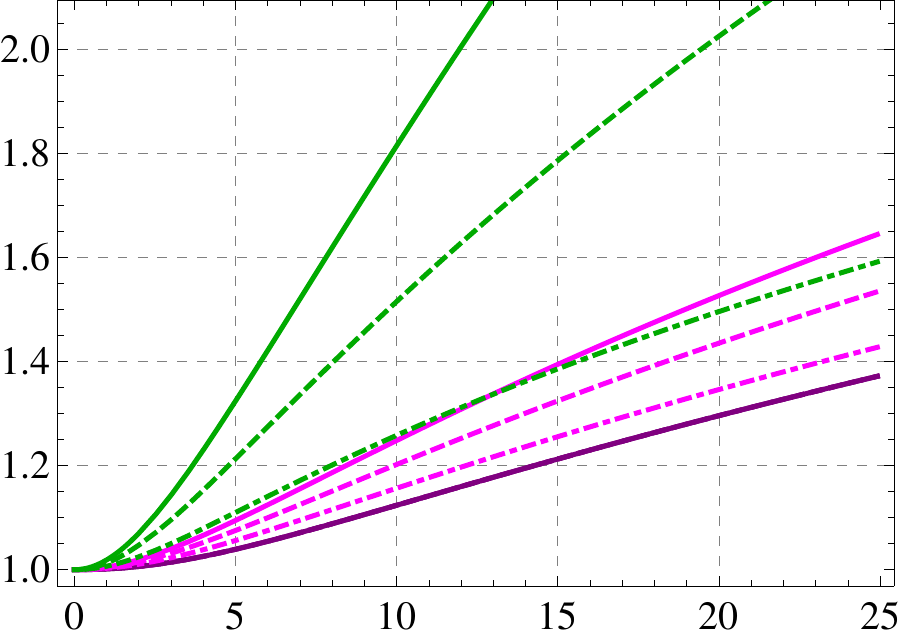}
& 
\qquad \includegraphics[width=5cm]{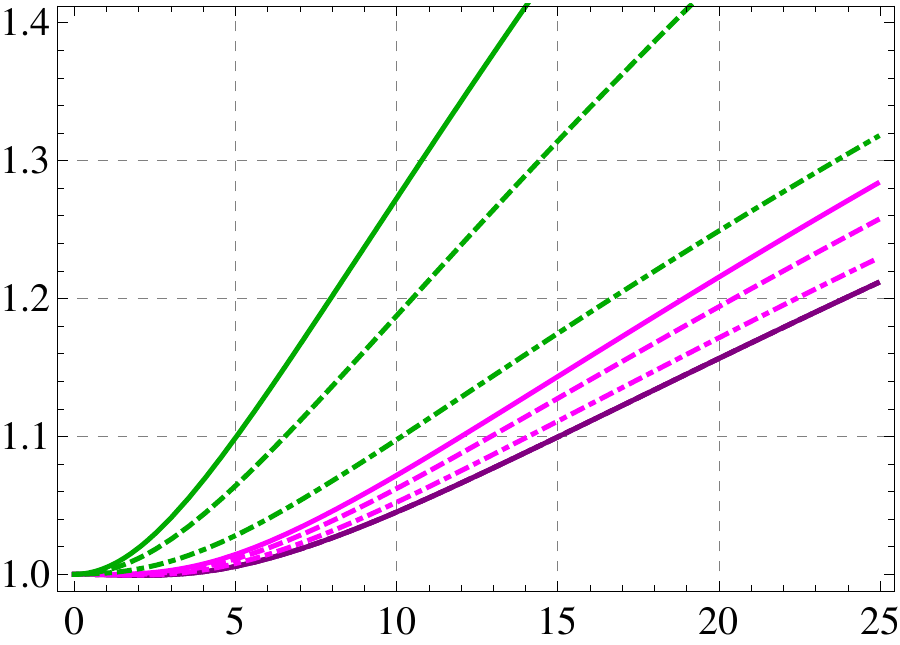}
\qquad
  \put(-510,25){\rotatebox{90}{$\chi_{(1)}/\chi_{(1)\mt{iso}}(T)$}}
         \put(-370,-10){$a/T$}
         \put(-333,25){\rotatebox{90}{$\chi_{(1)}/\chi_{(1)\mt{iso}}(T)$}}
         \put(-195,-10){$a/T$}
          \put(-160,25){\rotatebox{90}{$\chi_{(1)}/\chi_{(1)\mt{iso}}(T)$}}
         \put(-15,-10){$a/T$}
\\
(a) & \qquad(b) & \qquad(c)\\
& \\
\hspace{-0.9cm}
\includegraphics[width=5cm]{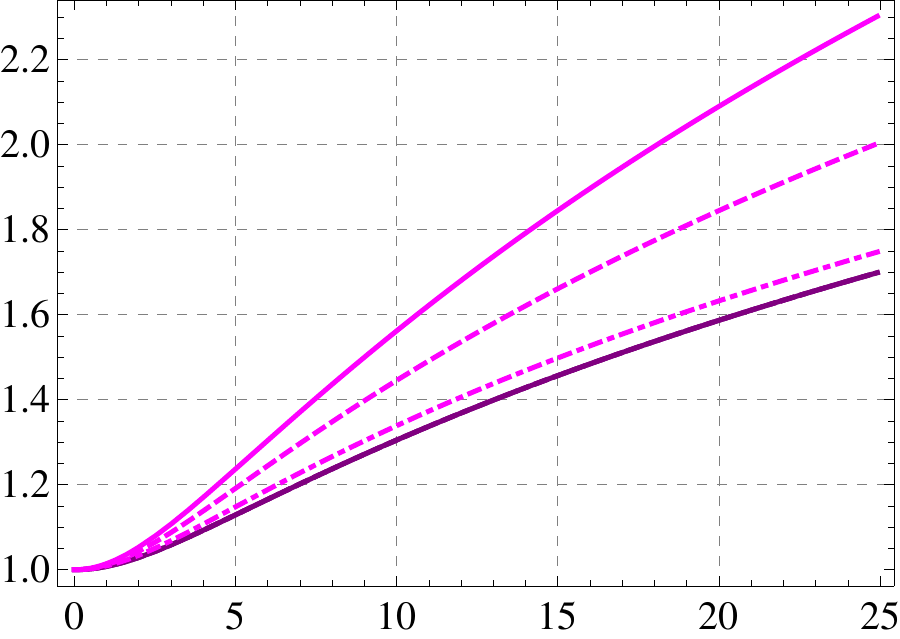} 
& 
\qquad \includegraphics[width=5cm]{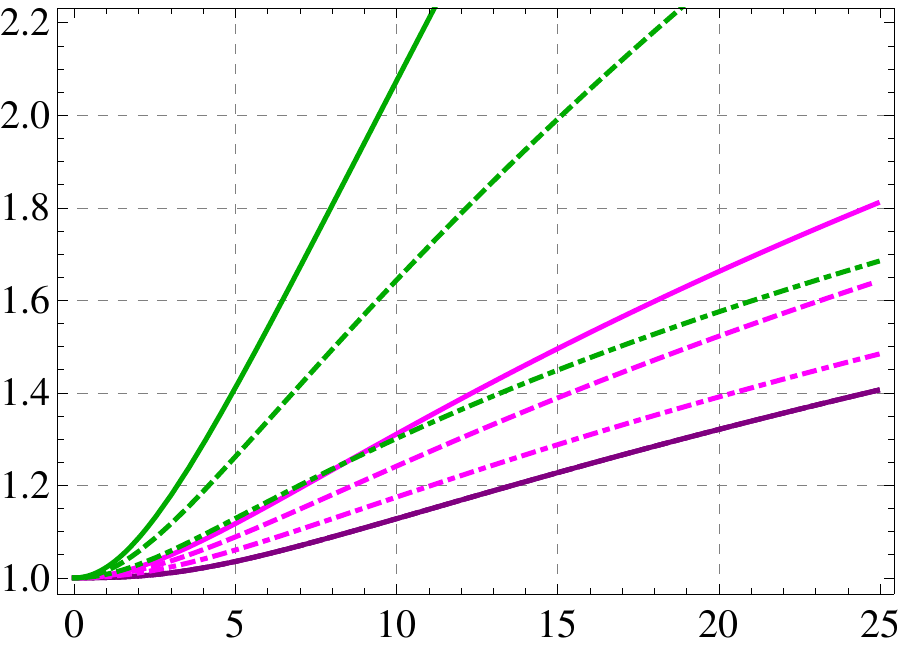}
& 
\qquad \includegraphics[width=5cm]{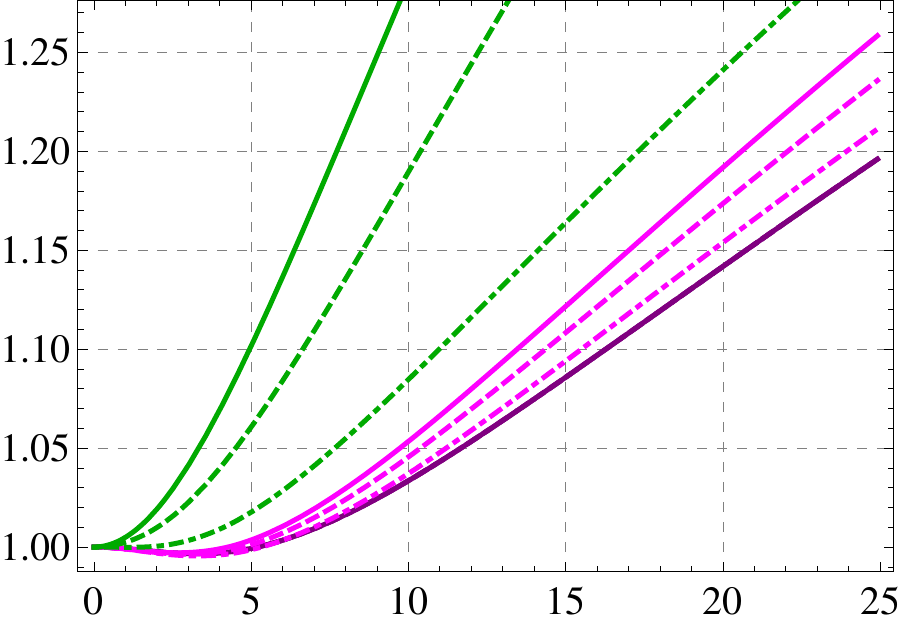}
\qquad
  \put(-510,25){\rotatebox{90}{$\chi_{(1)}/\chi_{(1)\mt{iso}}(T)$}}
         \put(-370,-10){$a/T$}
         \put(-333,25){\rotatebox{90}{$\chi_{(1)}/\chi_{(1)\mt{iso}}(T)$}}
         \put(-195,-10){$a/T$}
          \put(-160,25){\rotatebox{90}{$\chi_{(1)}/\chi_{(1)\mt{iso}}(T)$}}
         \put(-15,-10){$a/T$}
         \\
(d) & \qquad(e) & \qquad(f) \\
& \\
\hspace{-0.9cm}
\includegraphics[width=5cm]{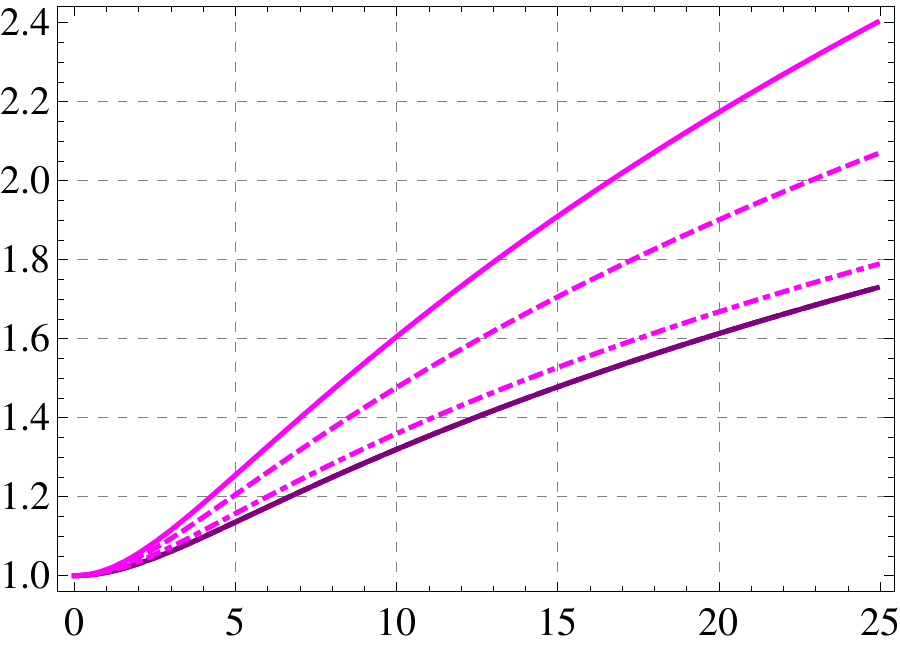} 
& 
\qquad \includegraphics[width=5cm]{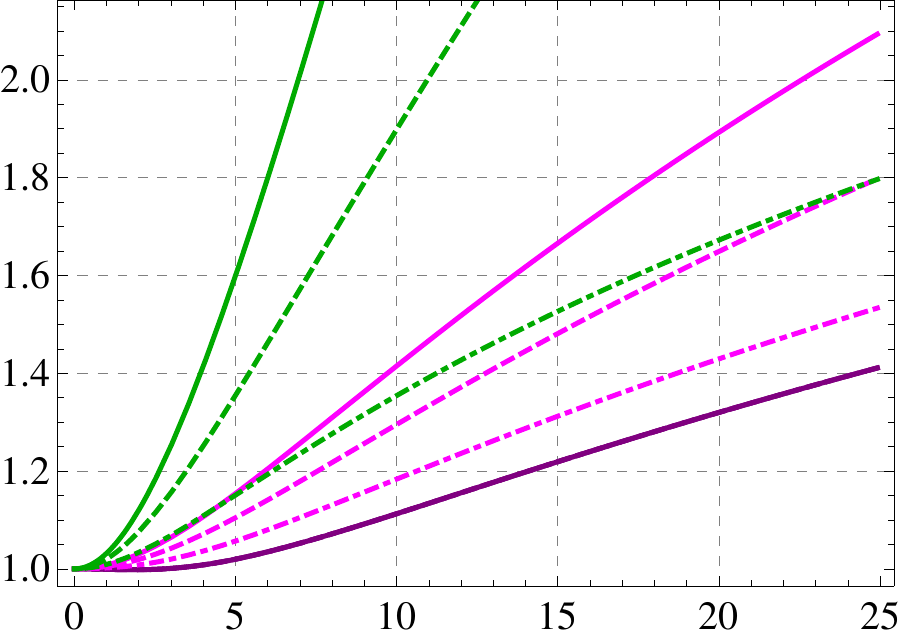}
& 
\qquad \includegraphics[width=5cm]{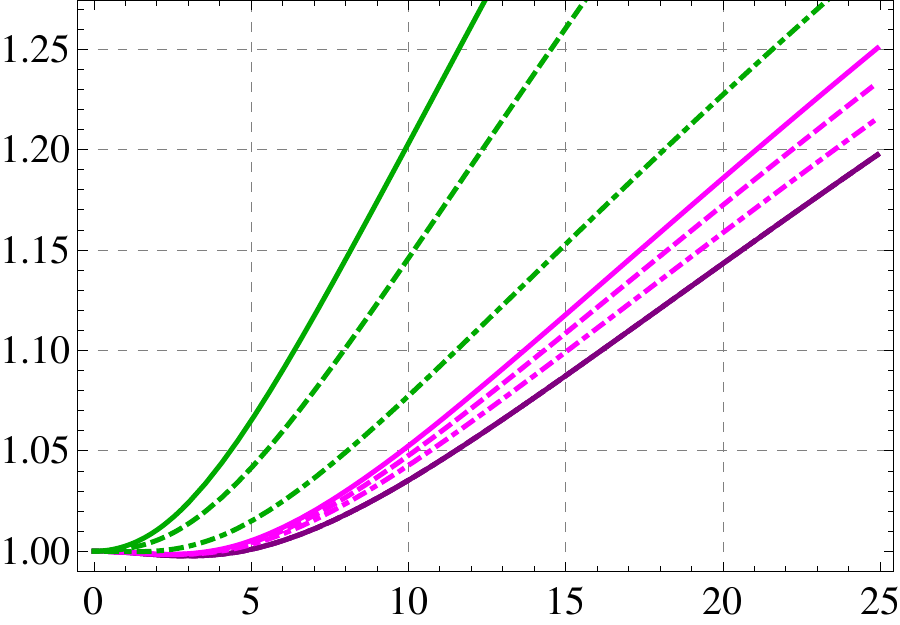}
\qquad
  \put(-510,25){\rotatebox{90}{$\chi_{(1)}/\chi_{(1)\mt{iso}}(T)$}}
         \put(-370,-10){$a/T$}
         \put(-333,25){\rotatebox{90}{$\chi_{(1)}/\chi_{(1)\mt{iso}}(T)$}}
         \put(-195,-10){$a/T$}
          \put(-160,25){\rotatebox{90}{$\chi_{(1)}/\chi_{(1)\mt{iso}}(T)$}}
         \put(-15,-10){$a/T$}
         \\
(g) & \qquad(h) & \qquad(i) \\
\end{tabular}
\end{center}
\caption{\small Plots of the spectral density $\chi_{(1)}$ normalized with respect to the isotropic result at fixed temperature $\chi_{(1)\mt{iso}}(T)$. Curves of different colors denote different values of $\textswab{q}$ as follows  $\textswab{q}=$0 (purple), 1/2 (magenta), 1 (green). The angles are $\vartheta=0$ (solid), $\pi/4$ (dashed), $\pi/2$ (dash-dotted). Columns correspond to different values of $\textswab{w}$: from left to right it is $\textswab{w}=0.5,1,1.5$. Rows correspond to different values of the quark mass: from top to bottom it is $\psi_\mt{H}=0,0.75,0.941$. Then, for instance, (f) corresponds to $\textswab{w}=1.5$, $\psi_\mt{H}=0.75$.}
\label{c4bc1}
\end{figure}

\begin{figure}
\begin{center}
\begin{tabular}{ccc}
\setlength{\unitlength}{1cm}
\hspace{-0.9cm}
\includegraphics[width=5cm]{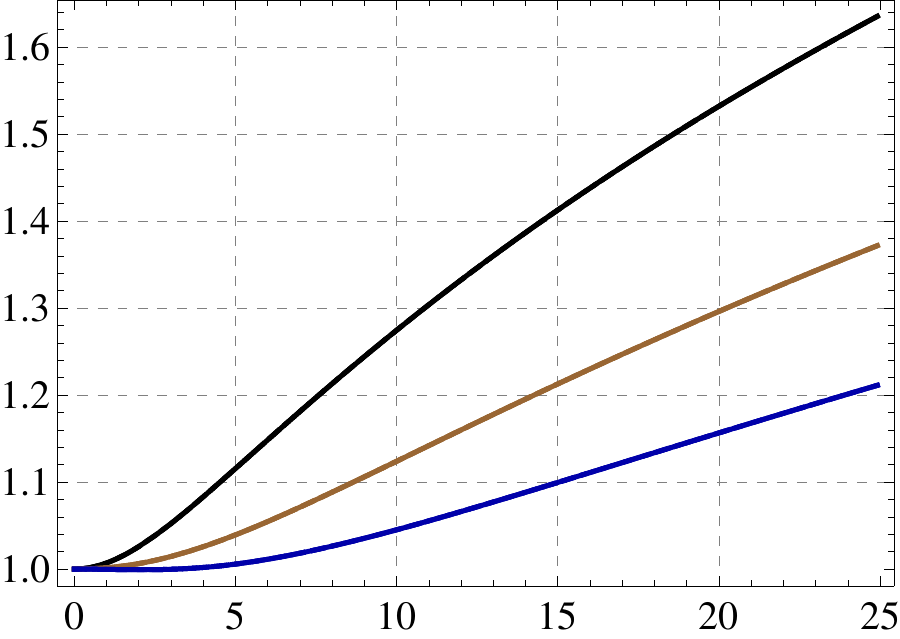} 
& 
\qquad \includegraphics[width=5cm]{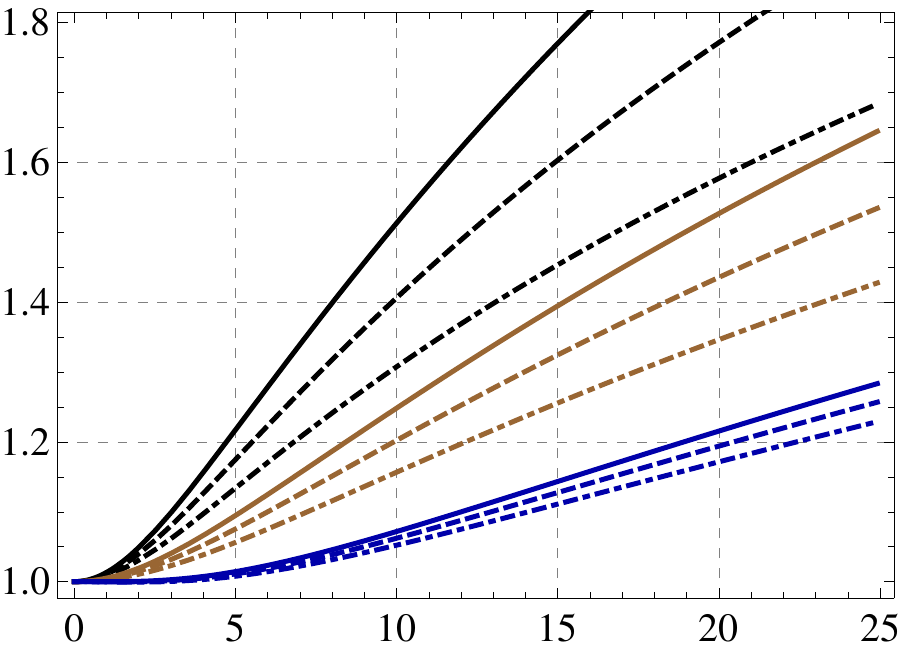}
& 
\qquad \includegraphics[width=5cm]{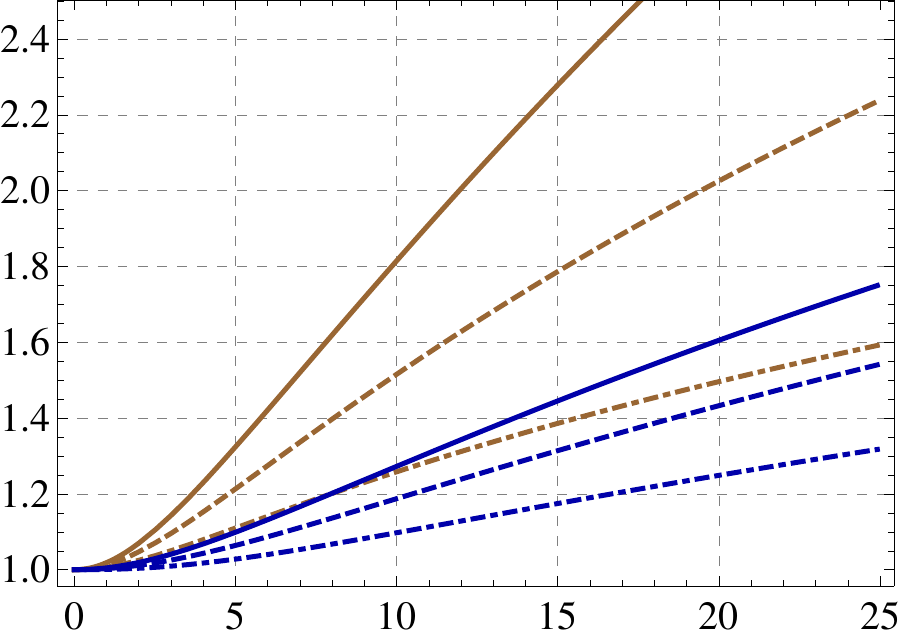}
\qquad
  \put(-510,25){\rotatebox{90}{$\chi_{(1)}/\chi_{(1)\mt{iso}}(T)$}}
         \put(-370,-10){$a/T$}
         \put(-333,25){\rotatebox{90}{$\chi_{(1)}/\chi_{(1)\mt{iso}}(T)$}}
         \put(-195,-10){$a/T$}
          \put(-160,25){\rotatebox{90}{$\chi_{(1)}/\chi_{(1)\mt{iso}}(T)$}}
         \put(-15,-10){$a/T$}
\\
(a) & \qquad(b) & \qquad(c)\\
& \\
\hspace{-0.9cm}
\includegraphics[width=5cm]{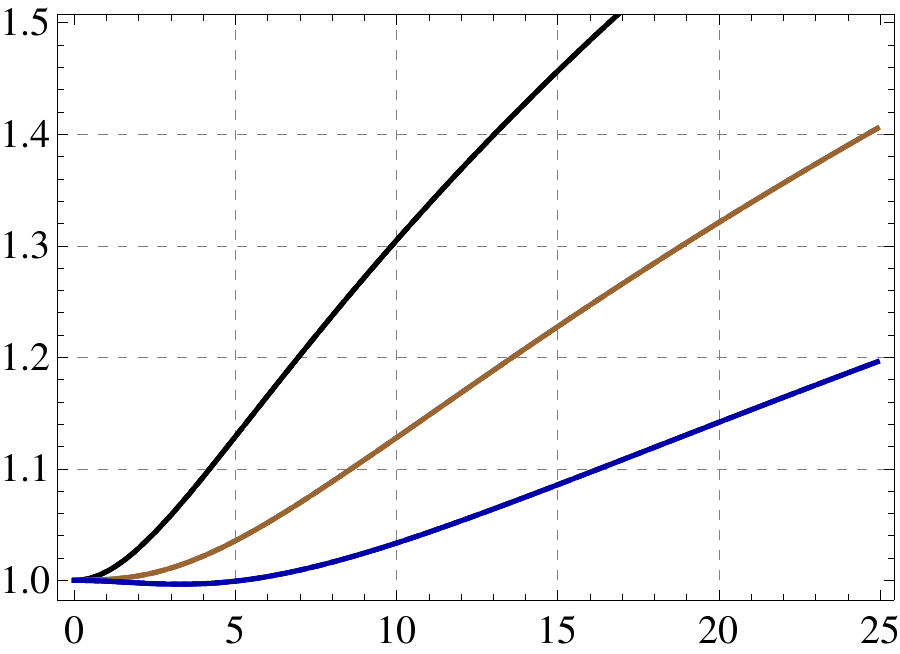} 
& 
\qquad \includegraphics[width=5cm]{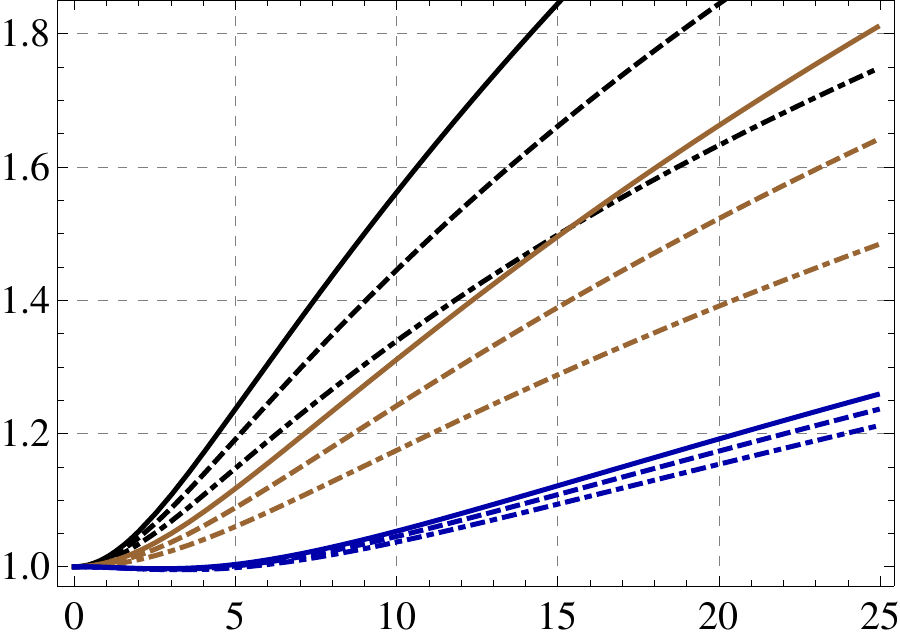}
& 
\qquad \includegraphics[width=5cm]{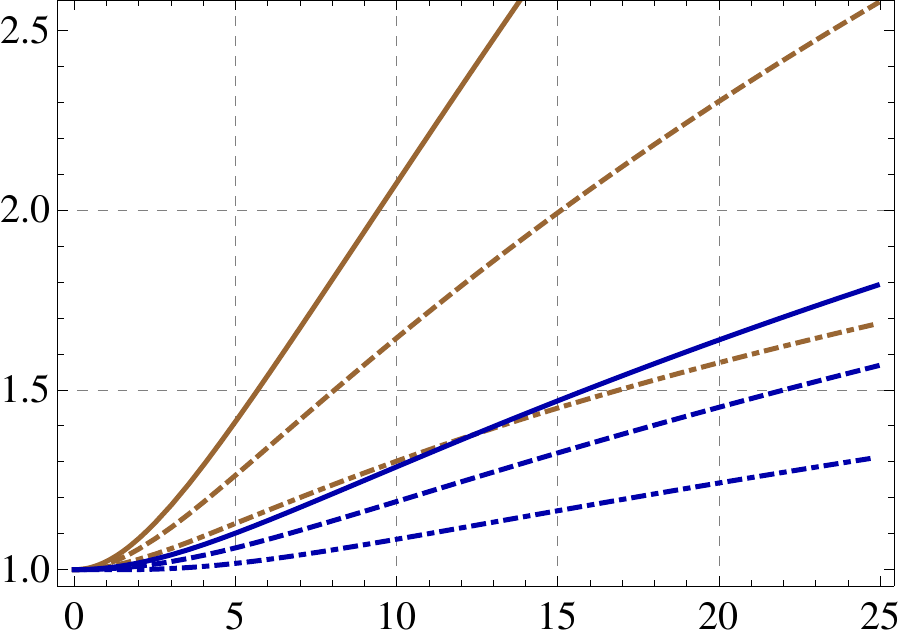}
\qquad
  \put(-510,25){\rotatebox{90}{$\chi_{(1)}/\chi_{(1)\mt{iso}}(T)$}}
         \put(-370,-10){$a/T$}
         \put(-333,25){\rotatebox{90}{$\chi_{(1)}/\chi_{(1)\mt{iso}}(T)$}}
         \put(-195,-10){$a/T$}
          \put(-160,25){\rotatebox{90}{$\chi_{(1)}/\chi_{(1)\mt{iso}}(T)$}}
         \put(-15,-10){$a/T$}
         \\
(d) & \qquad(e) & \qquad(f) \\
& \\
\hspace{-0.9cm}
\includegraphics[width=5cm]{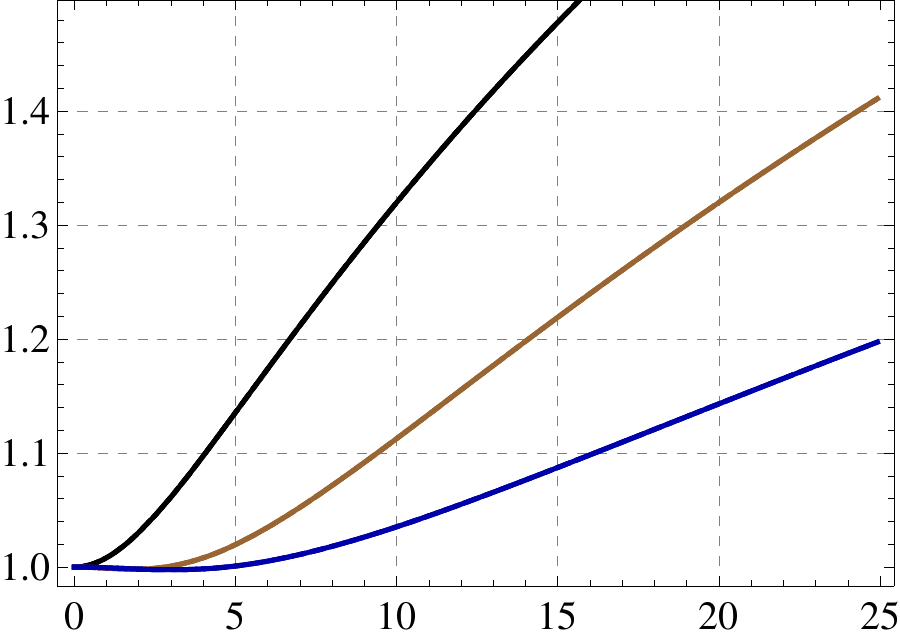} 
& 
\qquad \includegraphics[width=5cm]{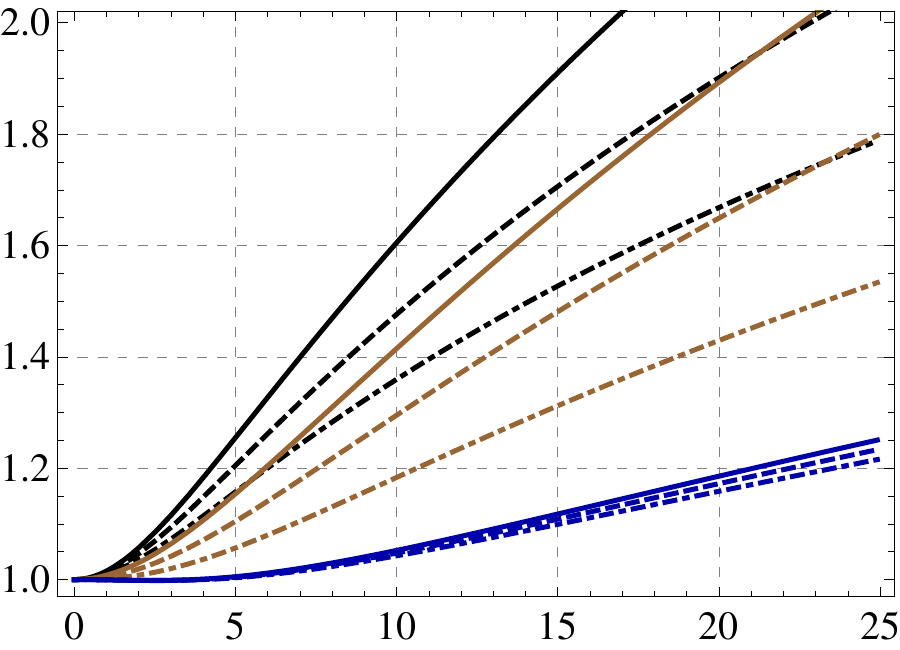}
& 
\qquad \includegraphics[width=5cm]{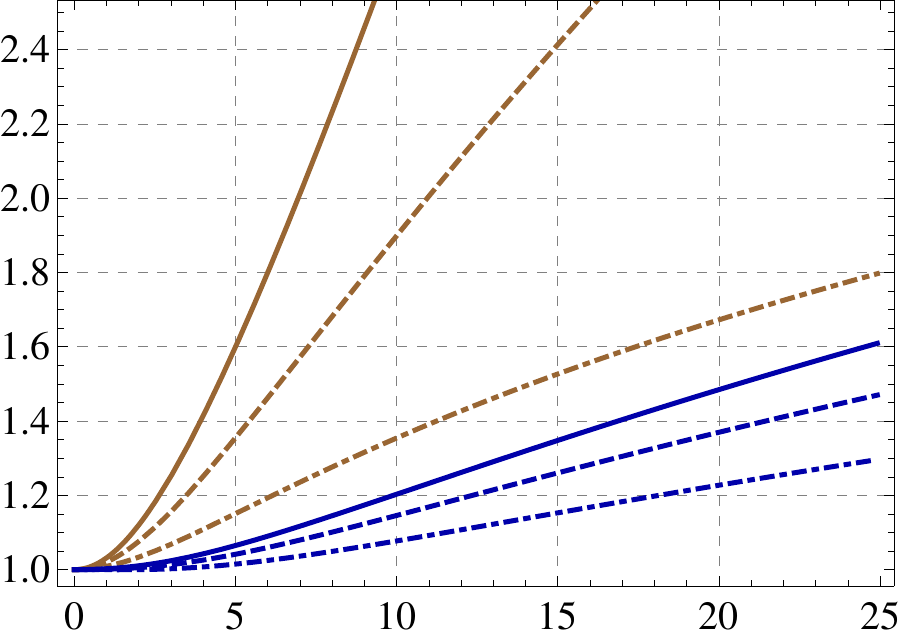}
\qquad
  \put(-510,25){\rotatebox{90}{$\chi_{(1)}/\chi_{(1)\mt{iso}}(T)$}}
         \put(-370,-10){$a/T$}
         \put(-333,25){\rotatebox{90}{$\chi_{(1)}/\chi_{(1)\mt{iso}}(T)$}}
         \put(-195,-10){$a/T$}
          \put(-160,25){\rotatebox{90}{$\chi_{(1)}/\chi_{(1)\mt{iso}}(T)$}}
         \put(-15,-10){$a/T$}
         \\
(g) & \qquad(h) & \qquad(i) \\
\end{tabular}
\end{center}
\caption{\small Plots of the spectral density $\chi_{(1)}$ normalized with respect to the isotropic result at fixed temperature $\chi_{(1)\mt{iso}}(T)$. Curves of different colors denote different values of $\textswab{w}$ as follows  $\textswab{w}=$1/2 (black), 1 (brown), 3/2 (blue). The angles are $\vartheta=0$ (solid), $\pi/4$ (dashed), $\pi/2$ (dash-dotted). Columns correspond to different values of $\textswab{q}$: from left to right it is $\textswab{q}=0,0.5,1$. Rows correspond to different values of the quark mass: from top to bottom it is $\psi_0=0,0.75,0.941$. Then, for instance, (f) corresponds to $\textswab{q}=1$, $\psi_\mt{H}=0.75$.}
\label{c4ac1}
\end{figure}

The equation to solve is (\ref{eomy1})) with $\vec{k}=q(\sin\vartheta,0,\cos\vartheta)$. Using the results obtained for different values of $q$, $\psi_\mt{H}$, and $\vartheta$ we compute the spectral density from
\be
\chi_{(1)}=  -\frac{4K_{\mt{D7}}}{\left|A_y(k, 0)\right|^2}\,\mbox{Im}\,  
\lim_{u \rightarrow \uh}  Q(u) A_y^*(k,u) A'_y(k, u) \,,
\label{chi1dd}
\ee
where $Q(u)$ was defined in (\ref{Q}).
The imaginary part of (\ref{chi1dd}) does not depend on $u$, as we shall prove in Appendix \ref{concur}. This justifies the fact that we evaluate the limit at $u=\uh$, where the numerics are under better control. The results are plotted in Fig.~\ref{c1w} for the spectral density $\chi_{(1)}$ as a function of $\wn$, and in Fig.~\ref{c1q} as a function of $\qn$.
In Figs.~\ref{c4bc1} and \ref{c4ac1} we plot $\chi_{(1)}$ as a function of the anisotropy $a/T$.


\subsection{Dilepton spectral density $\chi_{(2)}$}
\label{dilep2}

The gauge invariant fields satisfy  equations (\ref{eq:Ex})-(\ref{eq:Ez}). Solving such equations for $E_x$ and $E_z$ close to the boundary we find
\bea
E_{x}&=&E_{x}^{(0)}+\left( E_{x}^{(2)}\cos\vartheta + \frac{1}{2}E_{x}^{(0)} (q^2-k_0^2) \log u \right) u^{2}\cr
&& +\frac{1}{192}\Big(8\left(24 \psi_1^2 -5 a^2 + 3 (q^2-k_0^2)\right)E_x^{(2)}\cos\vartheta \cr
 &&
 \hskip 1cm -3\left(3 (q^2-k_0^2)^2+3 a^2 q^2-2 a^2 k_0^2+ a^2 q^2 \cos 2\vartheta\right)E_x^{(0)}\Big)\,u^4 \cr
&&
+\frac{1}{48}\left(24 \psi_1^2 -5 a^2 + 3 (q^2-k_0^2)\right)(q^2-k_0^2)u^4 \log u +O\left(u^{6}\right)\,,
\eea
and
\bea
E_{z}&=&E_{z}^{(0)}+ \frac{1}{2}E_{x}^{(0)}(q^2-k_0^2)\, u^2\,\mathrm{log}\,u 
\cr &&
 -\frac{3(q^2-k_0^2)[64\, E_z^{(4)}+3\,E_z^{(0)}(q^2-k_0^2)(a^2+q^2-k_0^2)]
 }{8 \left[ (q^2-k_0^2)(2\,a^2-3\,(q^2+8\,\psi_1^2-k_0^2))+3\,a^2\,q^2\,\mathrm{cos}^2\,\vartheta  \right]}   \,u^2  
 \cr &&
+ \frac{a^2\,E_{z}^{(0)}\,q^2\,(q^2-k_0^2)\,\mathrm{cos}^2\,\vartheta-a^2\,q^2\,\cos\vartheta\,\mathrm{sin}\,\vartheta \,[E_x^{(0)}(q^2-k_0^2)-8\,E_x^{(2)}\,\mathrm{cos}\,\vartheta]}{8 \left[ (q^2-k_0^2)(2\,a^2-3\,(q^2+8\,\psi_1^2-k_0^2))+3\,a^2\,q^2\,\mathrm{cos}^2\,\vartheta  \right]}\,u^2
\cr &&
+\left(E_z^{(4)}-\frac{1}{96}E_z^{(0)}(a^2(7q^2-4k_0^2)-6(q^2-k_0^2)(q^2+8\psi_1^2-k_0^2))\log u\right)u^4
\cr &&
+3 a^2 q^2 \left(E_x^{(0)} \sin 2\vartheta+ E_z^{(0)} \cos 2\vartheta\right)u^4\log u+O\left(u^{6}\right)\,,
\eea
where $E_x^{(0)}$, $E_x^{(2)}$, $E_z^{(0)}$, and $E_z^{(4)}$ are expansion coefficients which are not determined by the boundary equations, but that can be extracted from the numerical solutions, as we shall explain presently.

Using these expressions, we can write the boundary action as
\begin{equation}
S_{\epsilon}=-2K_\mt{D7}\int dt\, d\vec{x}\left[\mathcal{L}_{1}+\mathcal{L}_{2}+\mathcal{L}_{3}+\ldots
+O\left(u^{2}\right)\right]_{u=\epsilon}\,,
\label{bActiondil}
\end{equation}
where
\begin{align}
&\mathcal{L}_{1}=A_1\,E_x^{(0)\,2}+B_1\,E_z^{(0)\,2}+C_1\,E_x^{(0)}\,E_z^{(0)}\,,\cr
&\mathcal{L}_{2}=A_2\,E_x^{(0)}\,E_x^{(2)}+B_2\,E_x^{(0)}\,E_x^{(2)}+C_2\,E_x^{(0)}\,E_x^{(2)}+D_2\,E_x^{(0)}\,E_x^{(2)}\,,\cr
&\mathcal{L}_{3}=-\frac{\mathrm{log}\,u}{k_0^2}\left[(E_x^{(0)\,2}+E_z^{(0)\,2})\,k_0^2+(E_x^{(0)}\,\cos\vartheta - E_z^{(0)}\,\sin\vartheta)^2\,q^2 \right]\,,
\end{align}
and $A_i, B_i, C_i$ ($i=1,2$) and $D_2$  are given by (\ref{appformulas}) in Appendix \ref{app2}.
The contributions of $\mathcal{L}_{1}$ and $\mathcal{L}_{3}$ to the Green's functions are real, so they don't enter in the computation of  $\chi_{(2)}$. Defining
\begin{equation}
\nonumber
S_2=-2K_\mt{D7}\int dt\, d\vec{x}\,\mathcal{L}_{2}\,,
\end{equation}
we can write
 \begin{align}
&\frac{\delta^2 S_2}{\delta E_x^{(0)\,2}}=2 A_2 \frac{\delta E_x^{(2)} }{\delta E_x^{(0)}}+2 B_2 \frac{\delta E_z^{(4)} }{\delta E_x^{(0)}}\,, \cr
&\frac{\delta^2 S_2}{\delta E_z^{(0)\,2}}=2 C_2 \frac{\delta E_x^{(2)} }{\delta E_z^{(0)}}+2 D_2 \frac{\delta E_z^{(4)} }{\delta}\,, \cr
&\frac{\delta^2 S_2}{\delta E_x^{(0)} \delta E_z^{(0)} }= A_2 \frac{\delta E_x^{(2)} }{\delta E_z^{(0)}}+ B_2 \frac{\delta E_z^{(4)} }{\delta E_z^{(0)}}+ C_2 \frac{\delta E_x^{(2)} }{\delta E_x^{(0)}}+ D_2 \frac{\delta E_z^{(4)} }{\delta E_x^{(0)}}\,.
\end{align}
Using the explicit expressions for the coefficients, one can show that
\bea
\chi_{(2)} &=&  16\,K_\mt{D7}\,\mathrm{Im}\left( \cos\vartheta \, \frac{\delta E_x^{(2)}}{\delta E_x^{(0)}}+\frac{6 a^2 q^2 \cos^2\vartheta\,\sin\vartheta \,\frac{\delta E_x^{(2)}}{\delta E_z^{(0)}}+48(q^2-k_0^2)\,\frac{\delta E_z^{(4)}}{\delta E_z^{(0)}}}{6(q^2-k_0^2)(q^2-k_0^2+8\psi_1^2)-a^2(7q^2-4k_0^2)-3a^2q^2\cos 2\vartheta}\right)\,.\cr &&
\eea
When $q = k_0$ this expression reduces to (\ref{chi2simplified}), the expression used to calculate the photon production rate.

Having checked that there will be no divergent contributions for the correlators, we can now proceed as in \cite{photon} to find the way in which $E_x^{(2)}$ and $E_z^{(4)}$ vary with respect to $E_x^{(0)}$ and $E_z^{(0)}$. To calculate the functional derivative $\frac{\delta^2S_{\epsilon}}{\delta E^{(0)}_i\delta E^{(0)}_j}$, we can use the fields $E_x$ and $E_z$ to construct the column
\begin{eqnarray}
\textbf{E}\equiv\left(\begin{array}{cc}
E_x\\
E_z\\
\end{array}\right),
\label{Evector}
\end{eqnarray}
so that (\ref{eq:Ex}) and (\ref{eq:Ez}) can be written as the matrix equation
\begin{eqnarray}
\mathcal{M}^{-1}\left(\mathcal{M}\textbf{E}^{\prime}\right)^{\prime}+f(u)
\textbf{E}=0,
\label{matrixEOM}
\end{eqnarray}
where
\begin{eqnarray}
\mathcal{M}\equiv \frac{M\,g^{uu}}{k_0^2u^2\bar{k}^2}
\left(\begin{array}{cc}
(g^{tt}k_0^2+g^{zz}k_z^2)k_x^2 & -g^{xx}g^{zz}k_xk_z\\
-g^{xx}g^{zz}k_xk_z & (g^{tt}k_0^2+g^{xx}k_x^2)k_z^2\\
\end{array}\right)\,,
\qquad 
f(u)\equiv \frac{u^2\bar{k}^2}{g^{uu}}\, .
\end{eqnarray}
We can also write the boundary action (excluding the part with $A_yA_y^{\prime}$) as
\begin{eqnarray}
S_{\epsilon}=-2K_\mt{D7}\int dt \, d\vec x\left[\textbf{E}^\mt{T}\mathcal{M}\textbf{E}^{\prime}\right]_{u=\epsilon}.
\label{BdryAction}
\end{eqnarray}

Notice that if we can find two linearly independent solutions to (\ref{matrixEOM}), $\textbf{E}_{\bf 1}$ and $\textbf{E}_{\bf 2}$, such that at the boundary they reduce to $\textbf{E}_{\bf 1} \vert_\mt{bdry}=(1\;\; 0)^\mt{T}$ and $\textbf{E}_{\bf 2}\vert_\mt{bdry}=(0\;\; 1)^\mt{T}$, and we arrange them as the columns of a $2\times 2$ matrix $\mathcal{E}\equiv(\textbf{E}_{\bf 1}\;\; \textbf{E}_{\bf 2})$, then, given that (\ref{matrixEOM}) is linear, its general solution $\textbf{E}_\mt{sol}$ will be given by
\begin{eqnarray}
\textbf{E}_\mt{sol}=E_x^{(0)}\textbf{E}_{\bf 1}+E_z^{(0)}\textbf{E}_{\bf 2}=\mathcal{E}\left(\begin{array}{cc}
E_x^{(0)}\\
E_z^{(0)}\\
\end{array}\right).
\label{Egensolution}
\end{eqnarray}
Using (\ref{Egensolution}) we can write the boundary action (\ref{BdryAction}) as
\begin{eqnarray}
S_{\epsilon}=-2K_\mt{D7}\int dt\, d\vec x\left[(E_x^{(0)}\,E_z^{(0)})\mathcal{M}\mathcal{E}^\prime\left(\begin{array}{cc}
E_x^{(0)}\\
E_z^{(0)}\\
\end{array}\right)\right]_{u=\epsilon}\,,
\label{BdryActionMatrixtext}
\end{eqnarray}
where the fact that $\mathcal{E}$ becomes the identity matrix at the boundary has been used. From (\ref{BdryActionMatrixtext}) we see that the variation $\frac{\delta^2S_{\epsilon}}{\delta E^{(0)}_i\delta E^{(0)}_j}$, and hence the Green's function $G_{ij}^\mt{R}$, is given by the $ij$ component of the matrix $\mathcal{M}\mathcal{E}^\prime$.
As will be seen in Appendix \ref{concur}, this way of writing the variation of the action permits to express the imaginary part of the integrand in (\ref{BdryActionMatrixtext}), which is all we need to compute the spectral densities, in terms of $u$-independent quantities. The evaluation can then be done at the horizon, where the numerics are under better control. In Appendix \ref{concur} we also elaborate on how to construct the solutions necessary to carry out this procedure.

\begin{figure}
\begin{center}
\begin{tabular}{ccc}
\setlength{\unitlength}{1cm}
\hspace{-0.9cm}
\includegraphics[width=5cm]{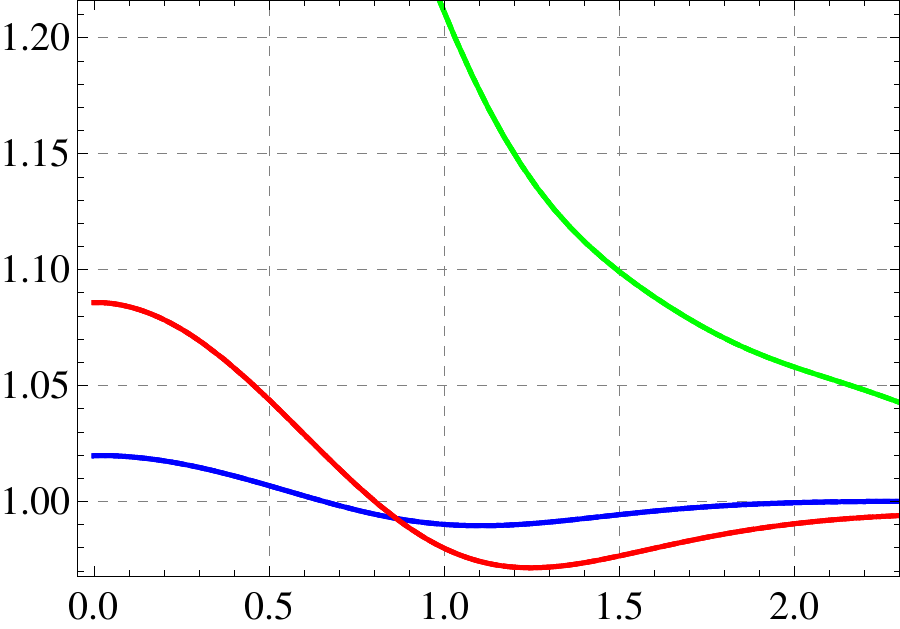} 
& 
\qquad \includegraphics[width=5cm]{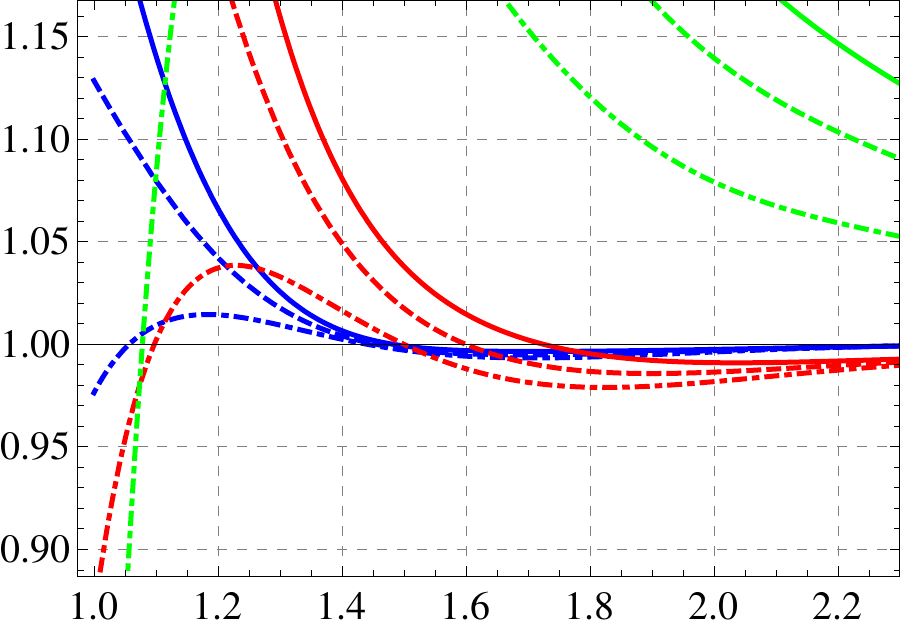}
& 
\qquad \includegraphics[width=5cm]{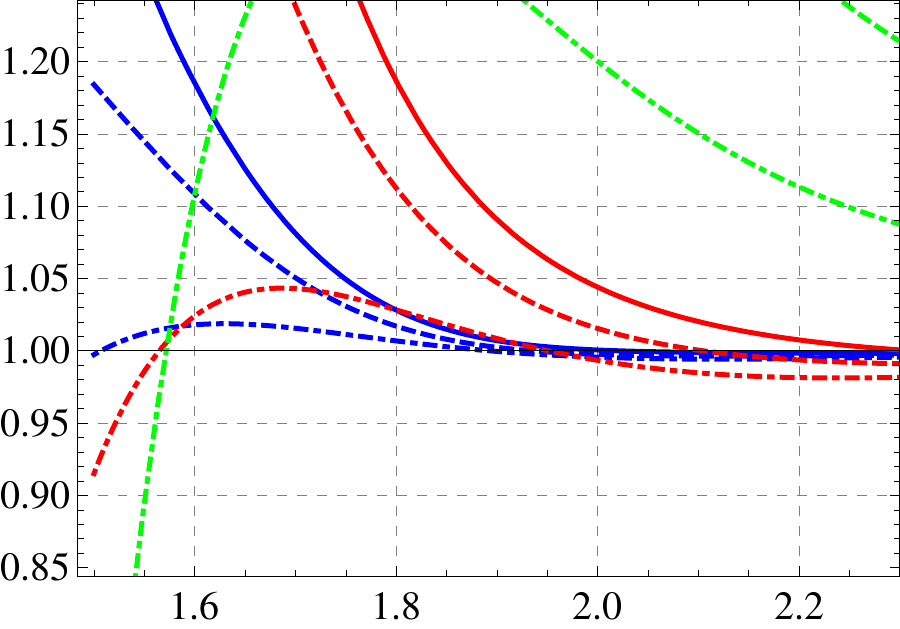}
\qquad
  \put(-510,25){\rotatebox{90}{$\chi_{(2)}/\chi_{(2)\mt{iso}}(T)$}}
         \put(-370,-10){$\textswab{w}$}
         \put(-333,25){\rotatebox{90}{$\chi_{(2)}/\chi_{(2)\mt{iso}}(T)$}}
         \put(-195,-10){$\textswab{w}$}
          \put(-160,25){\rotatebox{90}{$\chi_{(2)}/\chi_{(2)\mt{iso}}(T)$}}
         \put(-15,-10){$\textswab{w}$}
\\
(a) & \qquad(b) & \qquad(c)\\
& \\
\hspace{-0.9cm}
\includegraphics[width=5cm]{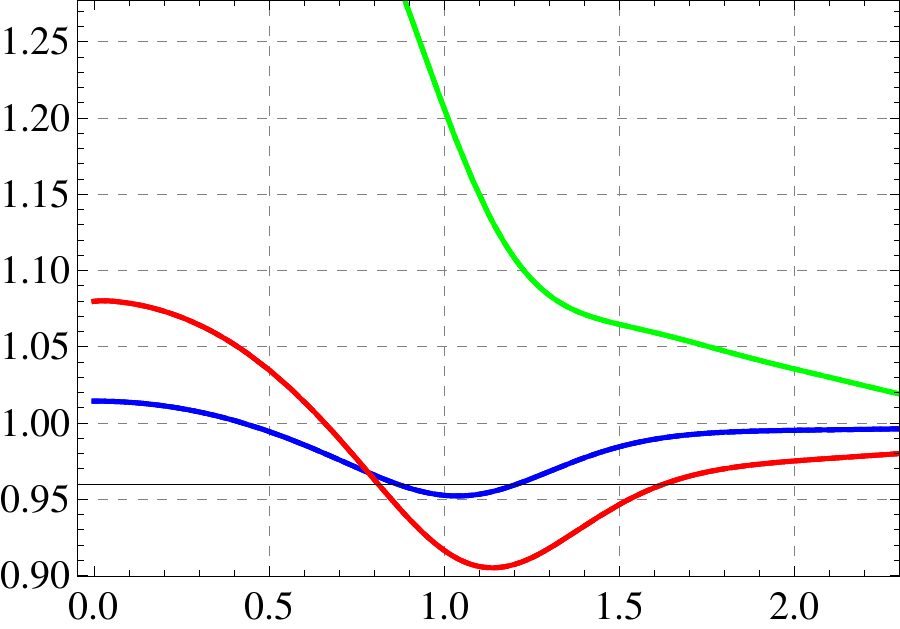} 
& 
\qquad \includegraphics[width=5cm]{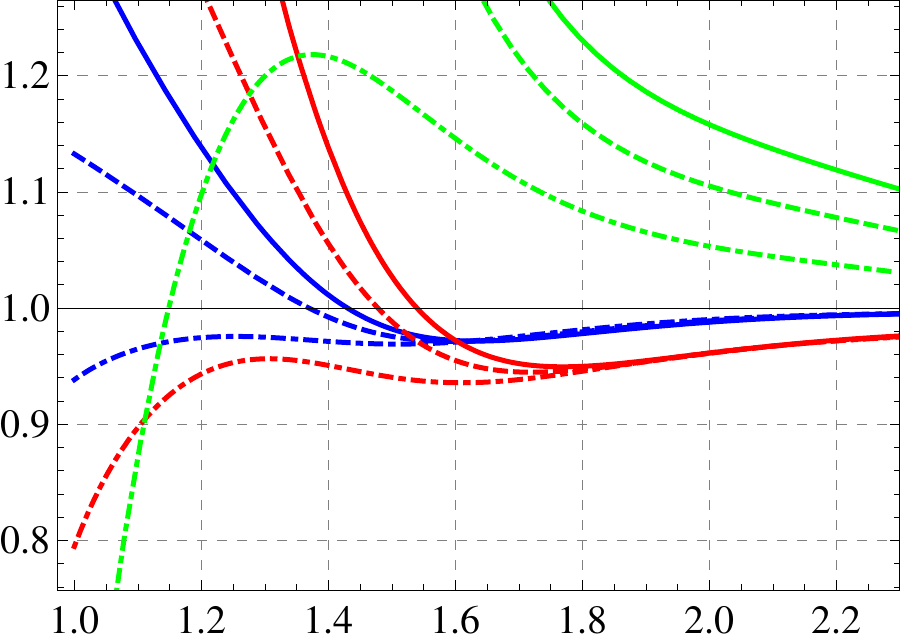}
& 
\qquad \includegraphics[width=5cm]{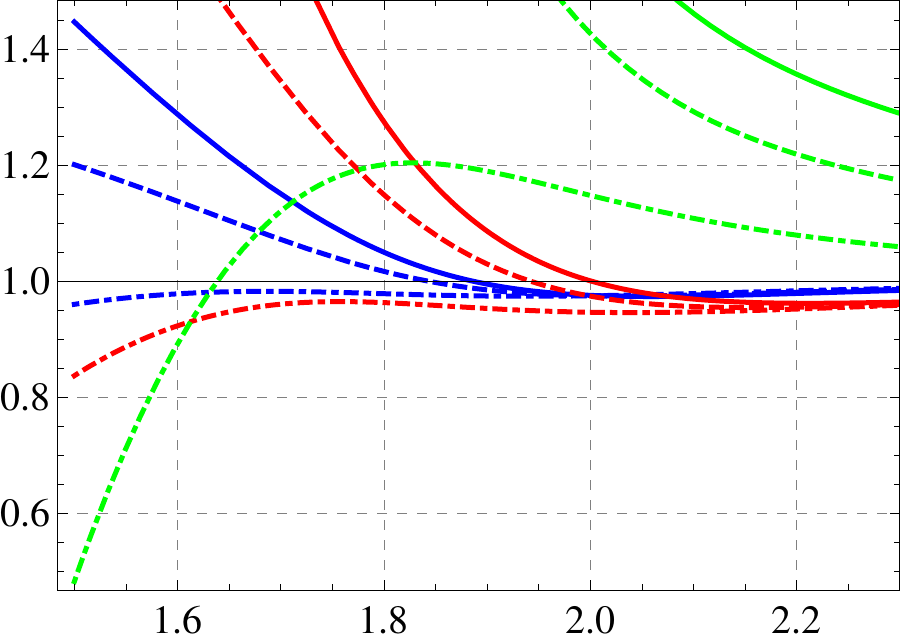}
\qquad
  \put(-510,25){\rotatebox{90}{$\chi_{(2)}/\chi_{(2)\mt{iso}}(T)$}}
         \put(-370,-10){$\textswab{w}$}
         \put(-333,25){\rotatebox{90}{$\chi_{(2)}/\chi_{(2)\mt{iso}}(T)$}}
         \put(-195,-10){$\textswab{w}$}
          \put(-160,25){\rotatebox{90}{$\chi_{(2)}/\chi_{(2)\mt{iso}}(T)$}}
         \put(-15,-10){$\textswab{w}$}
         \\
(d) & \qquad(e) & \qquad(f) \\
& \\
\hspace{-0.9cm}
\includegraphics[width=5cm]{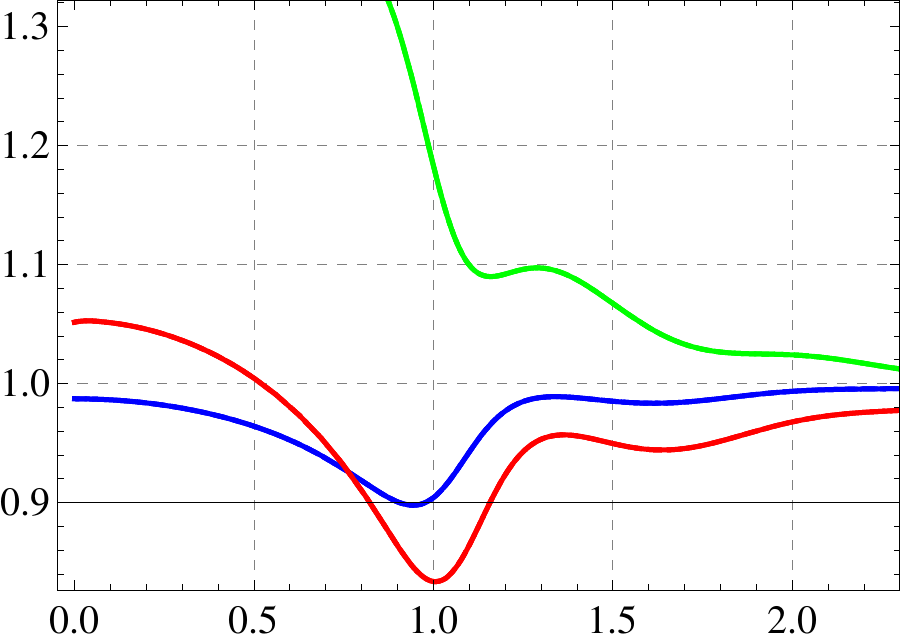} 
& 
\qquad \includegraphics[width=5cm]{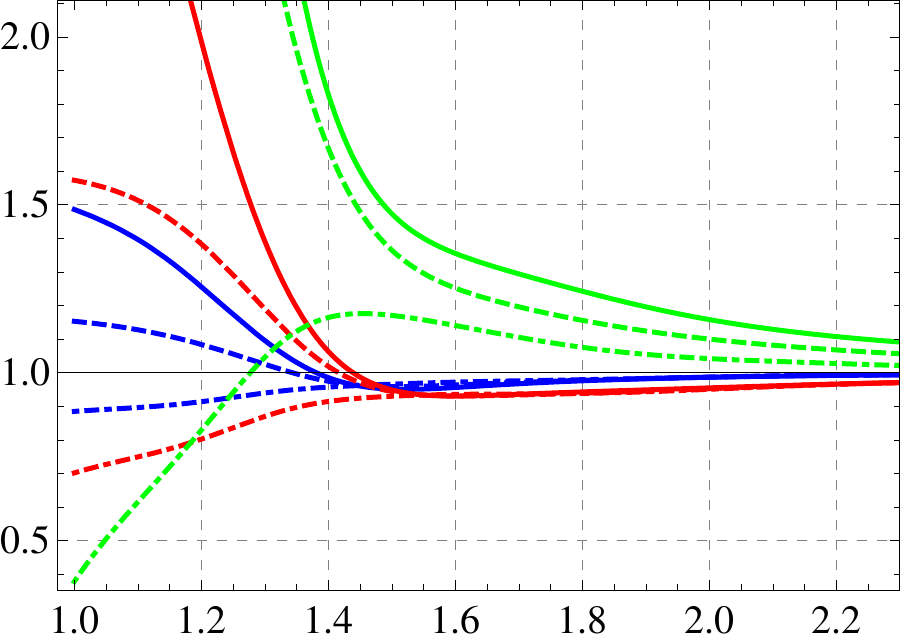}
& 
\qquad \includegraphics[width=5cm]{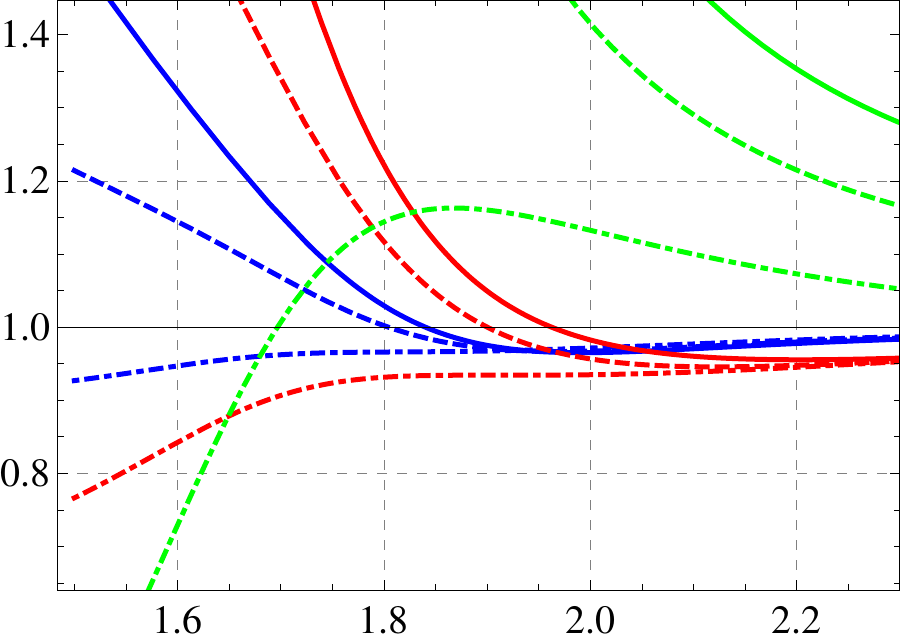}
\qquad
  \put(-510,25){\rotatebox{90}{$\chi_{(2)}/\chi_{(2)\mt{iso}}(T)$}}
         \put(-370,-10){$\textswab{w}$}
         \put(-333,25){\rotatebox{90}{$\chi_{(2)}/\chi_{(2)\mt{iso}}(T)$}}
         \put(-195,-10){$\textswab{w}$}
          \put(-160,25){\rotatebox{90}{$\chi_{(2)}/\chi_{(2)\mt{iso}}(T)$}}
         \put(-15,-10){$\textswab{w}$}
         \\
(g) & \qquad(h) & \qquad(i) \\
\end{tabular}
\end{center}
\caption{\small Plots of the spectral density $\chi_{(2)}$ normalized with respect to the isotropic result at fixed temperature $\chi_{(2)\mt{iso}}(T)$. Curves of different colors denote different values of $a/T$ as follows  $a/T=$4.41 (blue), 12.2 (red), 86 (green). The angles are $\vartheta=0$ (solid), $\pi/4$ (dashed), $\pi/2$ (dash-dotted). Columns correspond to different values of $\textswab{q}$: from left to right it is $\textswab{q}=0,1, 1.5$. Rows correspond to different values of the quark mass: from top to bottom it is $\psi_\mt{H}=0,0.75,0.941$. Then, for instance, (h) corresponds to $\textswab{q}=1$, $\psi_\mt{H}=0.941$. }
\label{c2w}
\end{figure}

\begin{figure}
\begin{center}
\begin{tabular}{ccc}
\setlength{\unitlength}{1cm}
\hspace{-0.9cm}
\includegraphics[width=5cm]{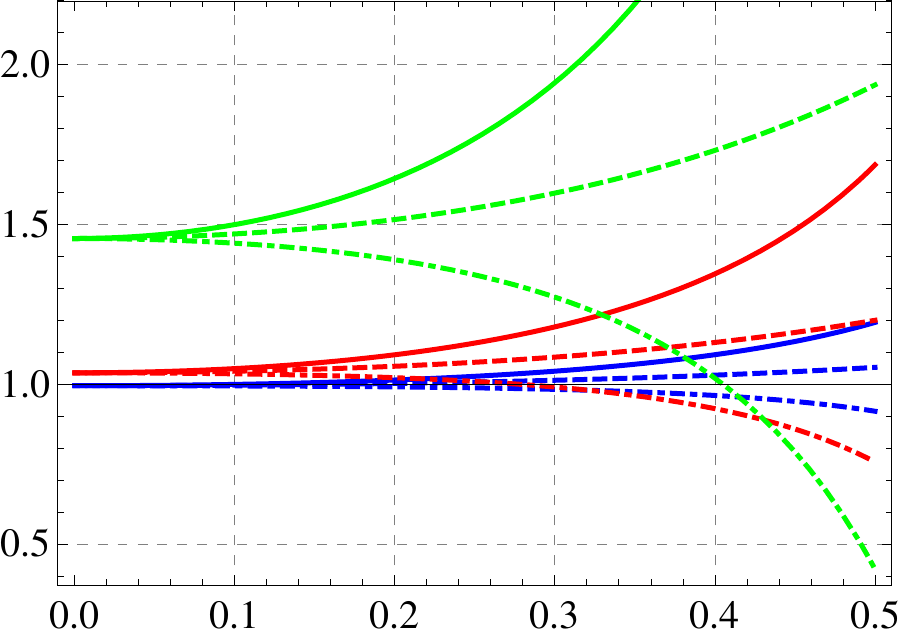} 
& 
\qquad \includegraphics[width=5cm]{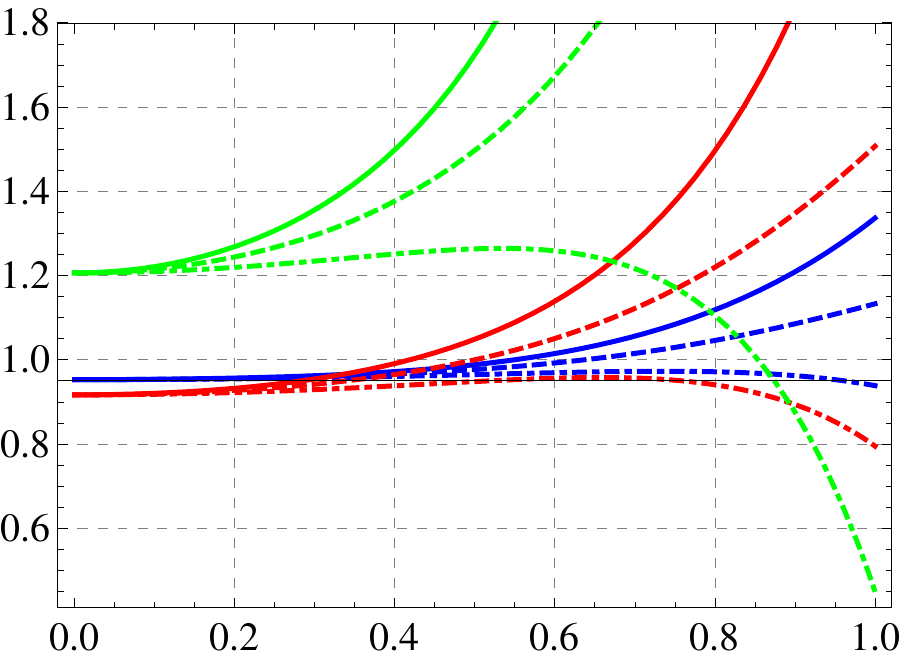}
& 
\qquad \includegraphics[width=5cm]{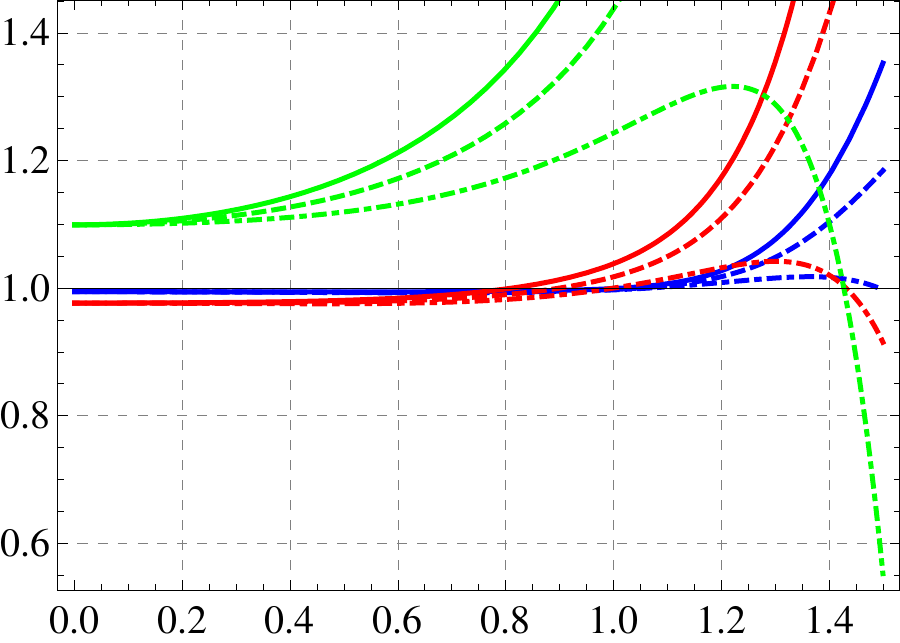}
\qquad
  \put(-510,25){\rotatebox{90}{$\chi_{(2)}/\chi_{(2)\mt{iso}}(T)$}}
         \put(-370,-10){$\textswab{q}$}
         \put(-333,25){\rotatebox{90}{$\chi_{(2)}/\chi_{(2)\mt{iso}}(T)$}}
         \put(-195,-10){$\textswab{q}$}
          \put(-160,25){\rotatebox{90}{$\chi_{(2)}/\chi_{(2)\mt{iso}}(T)$}}
         \put(-15,-10){$\textswab{q}$}
\\
(a) & \qquad(b) & \qquad(c)\\
& \\
\hspace{-0.9cm}
\includegraphics[width=5cm]{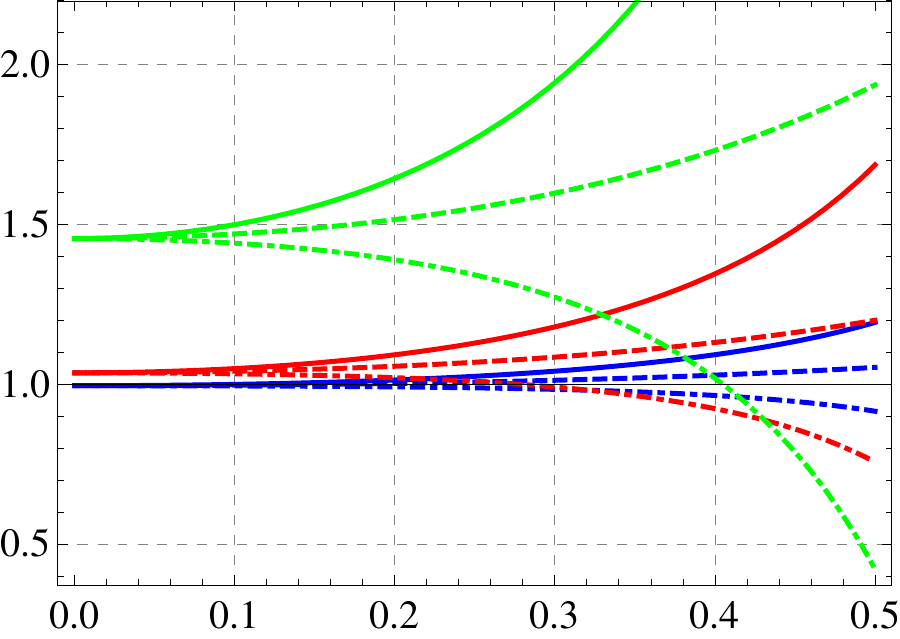} 
& 
\qquad \includegraphics[width=5cm]{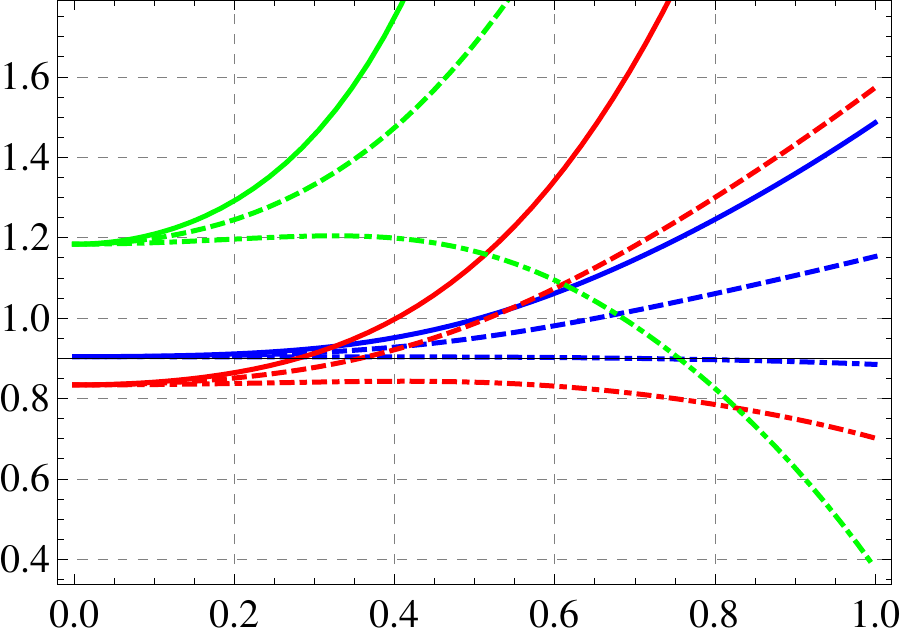}
& 
\qquad \includegraphics[width=5cm]{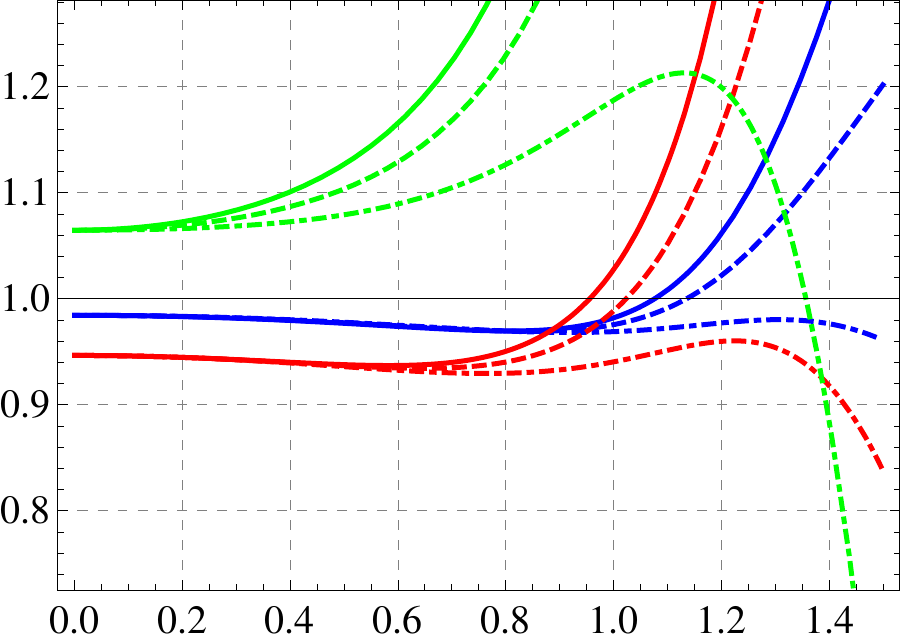}
\qquad
  \put(-510,25){\rotatebox{90}{$\chi_{(2)}/\chi_{(2)\mt{iso}}(T)$}}
         \put(-370,-10){$\textswab{q}$}
         \put(-333,25){\rotatebox{90}{$\chi_{(2)}/\chi_{(2)\mt{iso}}(T)$}}
         \put(-195,-10){$\textswab{q}$}
          \put(-160,25){\rotatebox{90}{$\chi_{(2)}/\chi_{(2)\mt{iso}}(T)$}}
         \put(-15,-10){$\textswab{q}$}
         \\
(d) & \qquad(e) & \qquad(f) \\
& \\
\hspace{-0.9cm}
\includegraphics[width=5cm]{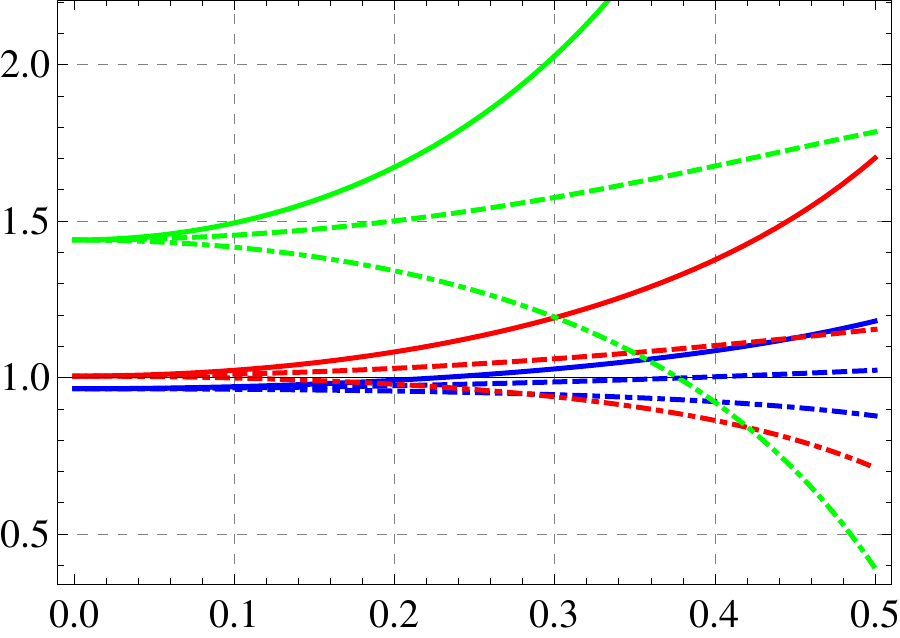} 
& 
\qquad \includegraphics[width=5cm]{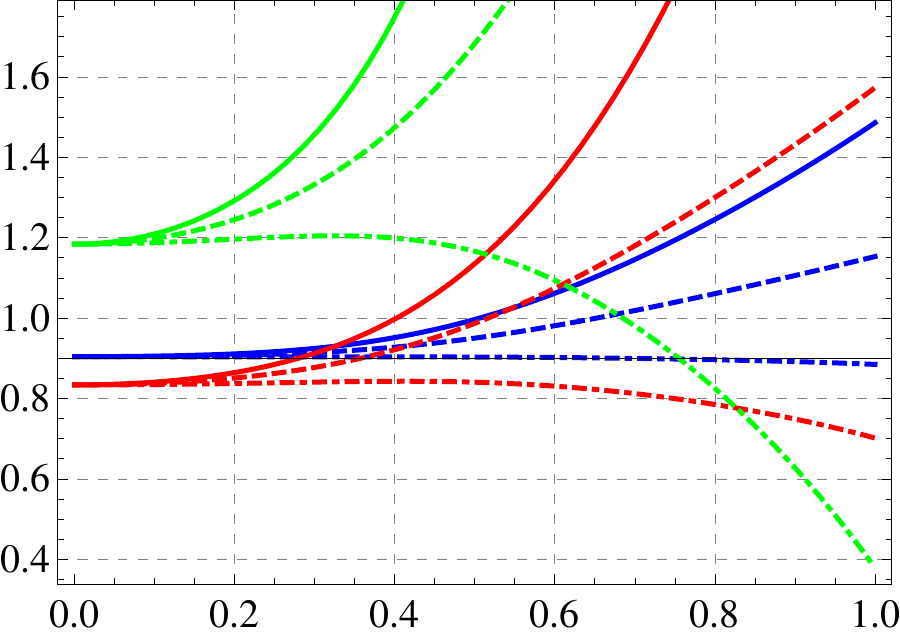}
& 
\qquad \includegraphics[width=5cm]{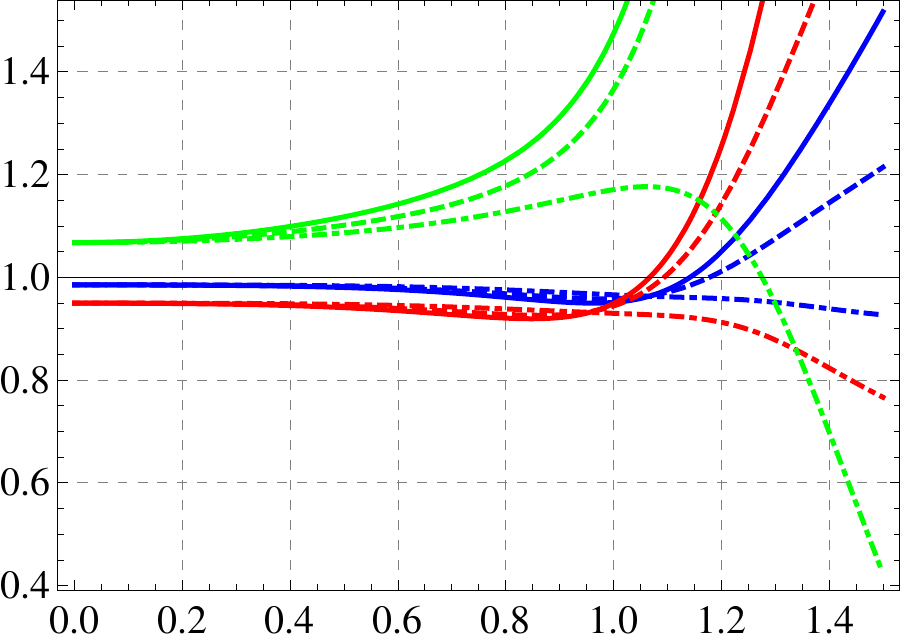}
\qquad
  \put(-510,25){\rotatebox{90}{$\chi_{(2)}/\chi_{(2)\mt{iso}}(T)$}}
         \put(-370,-10){$\textswab{q}$}
         \put(-333,25){\rotatebox{90}{$\chi_{(2)}/\chi_{(2)\mt{iso}}(T)$}}
         \put(-195,-10){$\textswab{q}$}
          \put(-160,25){\rotatebox{90}{$\chi_{(2)}/\chi_{(2)\mt{iso}}(T)$}}
         \put(-15,-10){$\textswab{q}$}
         \\
(g) & \qquad(h) & \qquad(i) \\
\end{tabular}
\end{center}
\caption{\small Plots of the spectral density $\chi_{(2)}$ normalized with respect to the isotropic result at fixed temperature $\chi_{(2)\mt{iso}}(T)$. Curves of different colors denote different values of $a/T$ as follows  $a/T=$4.41 (blue), 12.2 (red), 86 (green). The angles are $\vartheta=0$ (solid), $\pi/4$ (dashed), $\pi/2$ (dash-dotted). Columns correspond to different values of $\textswab{w}$: from left to right it is $\textswab{w}=0.5,1,1.5$. Rows correspond to different values of the quark mass: from top to bottom it is $\psi_\mt{H }=0,0.75,0.941$. Then, for instance, (f) corresponds to $\textswab{w}=1.5$, $\psi_\mt{H}=0.75$. }
\label{c2q}
\end{figure}

\begin{figure}
\begin{center}
\begin{tabular}{ccc}
\setlength{\unitlength}{1cm}
\hspace{-0.9cm}
\includegraphics[width=5cm]{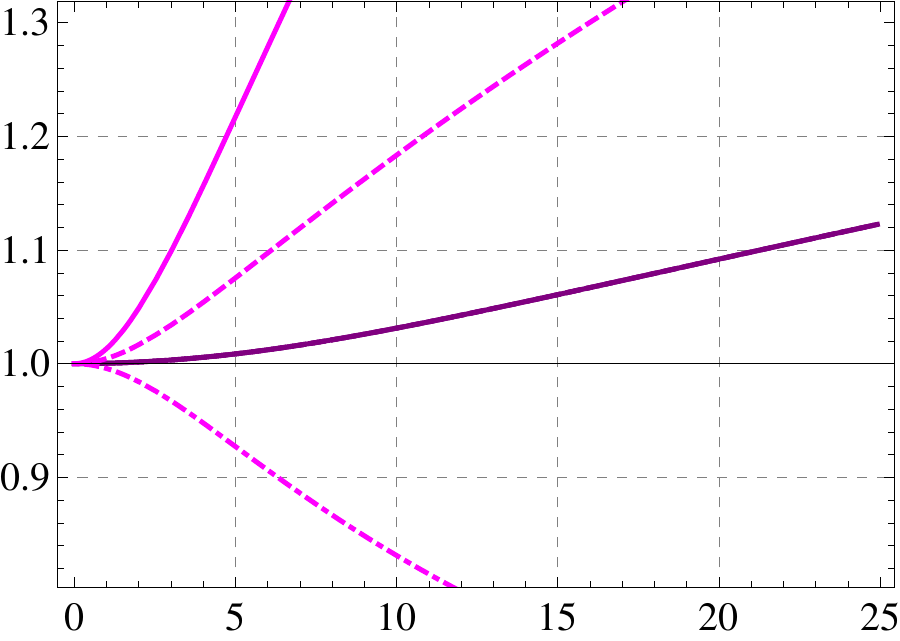} 
& 
\qquad \includegraphics[width=5cm]{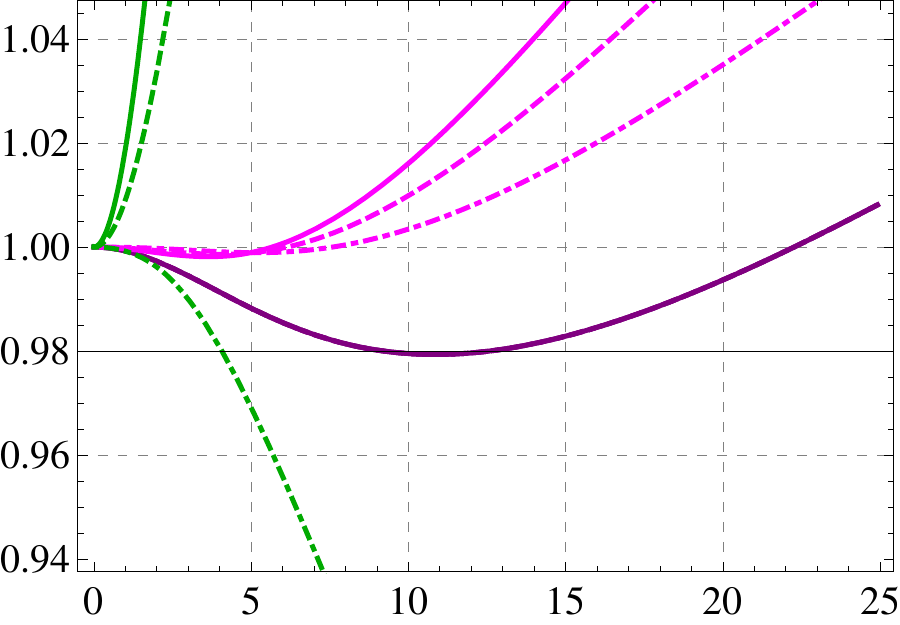}
& 
\qquad \includegraphics[width=5cm]{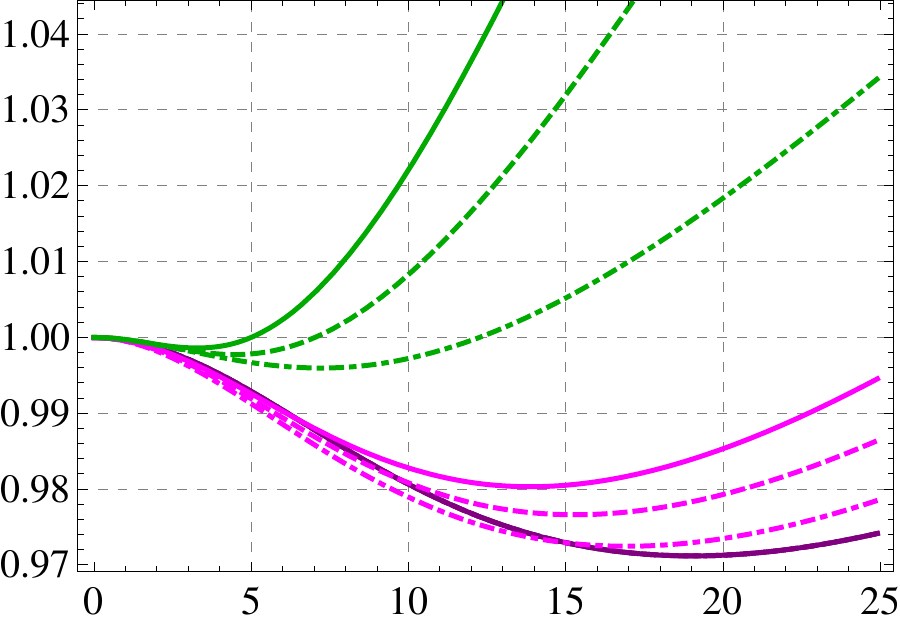}
\qquad
  \put(-510,25){\rotatebox{90}{$\chi_{(2)}/\chi_{(2)\mt{iso}}(T)$}}
         \put(-370,-10){$a/T$}
         \put(-333,25){\rotatebox{90}{$\chi_{(2)}/\chi_{(2)\mt{iso}}(T)$}}
         \put(-195,-10){$a/T$}
          \put(-160,25){\rotatebox{90}{$\chi_{(2)}/\chi_{(2)\mt{iso}}(T)$}}
         \put(-15,-10){$a/T$}
\\
(a) & \qquad(b) & \qquad(c)\\
& \\
\hspace{-0.9cm}
\includegraphics[width=5cm]{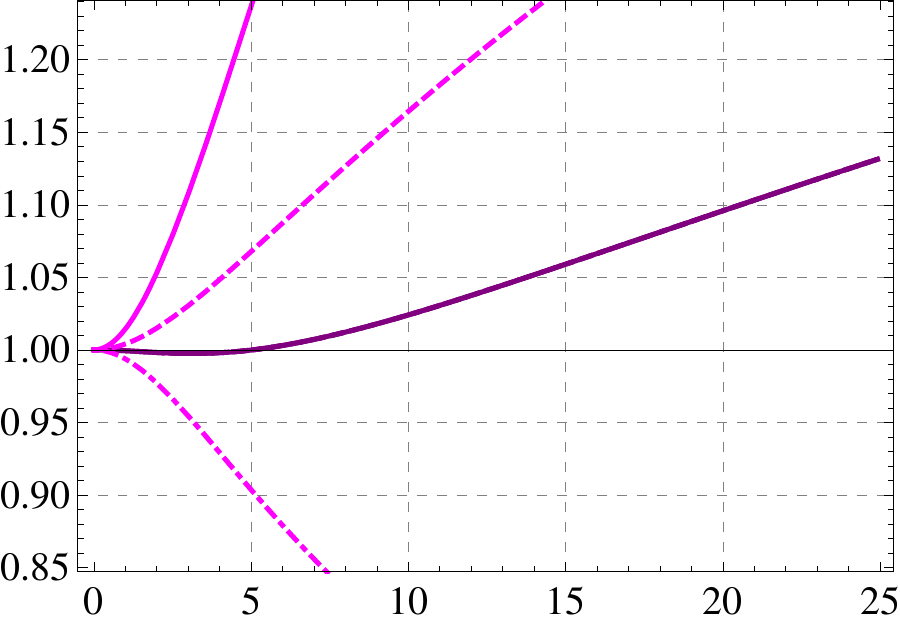} 
& 
\qquad \includegraphics[width=5cm]{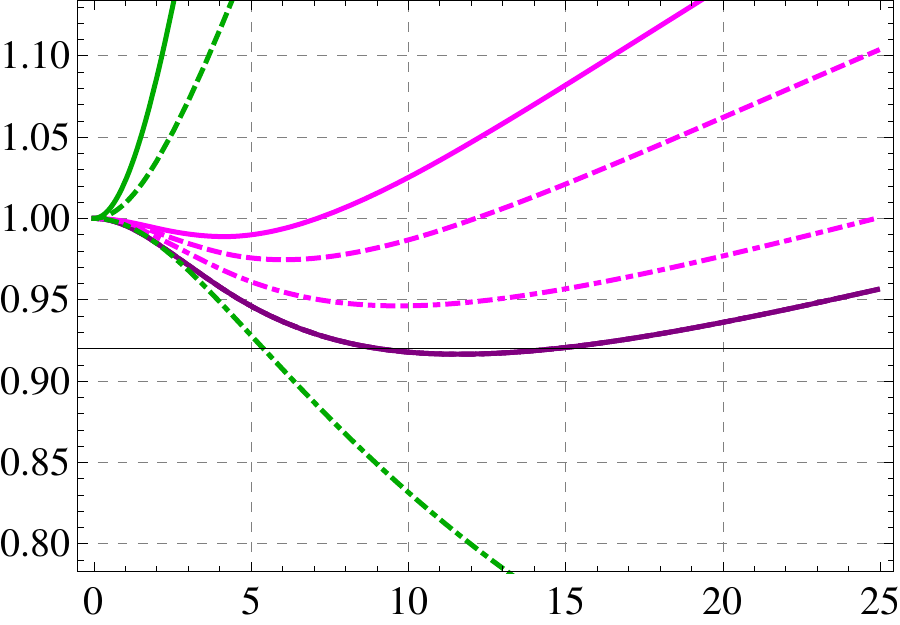}
& 
\qquad \includegraphics[width=5cm]{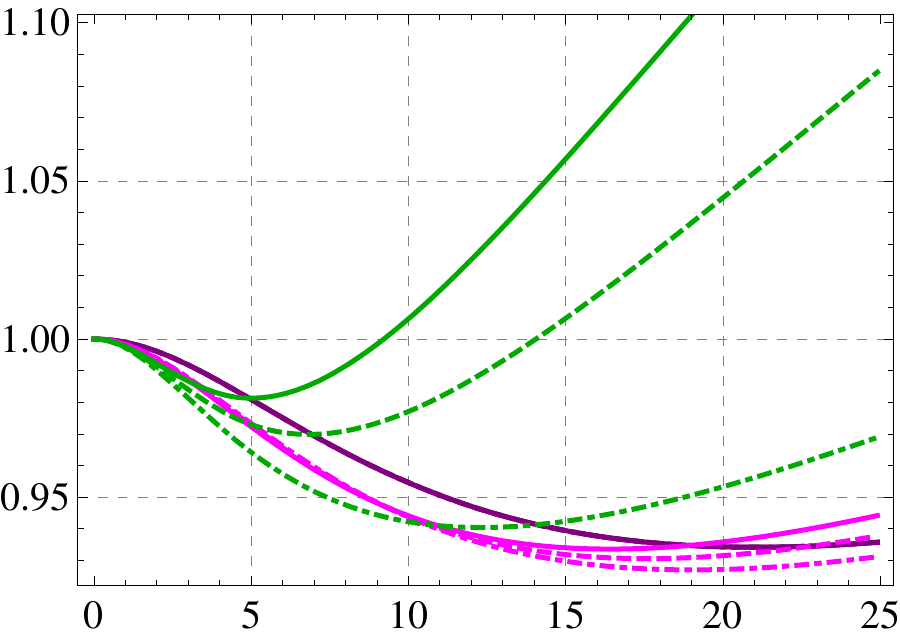}
\qquad
  \put(-510,25){\rotatebox{90}{$\chi_{(2)}/\chi_{(2)\mt{iso}}(T)$}}
         \put(-370,-10){$a/T$}
         \put(-333,25){\rotatebox{90}{$\chi_{(2)}/\chi_{(2)\mt{iso}}(T)$}}
         \put(-195,-10){$a/T$}
          \put(-160,25){\rotatebox{90}{$\chi_{(2)}/\chi_{(2)\mt{iso}}(T)$}}
         \put(-15,-10){$a/T$}
         \\
(d) & \qquad(e) & \qquad(f) \\
& \\
\hspace{-0.9cm}
\includegraphics[width=5cm]{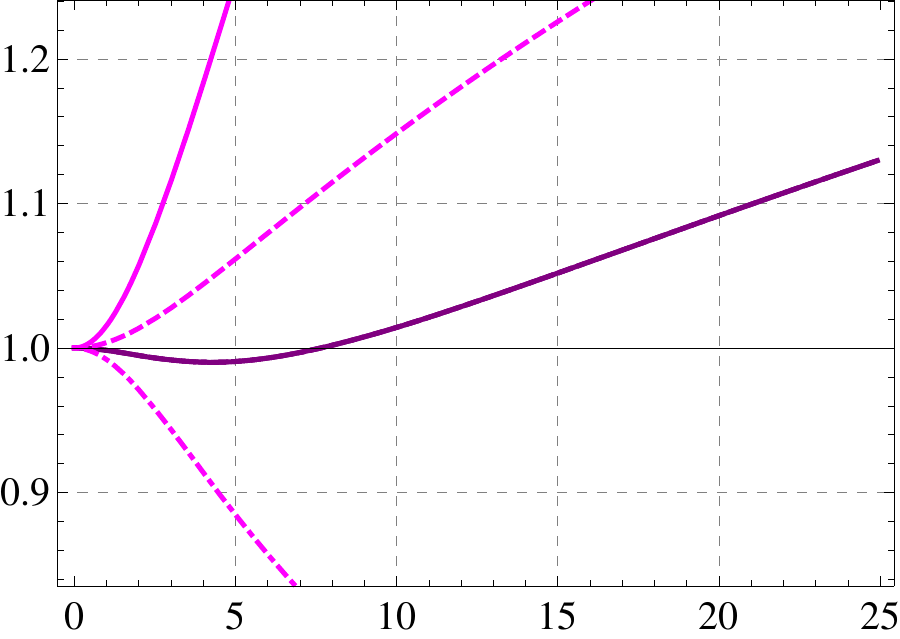} 
& 
\qquad \includegraphics[width=5cm]{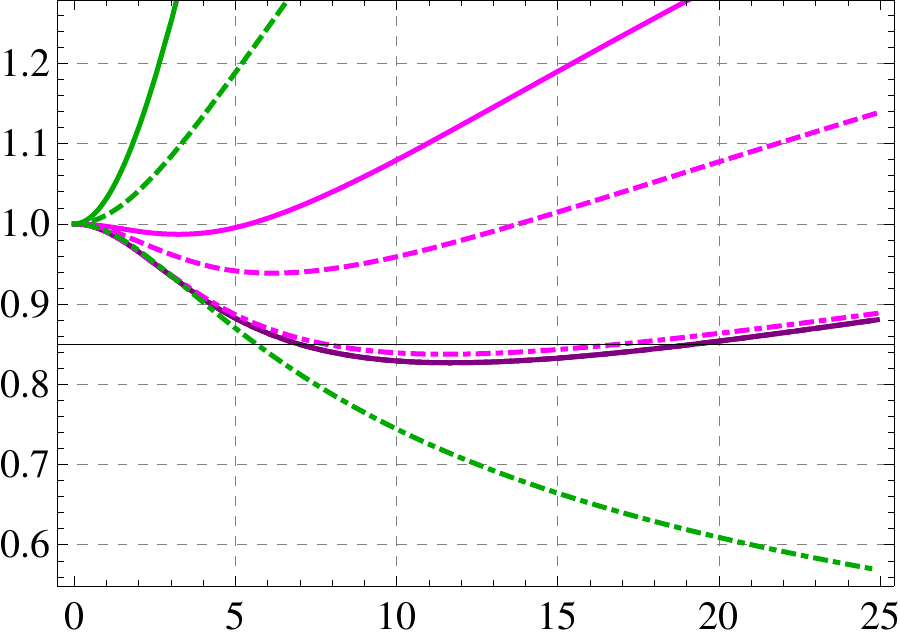}
& 
\qquad \includegraphics[width=5cm]{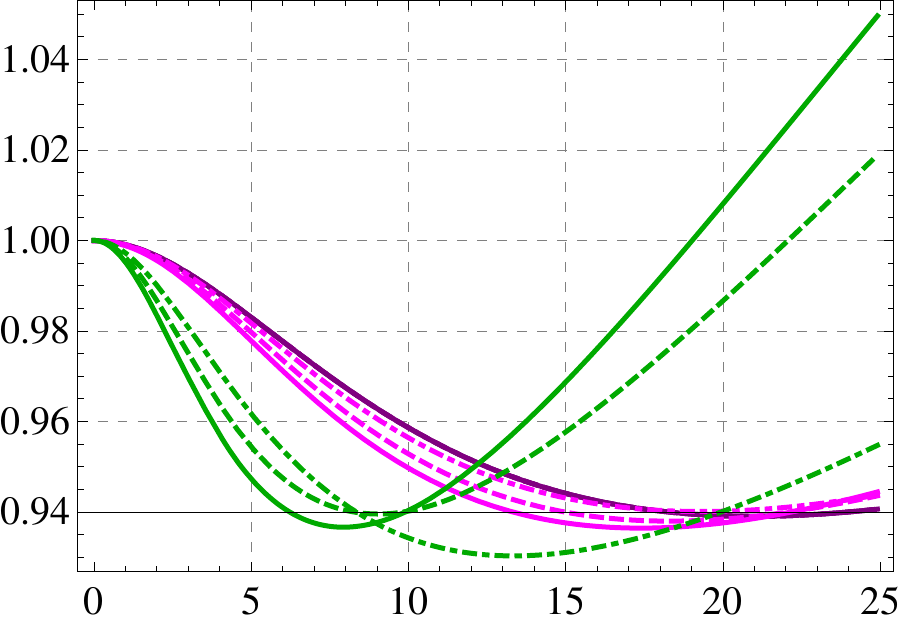}
\qquad
  \put(-510,25){\rotatebox{90}{$\chi_{(2)}/\chi_{(2)\mt{iso}}(T)$}}
         \put(-370,-10){$a/T$}
         \put(-333,25){\rotatebox{90}{$\chi_{(2)}/\chi_{(2)\mt{iso}}(T)$}}
         \put(-195,-10){$a/T$}
          \put(-160,25){\rotatebox{90}{$\chi_{(2)}/\chi_{(2)\mt{iso}}(T)$}}
         \put(-15,-10){$a/T$}
         \\
(g) & \qquad(h) & \qquad(i) \\
\end{tabular}
\end{center}
\caption{\small Plots of the spectral density $\chi_{(2)}$ normalized with respect to the isotropic result at fixed temperature $\chi_{(2)\mt{iso}}(T)$. Curves of different colors denote different values of $\textswab{q}$ as follows  $\textswab{q}=$0 (purple), 0.5 (magenta), 1 (green). The angles are $\vartheta=0$ (solid), $\pi/4$ (dashed), $\pi/2$ (dash-dotted). Columns correspond to different values of $\textswab{w}$: from left to right it is $\textswab{w}=0.5,1,1.5$. Rows correspond to different values of the quark mass: from top to bottom it is $\psi_\mt{H}=0,0.75,0.941$. Then, for instance, (f) corresponds to $\textswab{w}=1.5$, $\psi_\mt{H}=0.75$.}
\label{c4bc2}
\end{figure}

\begin{figure}
\begin{center}
\begin{tabular}{ccc}
\setlength{\unitlength}{1cm}
\hspace{-0.9cm}
\includegraphics[width=5cm]{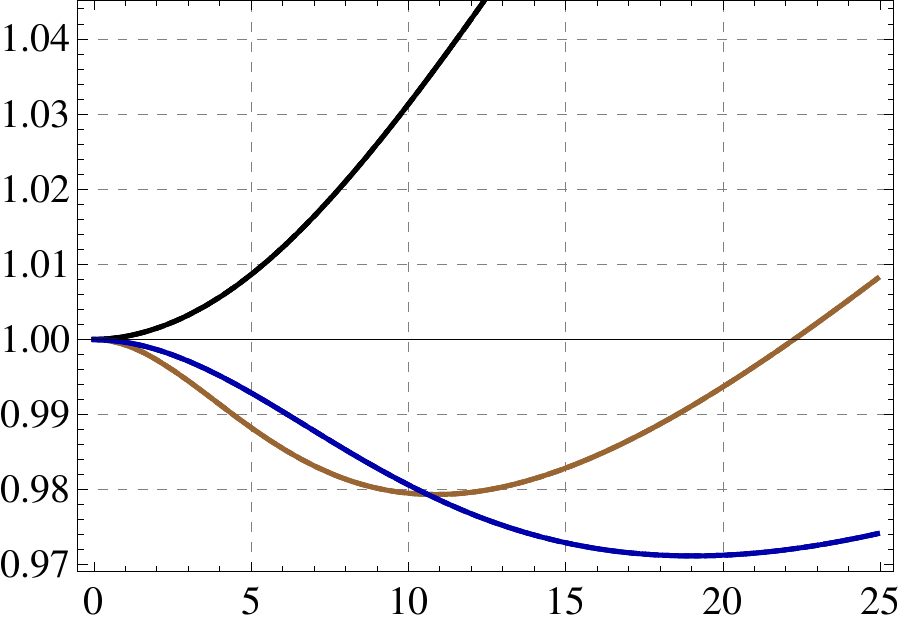} 
& 
\qquad \includegraphics[width=5cm]{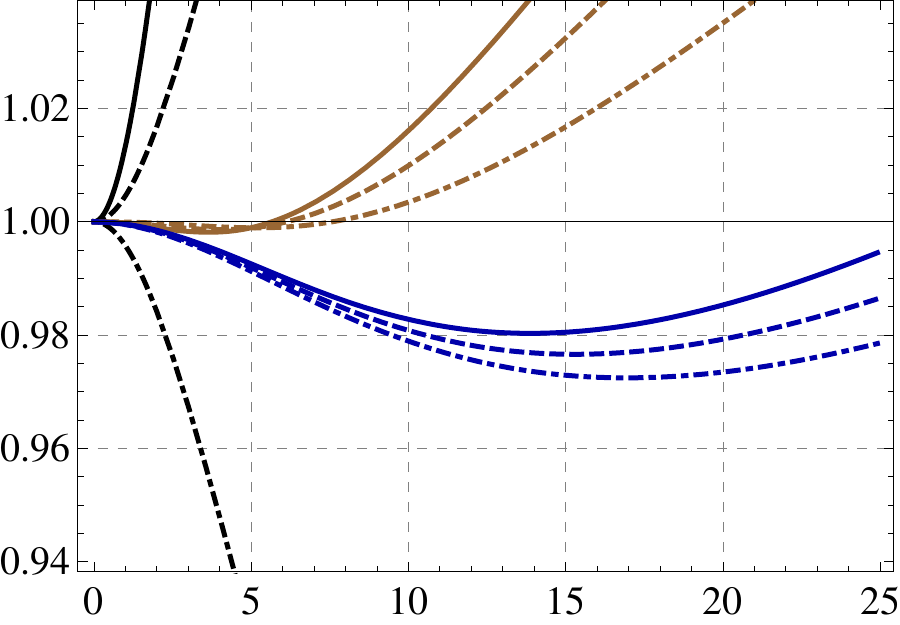}
& 
\qquad \includegraphics[width=5cm]{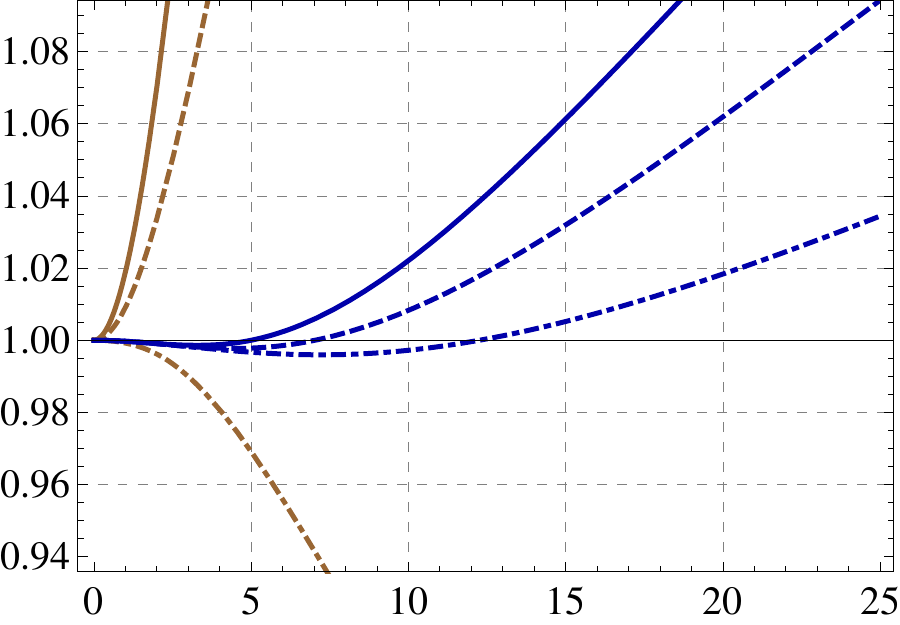}
\qquad
  \put(-510,25){\rotatebox{90}{$\chi_{(2)}/\chi_{(2)\mt{iso}}(T)$}}
         \put(-370,-10){$a/T$}
         \put(-333,25){\rotatebox{90}{$\chi_{(2)}/\chi_{(2)\mt{iso}}(T)$}}
         \put(-195,-10){$a/T$}
          \put(-160,25){\rotatebox{90}{$\chi_{(2)}/\chi_{(2)\mt{iso}}(T)$}}
         \put(-15,-10){$a/T$}
\\
(a) & \qquad(b) & \qquad(c)\\
& \\
\hspace{-0.9cm}
\includegraphics[width=5cm]{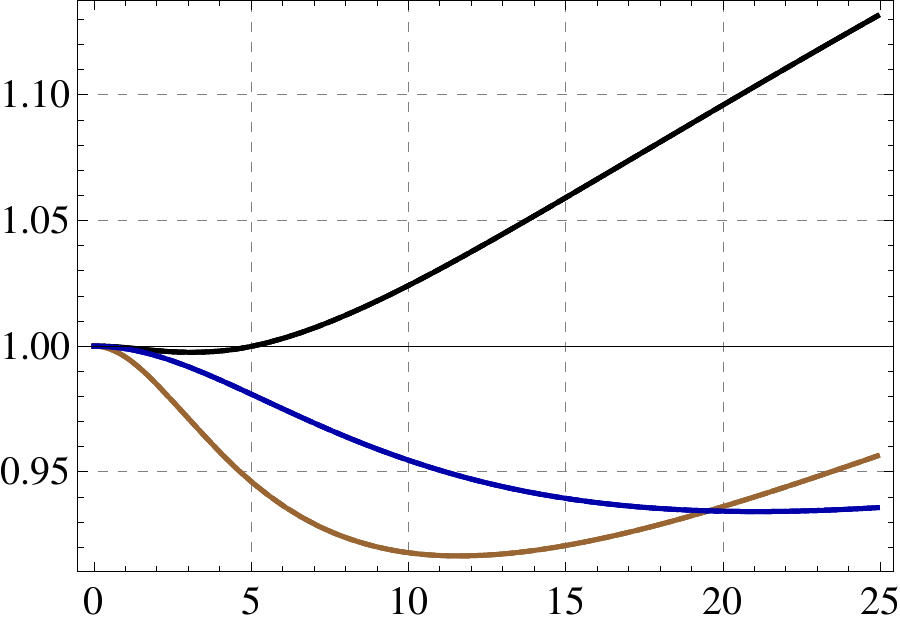} 
& 
\qquad \includegraphics[width=5cm]{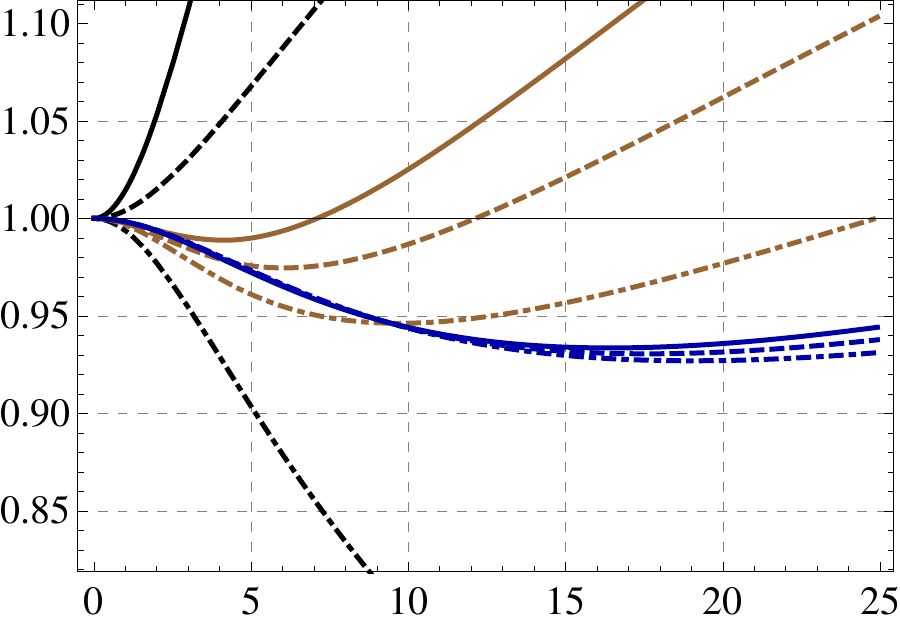}
& 
\qquad \includegraphics[width=5cm]{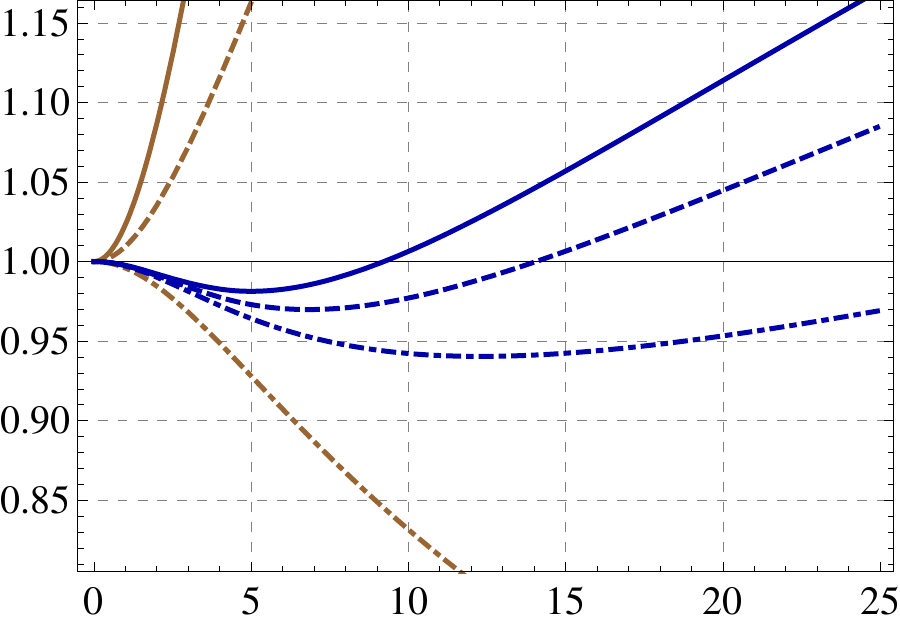}
\qquad
  \put(-510,25){\rotatebox{90}{$\chi_{(2)}/\chi_{(2)\mt{iso}}(T)$}}
         \put(-370,-10){$a/T$}
         \put(-333,25){\rotatebox{90}{$\chi_{(2)}/\chi_{(2)\mt{iso}}(T)$}}
         \put(-195,-10){$a/T$}
          \put(-160,25){\rotatebox{90}{$\chi_{(2)}/\chi_{(2)\mt{iso}}(T)$}}
         \put(-15,-10){$a/T$}
         \\
(d) & \qquad(e) & \qquad(f) \\
& \\
\hspace{-0.9cm}
\includegraphics[width=5cm]{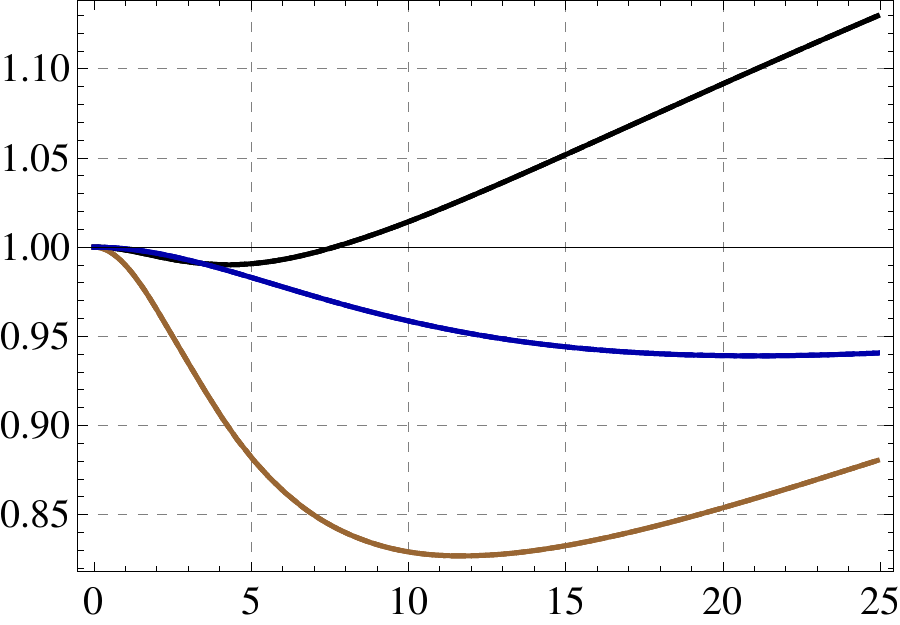} 
& 
\qquad \includegraphics[width=5cm]{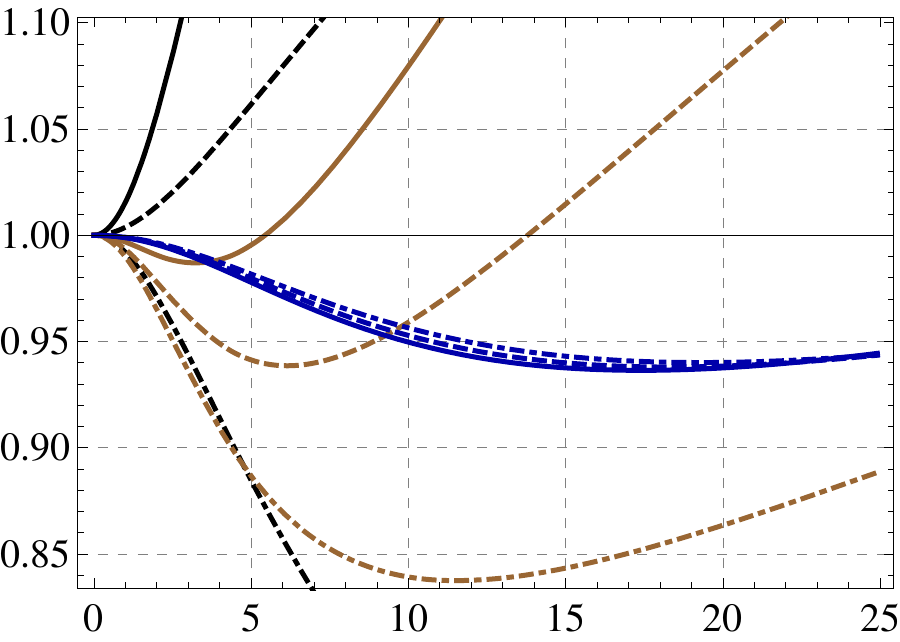}
& 
\qquad \includegraphics[width=5cm]{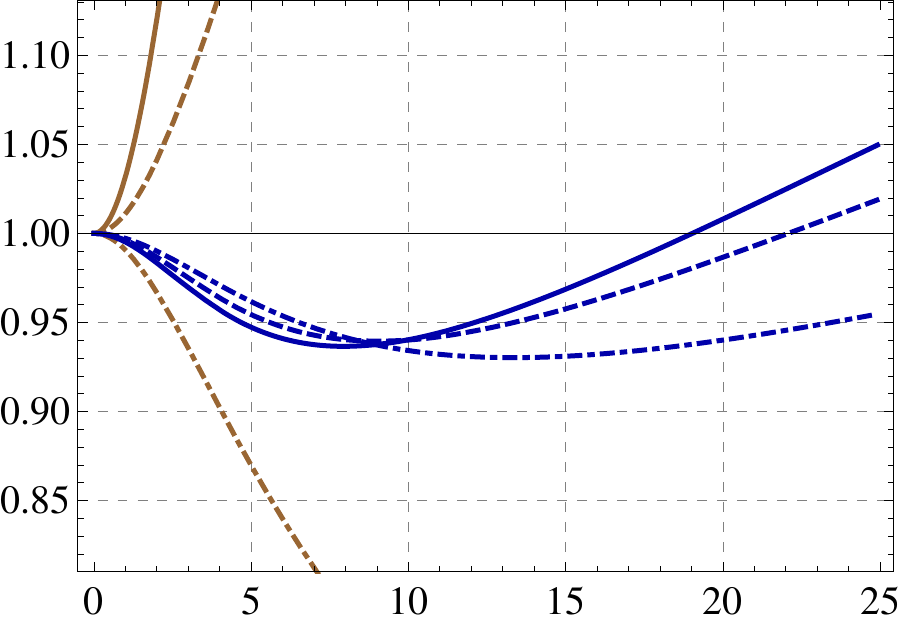}
\qquad
  \put(-510,25){\rotatebox{90}{$\chi_{(2)}/\chi_{(2)\mt{iso}}(T)$}}
         \put(-370,-10){$a/T$}
         \put(-333,25){\rotatebox{90}{$\chi_{(2)}/\chi_{(2)\mt{iso}}(T)$}}
         \put(-195,-10){$a/T$}
          \put(-160,25){\rotatebox{90}{$\chi_{(2)}/\chi_{(2)\mt{iso}}(T)$}}
         \put(-15,-10){$a/T$}
         \\
(g) & \qquad(h) & \qquad(i) \\
\end{tabular}
\end{center}
\caption{\small Plots of the spectral density $\chi_{(2)}$ normalized with respect to the isotropic result at fixed temperature $\chi_{(2)\mt{iso}}(T)$. Curves of different colors denote different values of $\textswab{w}$ as follows  $\textswab{w}=$0.5 (black), 1 (brown), 1.5 (blue). The angles are $\vartheta=0$ (solid), $\pi/4$ (dashed), $\pi/2$ (dash-dotted). Columns correspond to different values of $\textswab{q}$: from left to right it is $\textswab{q}=0,0.5,1$. Rows correspond to different values of the quark mass: from top to bottom it is $\psi_0=0,0.75,0.941$. Then, for instance, (f) corresponds to $\textswab{q}=1$, $\psi_\mt{H}=0.75$.}
\label{c4ac2}
\end{figure}

With this ground work in place, we use these expressions to numerically obtain the dilepton production rate for different values of $q$, $\psi_\mt{H}$, and $\vartheta$.  In Fig.~\ref{c2w} we plot the spectral density $\chi_{(2)}$ as a function of $\wn$, normalized with the corresponding spectral density $\chi_{(2)\mt{iso}}$ for an isotropic plasma at the same temperature.
The same quantity as a function of $\qn$ is plotted in Fig.~\ref{c2q}.
In Figs.~\ref{c4bc2} and \ref{c4ac2} we plot $\chi_{(2)}$ as a function of the anisotropy $a/T$.


\subsection{Total dilepton production rate}

\begin{figure}
\begin{center}
\begin{tabular}{ccc}
\setlength{\unitlength}{1cm}
\hspace{-0.9cm}
\includegraphics[width=5cm]{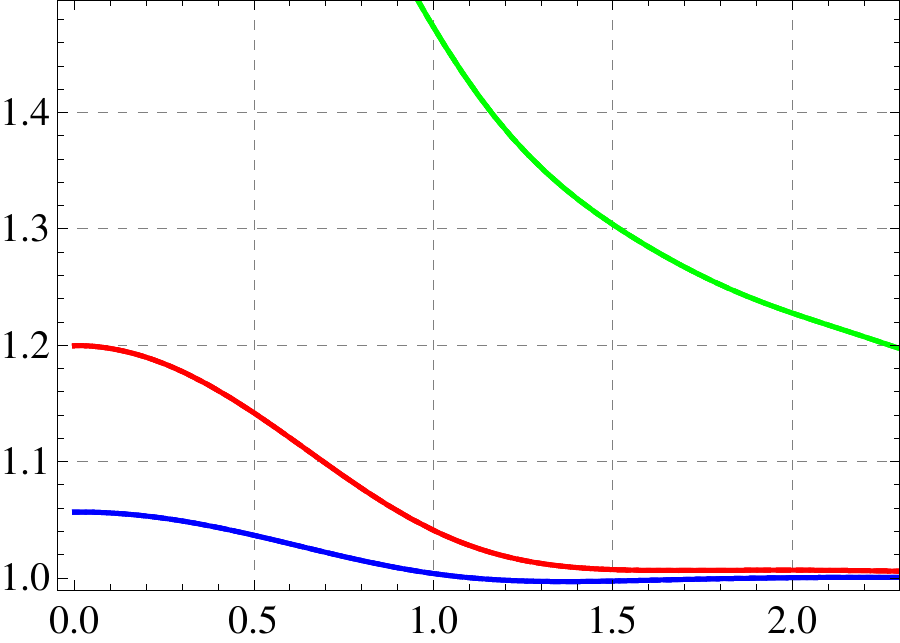} 
& 
\qquad \includegraphics[width=5cm]{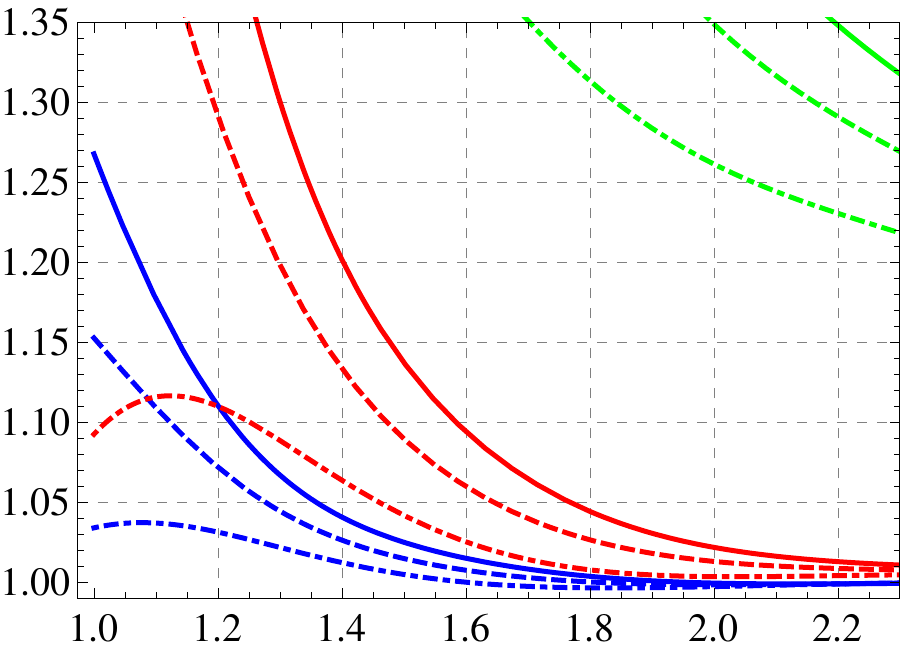}
& 
\qquad \includegraphics[width=5cm]{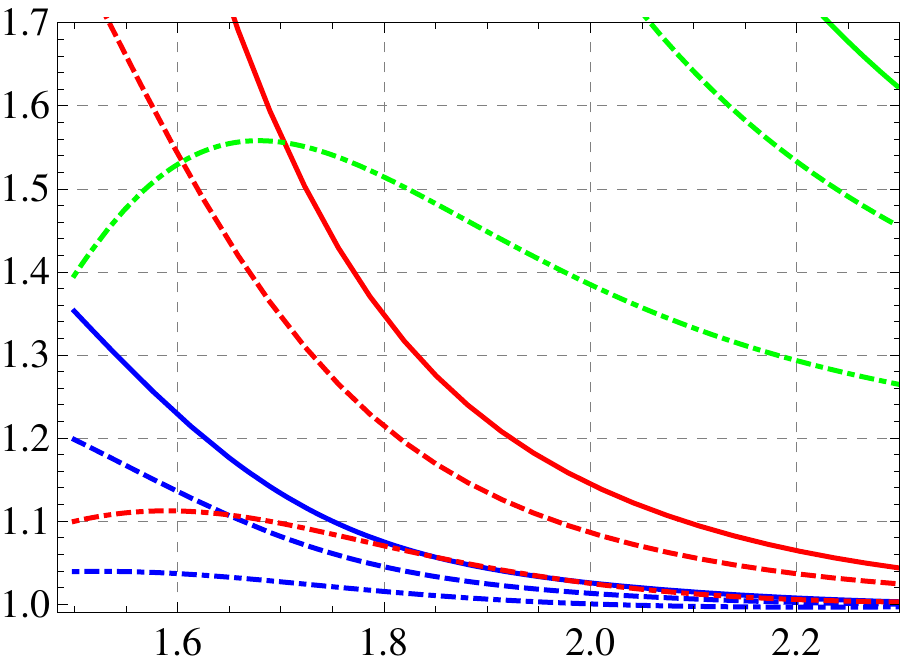}
\qquad
  \put(-510,25){\rotatebox{90}{$\chi_\mu^\mu/\chi_{\mu\mt{iso}}^\mu(T)$}}
         \put(-370,-10){$\textswab{w}$}
         \put(-333,25){\rotatebox{90}{$\chi_\mu^\mu/\chi_{\mu\mt{iso}}^\mu(T)$}}
         \put(-195,-10){$\textswab{w}$}
          \put(-160,25){\rotatebox{90}{$\chi_\mu^\mu/\chi_{\mu\mt{iso}}^\mu(T)$}}
         \put(-15,-10){$\textswab{w}$}
\\
(a) & \qquad(b) & \qquad(c)\\
& \\
\hspace{-0.9cm}
\includegraphics[width=5cm]{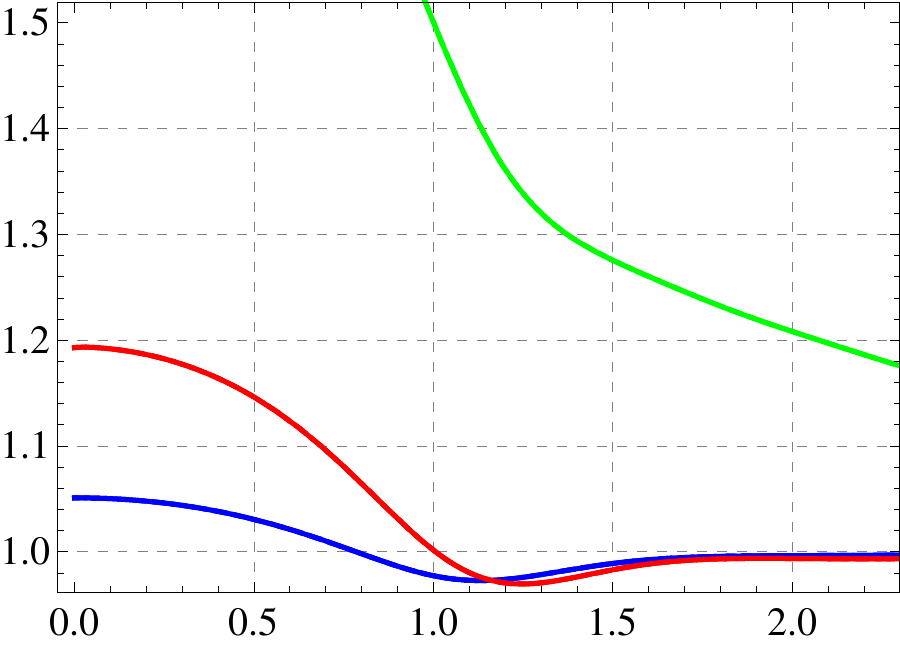} 
& 
\qquad \includegraphics[width=5cm]{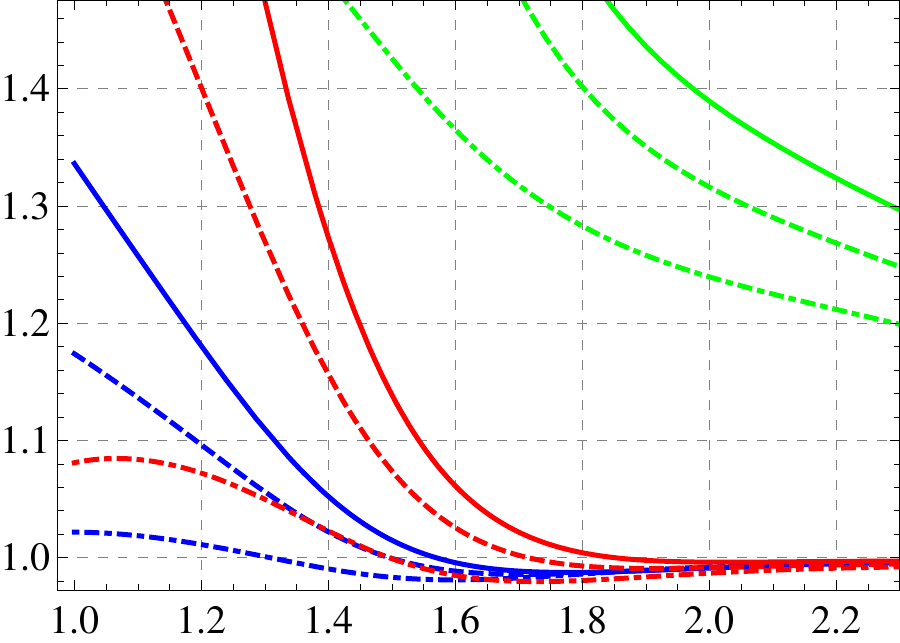}
& 
\qquad \includegraphics[width=5cm]{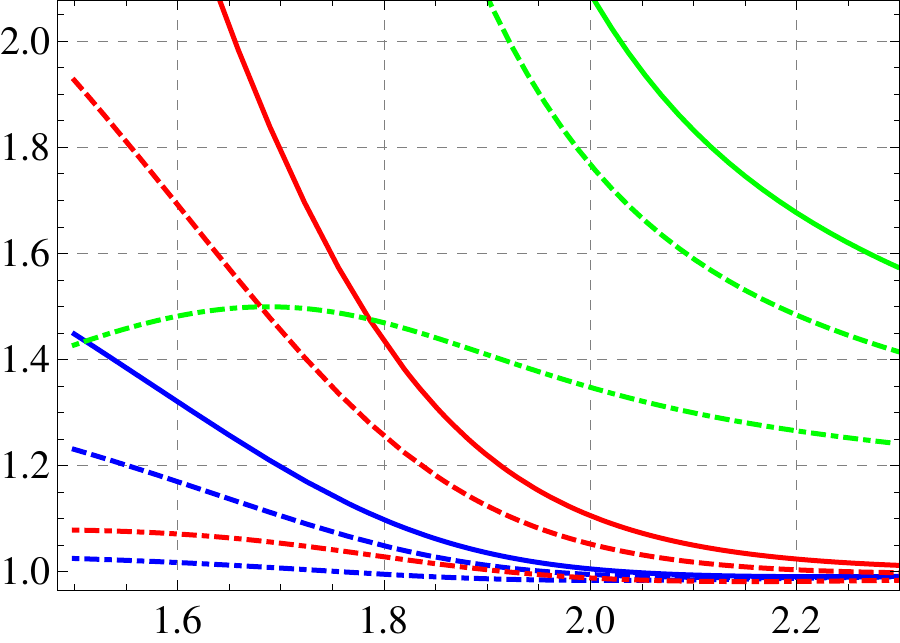}
\qquad
  \put(-510,25){\rotatebox{90}{$\chi_\mu^\mu/\chi_{\mu\mt{iso}}^\mu(T)$}}
         \put(-370,-10){$\textswab{w}$}
         \put(-333,25){\rotatebox{90}{$\chi_\mu^\mu/\chi_{\mu\mt{iso}}^\mu(T)$}}
         \put(-195,-10){$\textswab{w}$}
          \put(-160,25){\rotatebox{90}{$\chi_\mu^\mu/\chi_{\mu\mt{iso}}^\mu(T)$}}
         \put(-15,-10){$\textswab{w}$}
         \\
(d) & \qquad(e) & \qquad(f) \\
& \\
\hspace{-0.9cm}
\includegraphics[width=5cm]{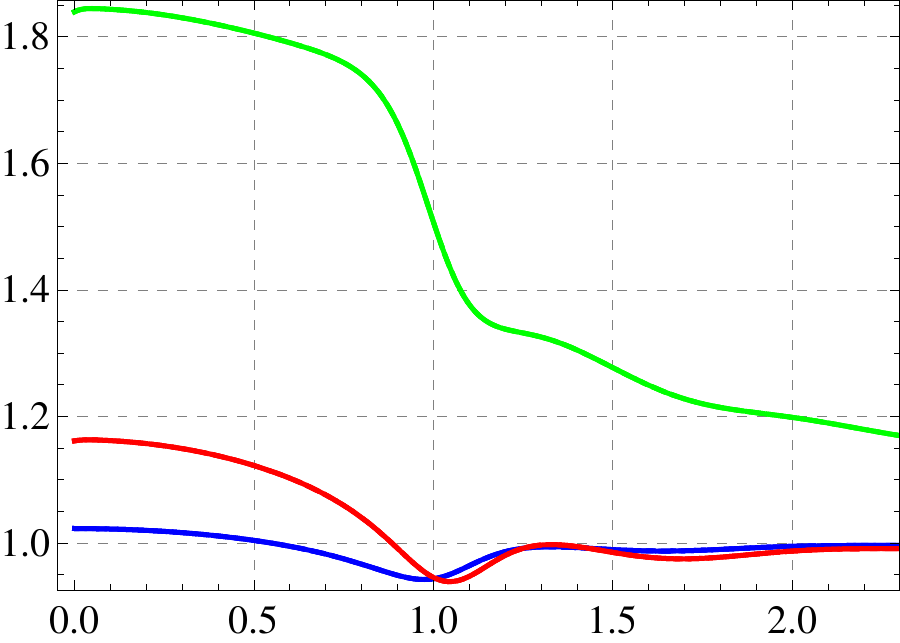} 
& 
\qquad \includegraphics[width=5cm]{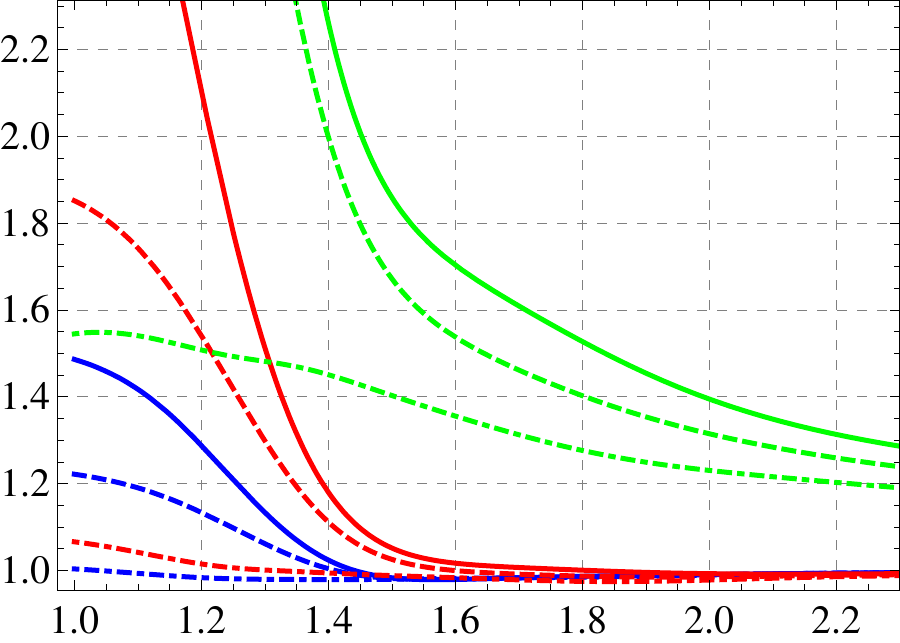}
& 
\qquad \includegraphics[width=5cm]{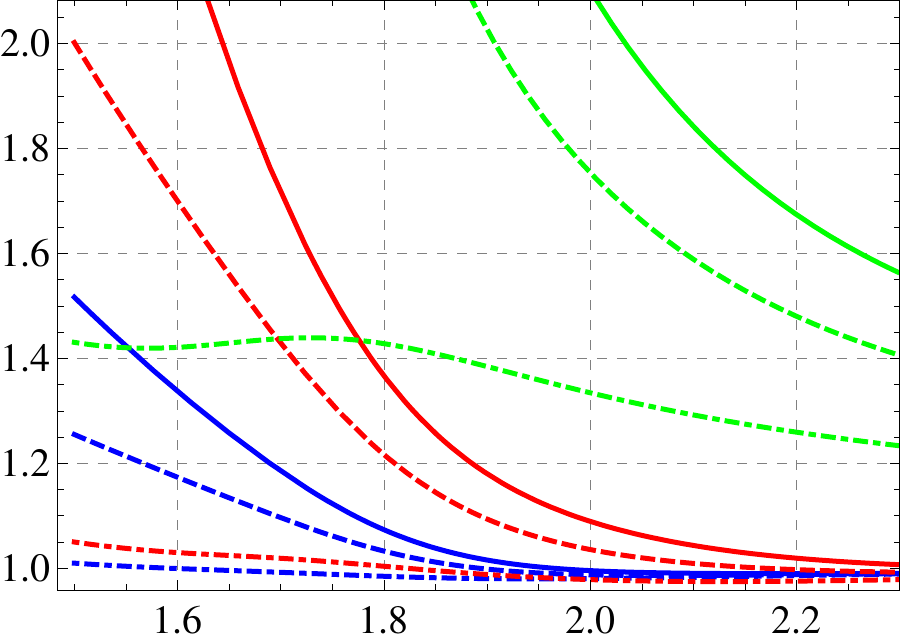}
\qquad
  \put(-510,25){\rotatebox{90}{$\chi_\mu^\mu/\chi_{\mu\mt{iso}}^\mu(T)$}}
         \put(-370,-10){$\textswab{w}$}
         \put(-333,25){\rotatebox{90}{$\chi_\mu^\mu/\chi_{\mu\mt{iso}}^\mu(T)$}}
         \put(-195,-10){$\textswab{w}$}
          \put(-160,25){\rotatebox{90}{$\chi_\mu^\mu/\chi_{\mu\mt{iso}}^\mu(T)$}}
         \put(-15,-10){$\textswab{w}$}
         \\
(g) & \qquad(h) & \qquad(i) \\
\end{tabular}
\end{center}
\caption{\small Plots of the spectral density $\chi_\mu^\mu$ normalized with respect to the isotropic result at fixed temperature $\chi_{\mu\mt{iso}}^\mu(T)$. Curves of different colors denote different values of $a/T$ as follows  $a/T=$4.41 (blue), 12.2 (red), 86 (green). The angles are $\vartheta=0$ (solid), $\pi/4$ (dashed), $\pi/2$ (dash-dotted). Columns correspond to different values of $\textswab{q}$: from left to right it is $\textswab{q}=0,1,1.5$. Rows correspond to different values of the quark mass: from top to bottom it is $\psi_\mt{H}=0,0.75,0.941$. Then, for instance, (h) corresponds to $\textswab{q}=1$, $\psi_\mt{H}=0.941$. }
\label{ctw}
\end{figure}

\begin{figure}
\begin{center}
\begin{tabular}{ccc}
\setlength{\unitlength}{1cm}
\hspace{-0.9cm}
\includegraphics[width=5cm]{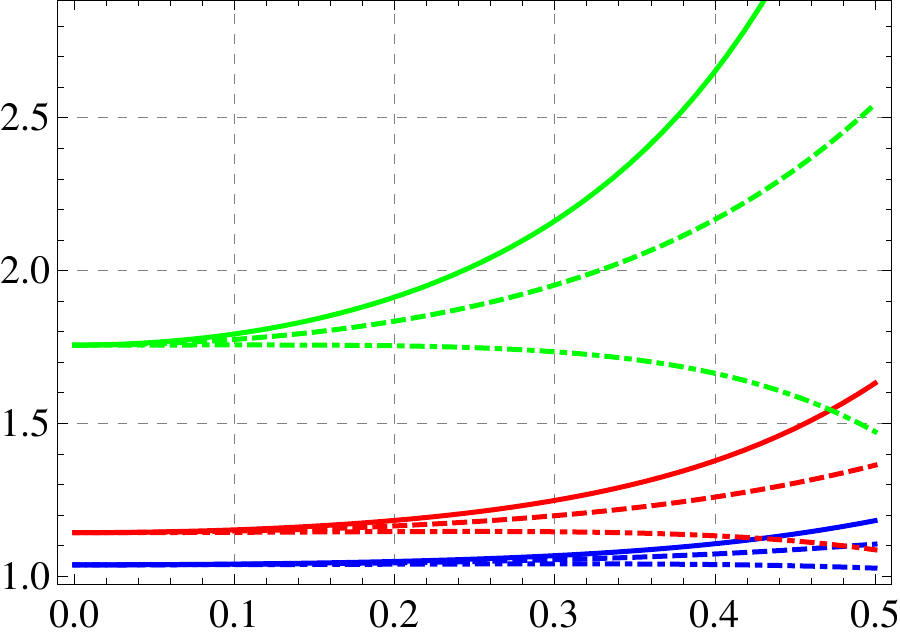} 
& 
\qquad \includegraphics[width=5cm]{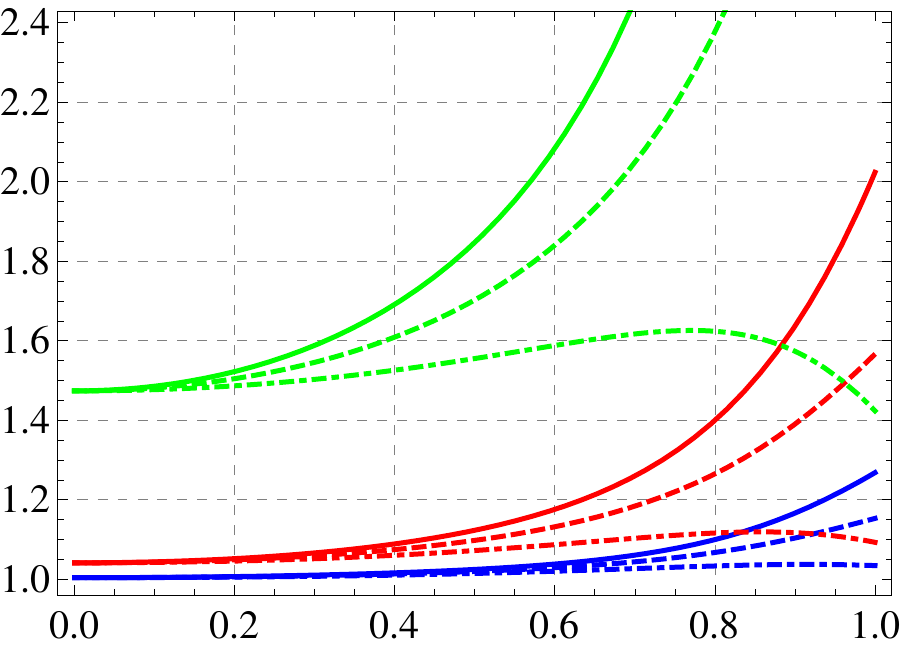}
& 
\qquad \includegraphics[width=5cm]{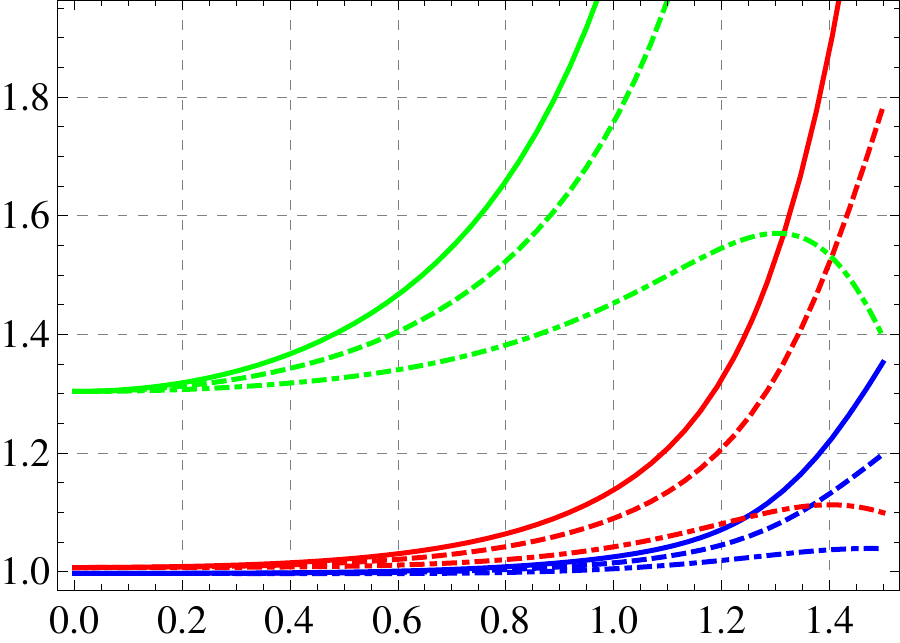}
\qquad
  \put(-510,25){\rotatebox{90}{$\chi_\mu^\mu/\chi_{\mu\mt{iso}}^\mu(T)$}}
         \put(-370,-10){$\textswab{q}$}
         \put(-333,25){\rotatebox{90}{$\chi_\mu^\mu/\chi_{\mu\mt{iso}}^\mu(T)$}}
         \put(-195,-10){$\textswab{q}$}
          \put(-160,25){\rotatebox{90}{$\chi_\mu^\mu/\chi_{\mu\mt{iso}}^\mu(T)$}}
         \put(-15,-10){$\textswab{q}$}
\\
(a) & \qquad(b) & \qquad(c)\\
& \\
\hspace{-0.9cm}
\includegraphics[width=5cm]{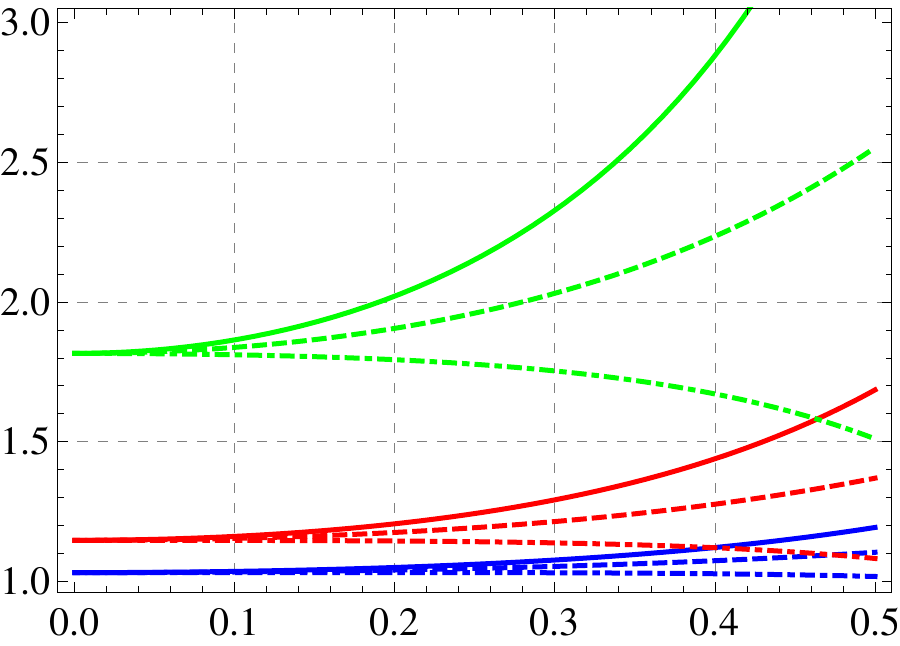} 
& 
\qquad \includegraphics[width=5cm]{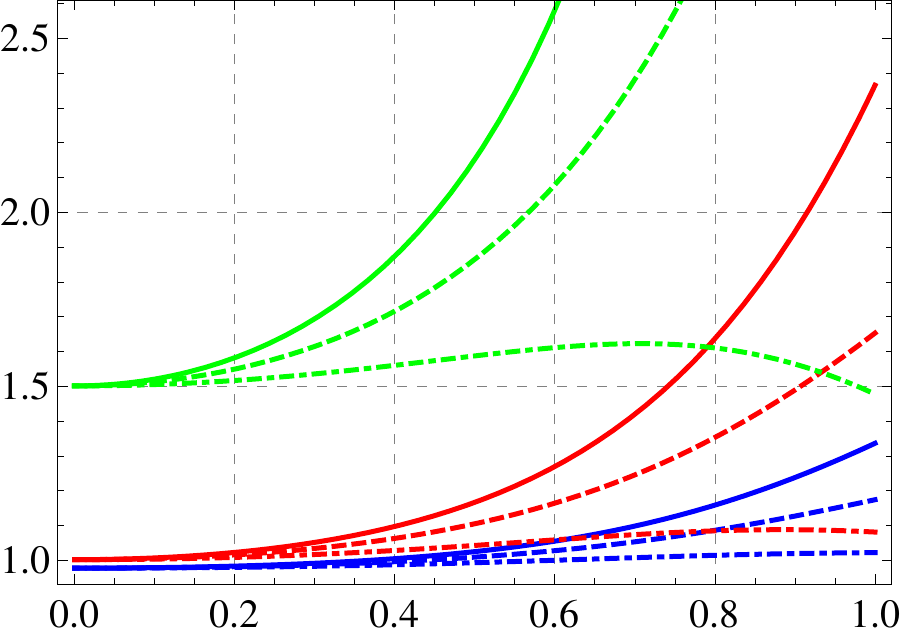}
& 
\qquad \includegraphics[width=5cm]{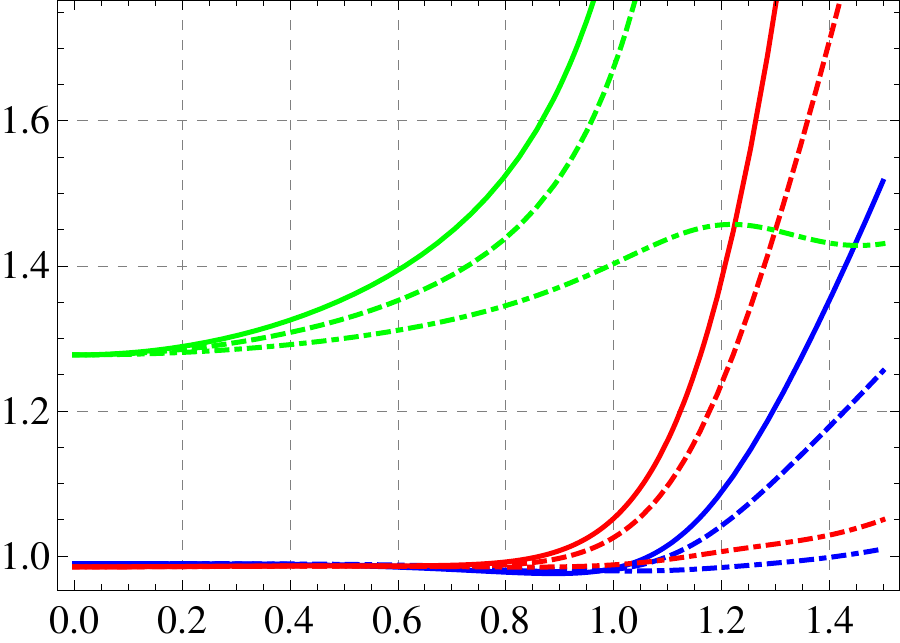}
\qquad
  \put(-510,25){\rotatebox{90}{$\chi_\mu^\mu/\chi_{\mu\mt{iso}}^\mu(T)$}}
         \put(-370,-10){$\textswab{q}$}
         \put(-333,25){\rotatebox{90}{$\chi_\mu^\mu/\chi_{\mu\mt{iso}}^\mu(T)$}}
         \put(-195,-10){$\textswab{q}$}
          \put(-160,25){\rotatebox{90}{$\chi_\mu^\mu/\chi_{\mu\mt{iso}}^\mu(T)$}}
         \put(-15,-10){$\textswab{q}$}
         \\
(d) & \qquad(e) & \qquad(f) \\
& \\
\hspace{-0.9cm}
\includegraphics[width=5cm]{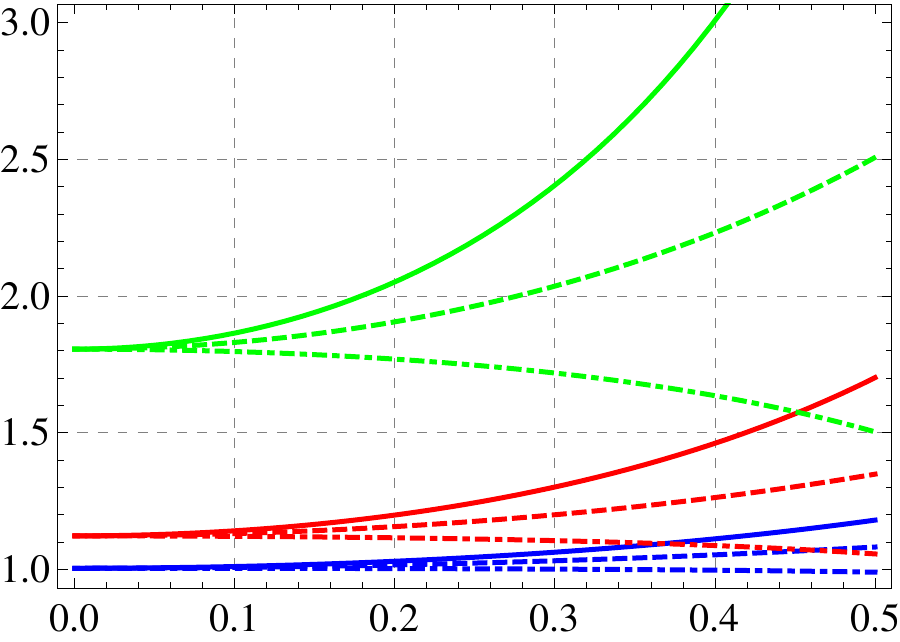} 
& 
\qquad \includegraphics[width=5cm]{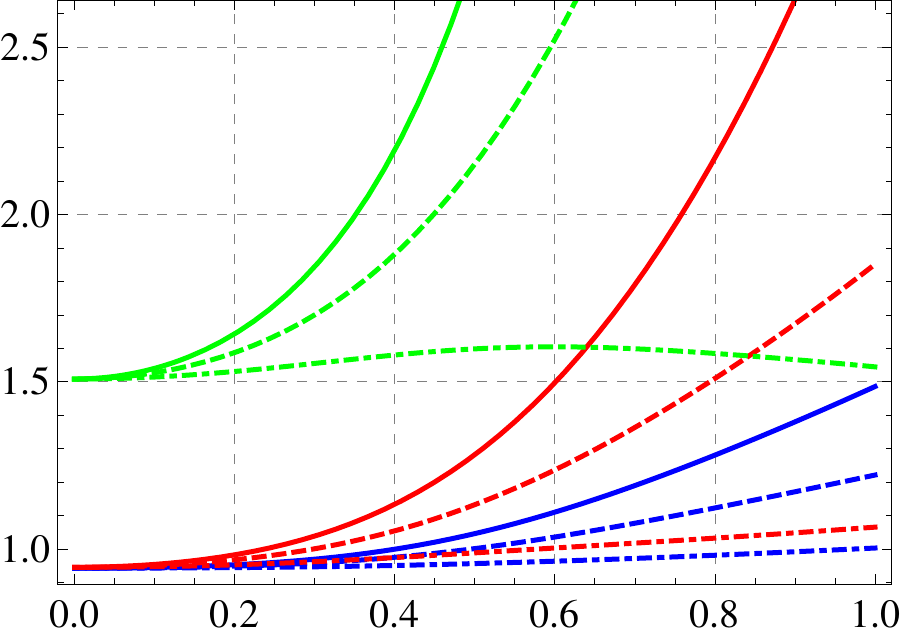}
& 
\qquad \includegraphics[width=5cm]{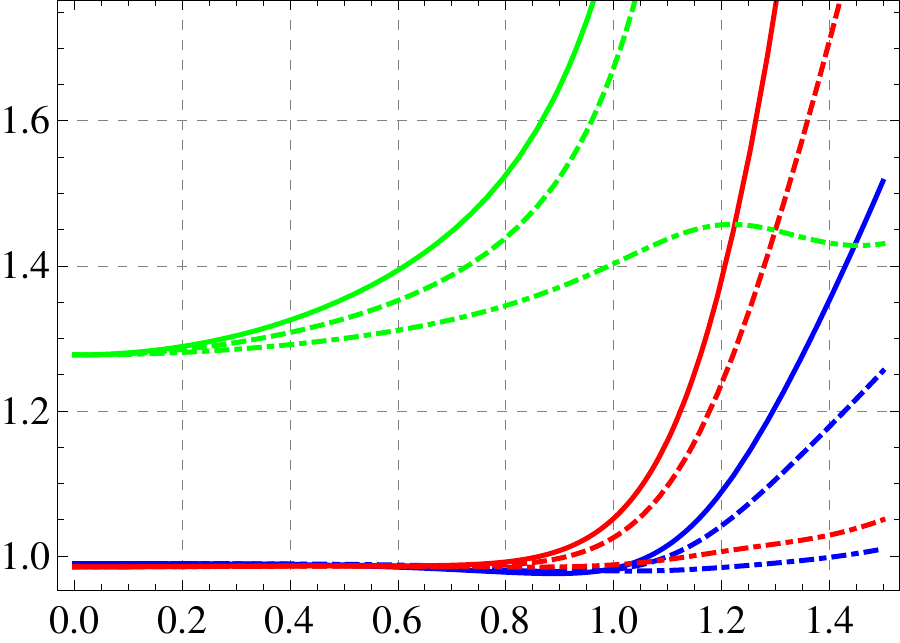}
\qquad
  \put(-510,25){\rotatebox{90}{$\chi_\mu^\mu/\chi_{\mu\mt{iso}}^\mu(T)$}}
         \put(-370,-10){$\textswab{q}$}
         \put(-333,25){\rotatebox{90}{$\chi_\mu^\mu/\chi_{\mu\mt{iso}}^\mu(T)$}}
         \put(-195,-10){$\textswab{q}$}
          \put(-160,25){\rotatebox{90}{$\chi_\mu^\mu/\chi_{\mu\mt{iso}}^\mu(T)$}}
         \put(-15,-10){$\textswab{q}$}
         \\
(g) & \qquad(h) & \qquad(i) \\
\end{tabular}
\end{center}
\caption{\small Plots of the spectral density $\chi_\mu^\mu$ normalized with respect to the isotropic result at fixed temperature $\chi_{\mu\mt{iso}}^\mu(T)$. Curves of different colors denote different values of $a/T$ as follows  $a/T=$4.41 (blue), 12.2 (red), 86 (green). The angles are $\vartheta=0$ (solid), $\pi/4$ (dashed), $\pi/2$ (dash-dotted). Columns correspond to different values of $\textswab{w}$: from left to right it is $\textswab{w}=0.5,1,1.5$. Rows correspond to different values of the quark mass: from top to bottom it is $\psi_\mt{H}=0,0.75,0.941$. Then, for instance, (f) corresponds to $\textswab{w}=1.5$, $\psi_\mt{H}=0.75$. }
\label{ctq}
\end{figure}

\begin{figure}
\begin{center}
\begin{tabular}{ccc}
\setlength{\unitlength}{1cm}
\hspace{-0.9cm}
\includegraphics[width=5cm]{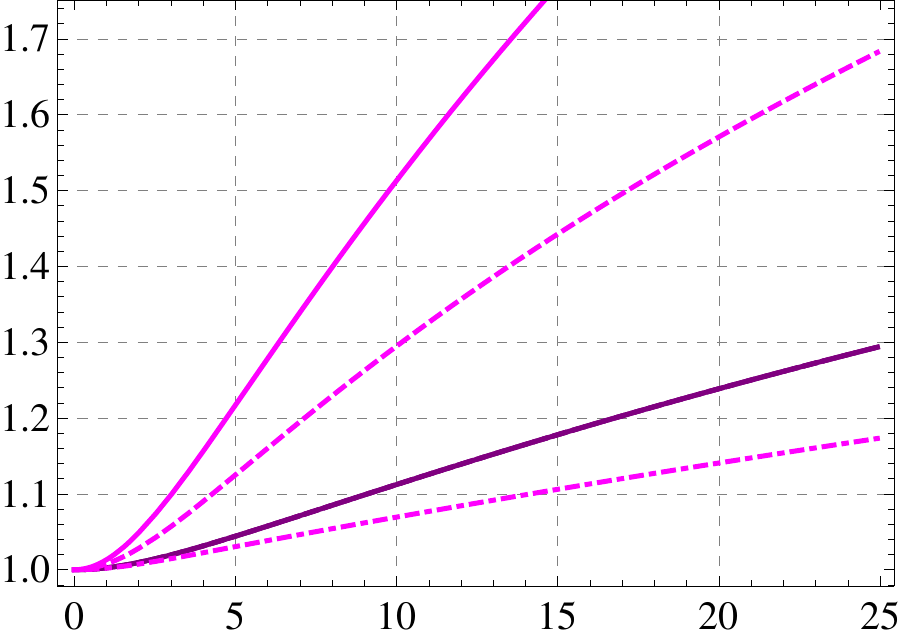} 
& 
\qquad \includegraphics[width=5cm]{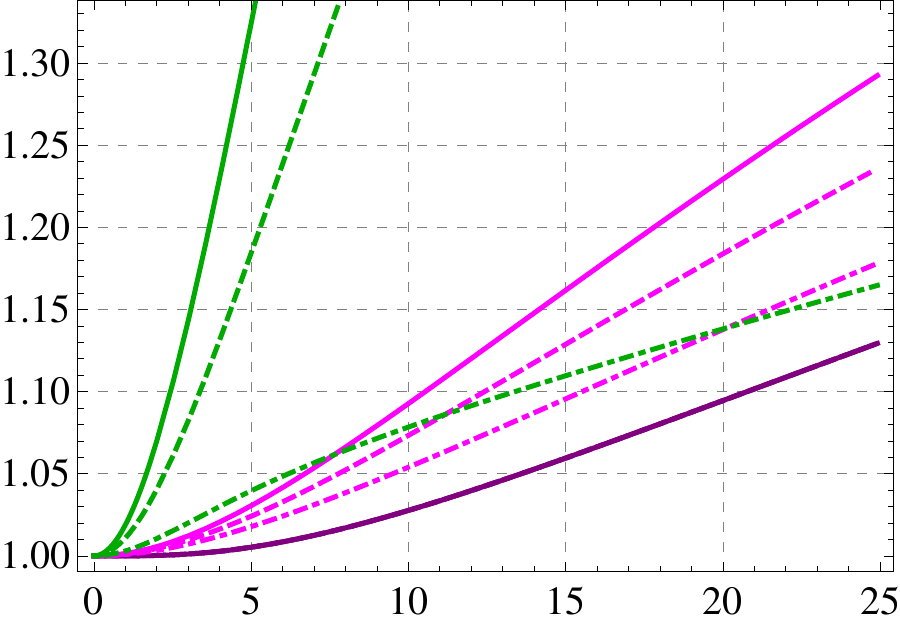}
& 
\qquad \includegraphics[width=5cm]{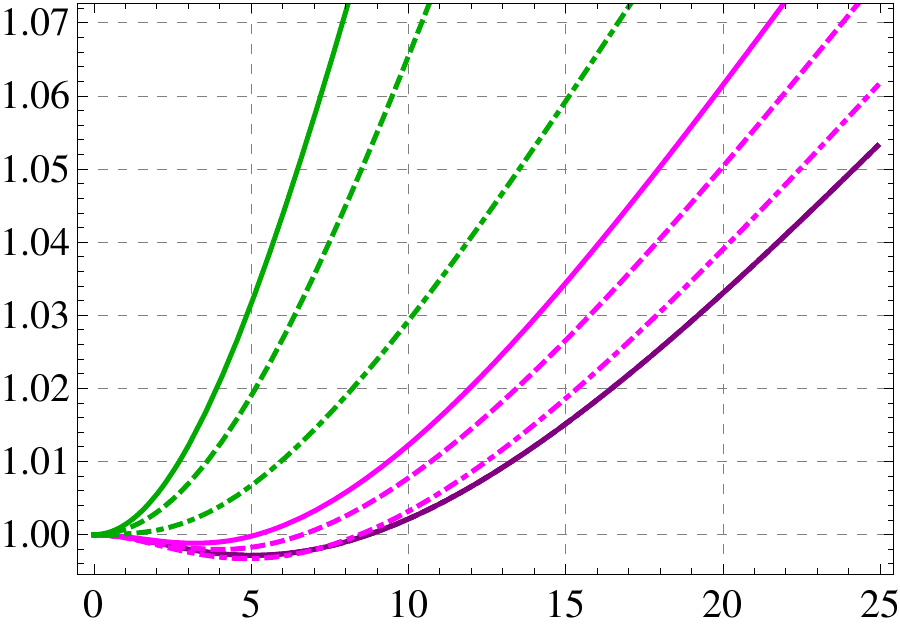}
\qquad
  \put(-510,25){\rotatebox{90}{$\chi_\mu^\mu/\chi_{\mu\mt{iso}}^\mu(T)$}}
         \put(-370,-10){$a/T$}
         \put(-333,25){\rotatebox{90}{$\chi_\mu^\mu/\chi_{\mu\mt{iso}}^\mu(T)$}}
         \put(-195,-10){$a/T$}
          \put(-160,25){\rotatebox{90}{$\chi_\mu^\mu/\chi_{\mu\mt{iso}}^\mu(T)$}}
         \put(-15,-10){$a/T$}
\\
(a) & \qquad(b) & \qquad(c)\\
& \\
\hspace{-0.9cm}
\includegraphics[width=5cm]{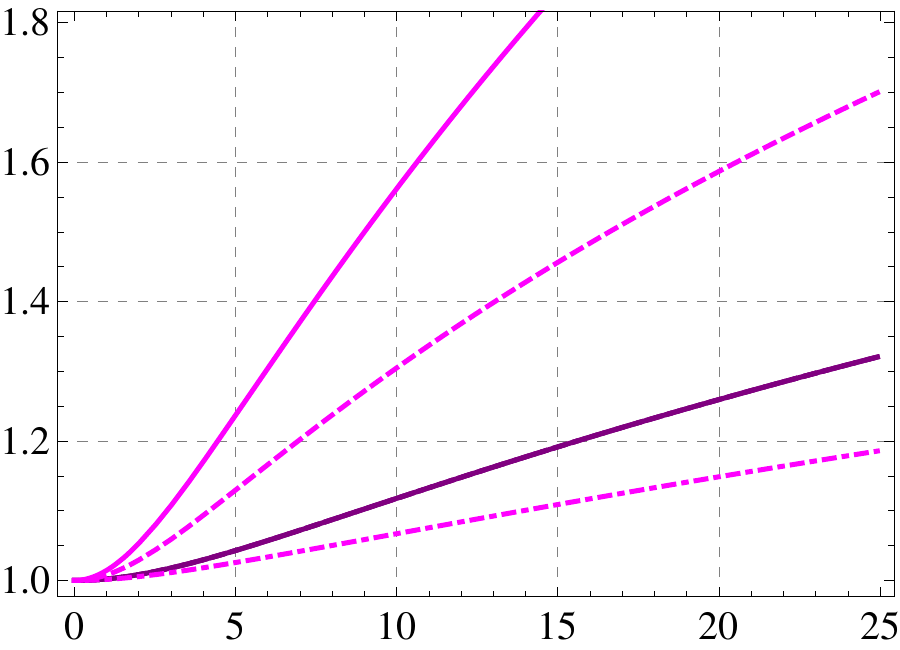} 
& 
\qquad \includegraphics[width=5cm]{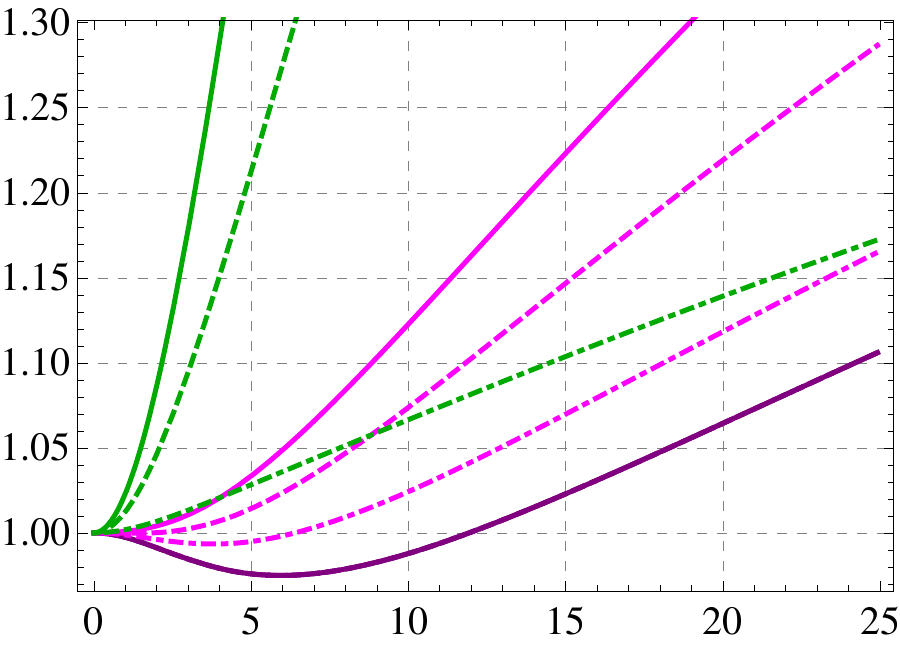}
& 
\qquad \includegraphics[width=5cm]{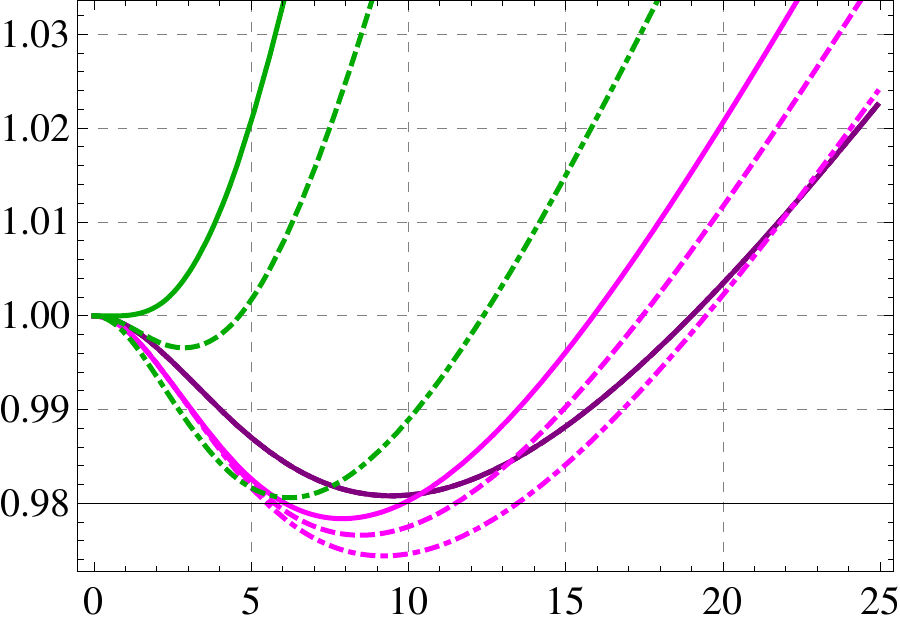}
\qquad
  \put(-510,25){\rotatebox{90}{$\chi_\mu^\mu/\chi_{\mu\mt{iso}}^\mu(T)$}}
         \put(-370,-10){$a/T$}
         \put(-333,25){\rotatebox{90}{$\chi_\mu^\mu/\chi_{\mu\mt{iso}}^\mu(T)$}}
         \put(-195,-10){$a/T$}
          \put(-160,25){\rotatebox{90}{$\chi_\mu^\mu/\chi_{\mu\mt{iso}}^\mu(T)$}}
         \put(-15,-10){$a/T$}
         \\
(d) & \qquad(e) & \qquad(f) \\
& \\
\hspace{-0.9cm}
\includegraphics[width=5cm]{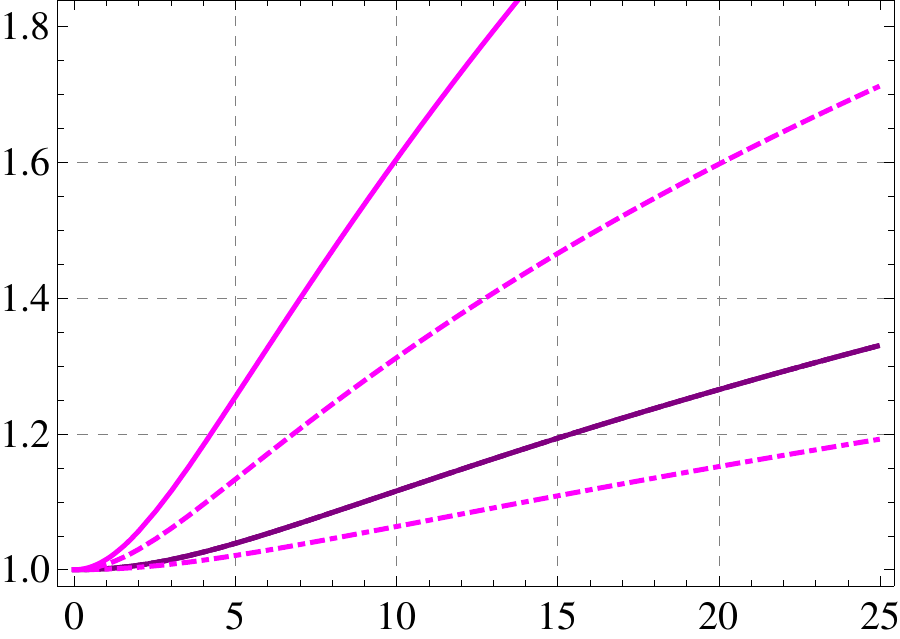} 
& 
\qquad \includegraphics[width=5cm]{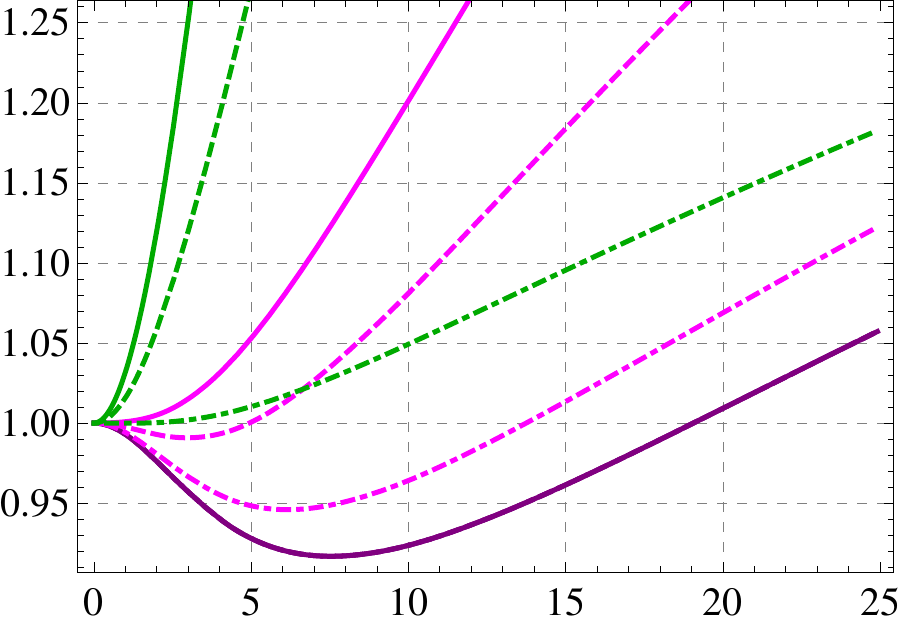}
& 
\qquad \includegraphics[width=5cm]{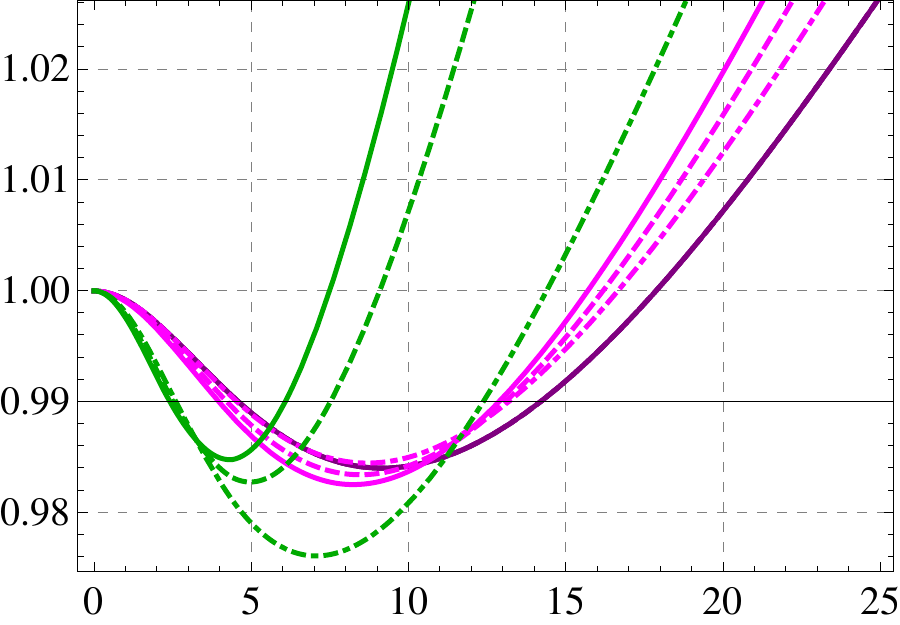}
\qquad
  \put(-510,25){\rotatebox{90}{$\chi_\mu^\mu/\chi_{\mu\mt{iso}}^\mu(T)$}}
         \put(-370,-10){$a/T$}
         \put(-333,25){\rotatebox{90}{$\chi_\mu^\mu/\chi_{\mu\mt{iso}}^\mu(T)$}}
         \put(-195,-10){$a/T$}
          \put(-160,25){\rotatebox{90}{$\chi_\mu^\mu/\chi_{\mu\mt{iso}}^\mu(T)$}}
         \put(-15,-10){$a/T$}
         \\
(g) & \qquad(h) & \qquad(i) \\
\end{tabular}
\end{center}
\caption{\small Plots of the spectral density $\chi_\mu^\mu$ normalized with respect to the isotropic result at fixed temperature $\chi_{\mu\mt{iso}}^\mu(T)$. Curves of different colors denote different values of $\textswab{q}$ as follows  $\textswab{q}=$0 (purple), 0.5 (magenta), 1 (green). The angles are $\vartheta=0$ (solid), $\pi/4$ (dashed), $\pi/2$ (dash-dotted). Columns correspond to different values of $\textswab{w}$: from left to right it is $\textswab{w}=0.5,1,1.5$. Rows correspond to different values of the quark mass: from top to bottom it is $\psi_\mt{H}=0,0.75,0.941$. Then, for instance, (f) corresponds to $\textswab{w}=1.5$, $\psi_\mt{H}=0.75$.}
\label{c4bct}
\end{figure}

\begin{figure}
\begin{center}
\begin{tabular}{ccc}
\setlength{\unitlength}{1cm}
\hspace{-0.9cm}
\includegraphics[width=5cm]{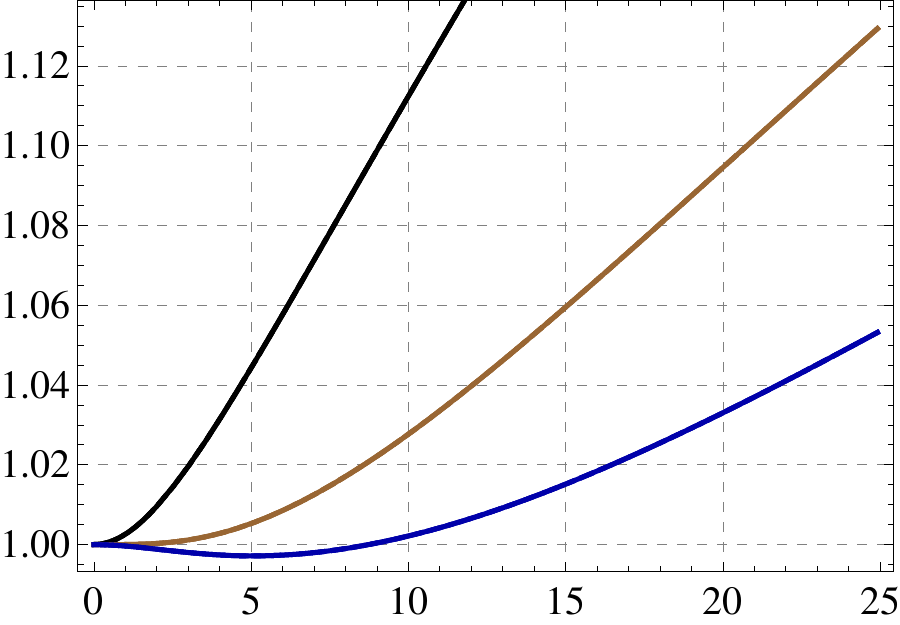} 
& 
\qquad \includegraphics[width=5cm]{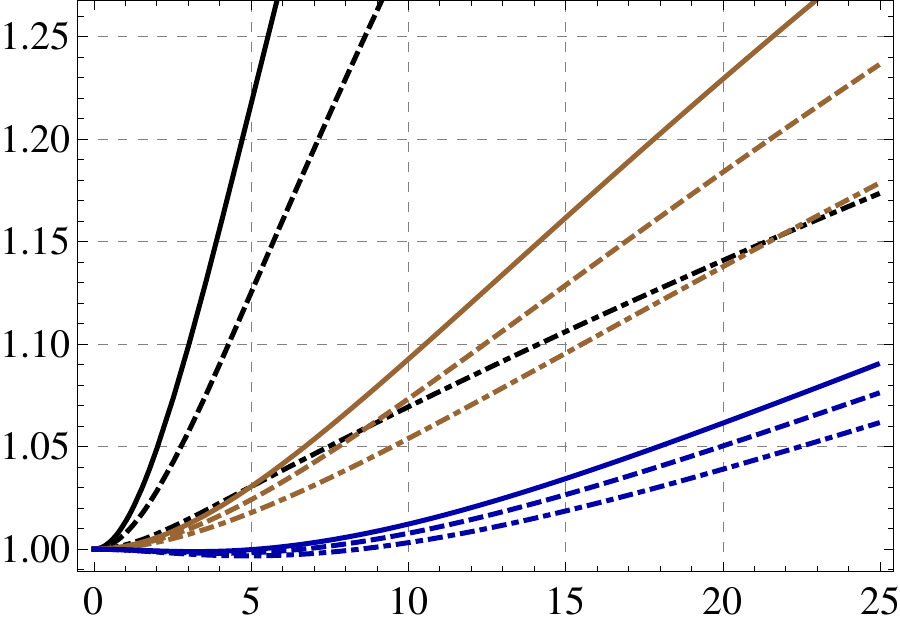}
& 
\qquad \includegraphics[width=5cm]{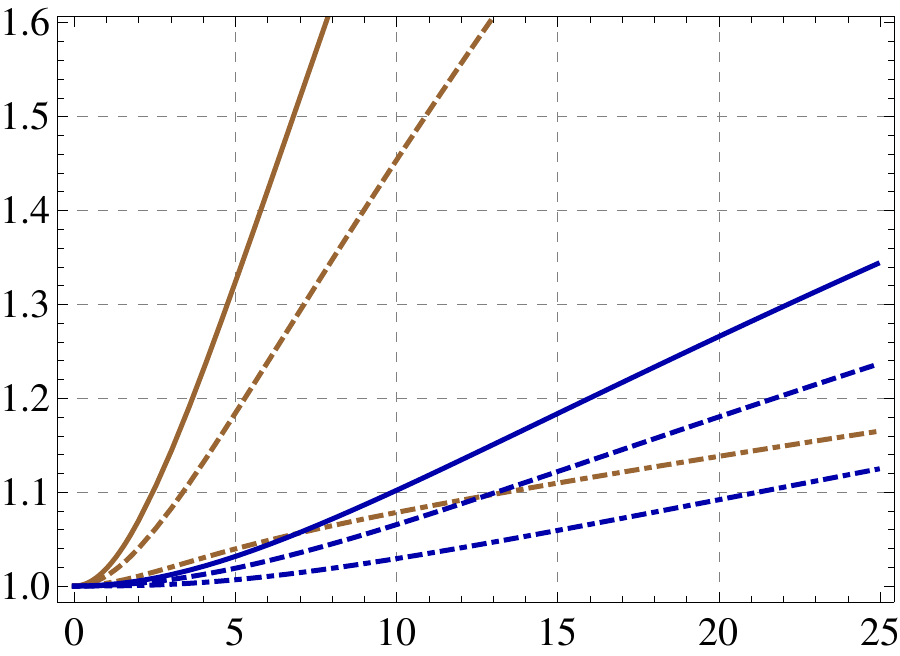}
\qquad
  \put(-510,25){\rotatebox{90}{$\chi_\mu^\mu/\chi^\mu_{\mu\mt{iso}}(T)$}}
         \put(-370,-10){$a/T$}
         \put(-333,25){\rotatebox{90}{$\chi_\mu^\mu/\chi^\mu_{\mu\mt{iso}}(T)$}}
         \put(-195,-10){$a/T$}
          \put(-160,25){\rotatebox{90}{$\chi_\mu^\mu/\chi^\mu_{\mu\mt{iso}}(T)$}}
         \put(-15,-10){$a/T$}
\\
(a) & \qquad(b) & \qquad(c)\\
& \\
\hspace{-0.9cm}
\includegraphics[width=5cm]{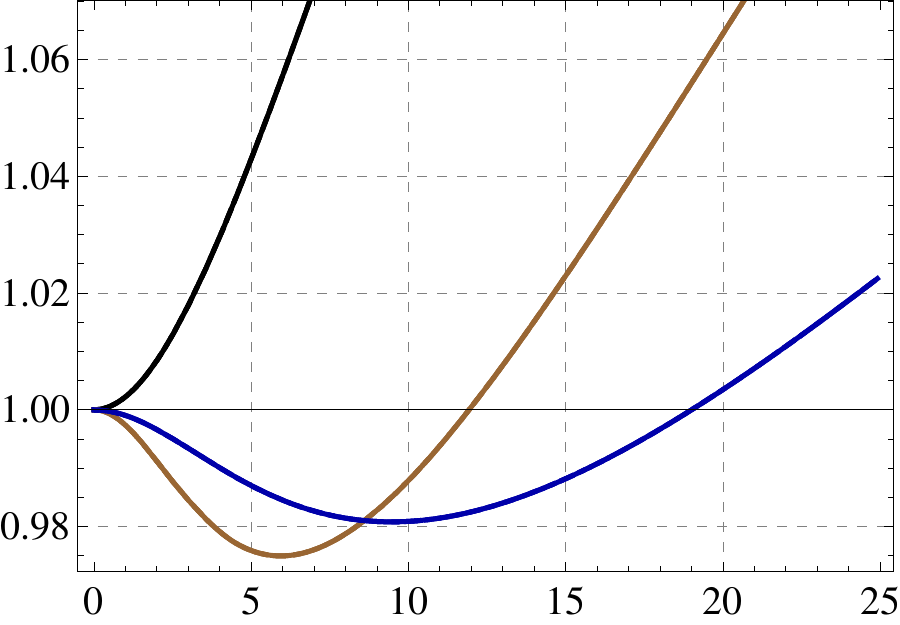} 
& 
\qquad \includegraphics[width=5cm]{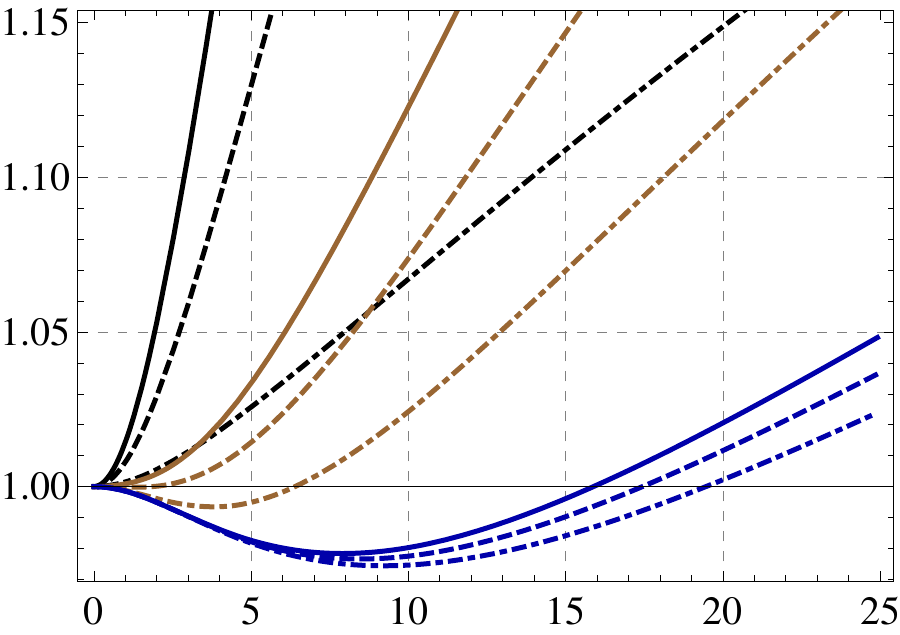}
& 
\qquad \includegraphics[width=5cm]{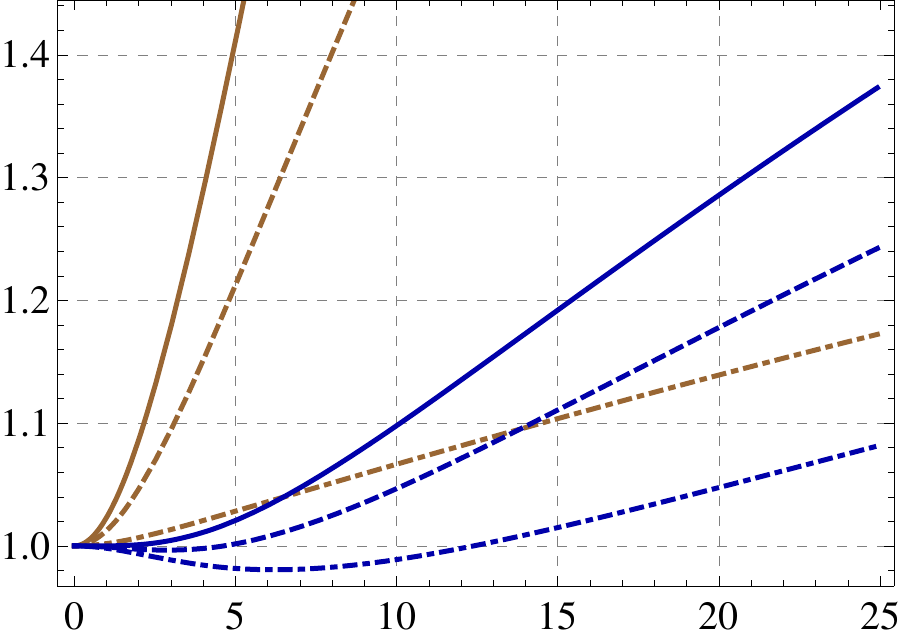}
\qquad
  \put(-510,25){\rotatebox{90}{$\chi_\mu^\mu/\chi^\mu_{\mu\mt{iso}}(T)$}}
         \put(-370,-10){$a/T$}
         \put(-333,25){\rotatebox{90}{$\chi_\mu^\mu/\chi^\mu_{\mu\mt{iso}}(T)$}}
         \put(-195,-10){$a/T$}
          \put(-160,25){\rotatebox{90}{$\chi_\mu^\mu/\chi^\mu_{\mu\mt{iso}}(T)$}}
         \put(-15,-10){$a/T$}
         \\
(d) & \qquad(e) & \qquad(f) \\
& \\
\hspace{-0.9cm}
\includegraphics[width=5cm]{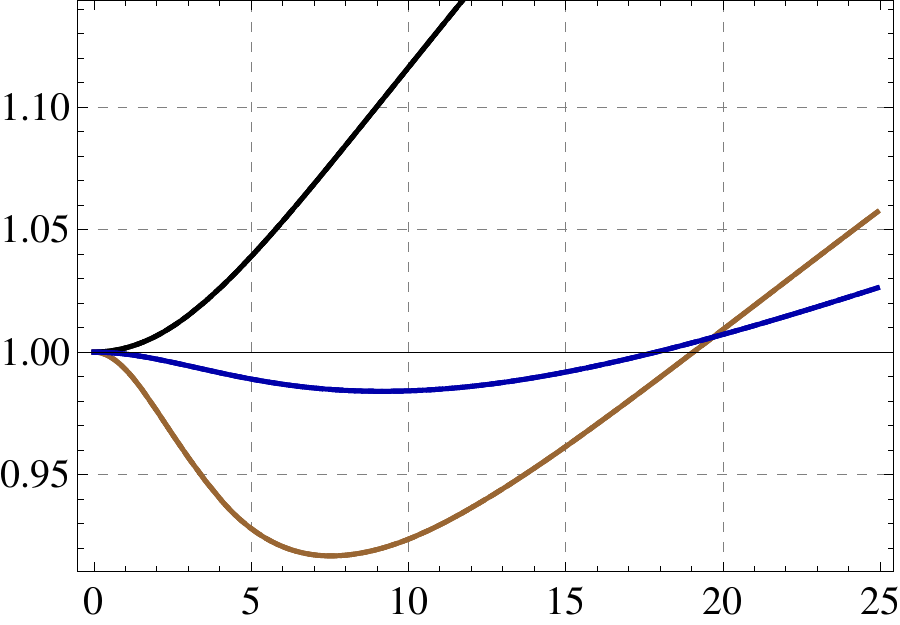} 
& 
\qquad \includegraphics[width=5cm]{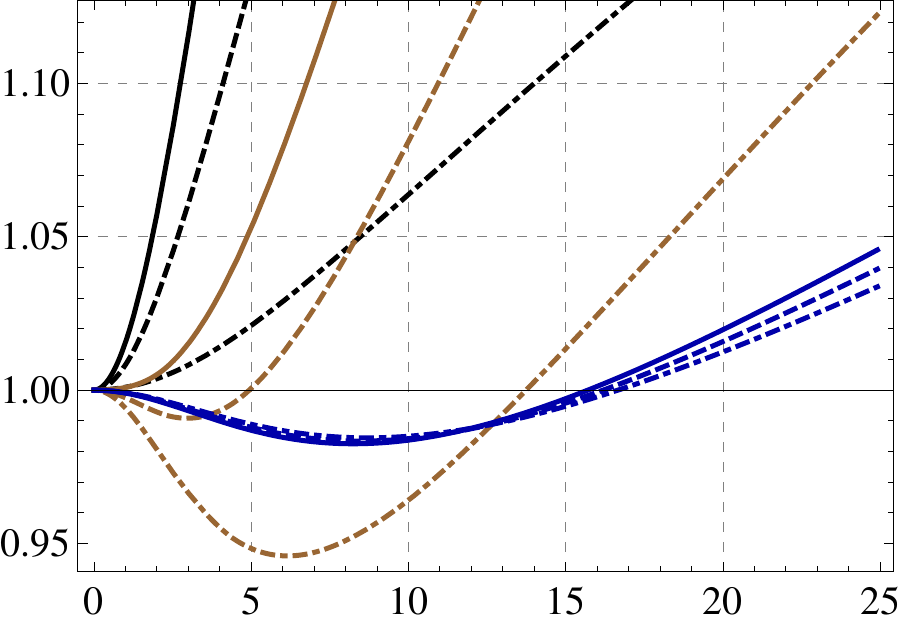}
& 
\qquad \includegraphics[width=5cm]{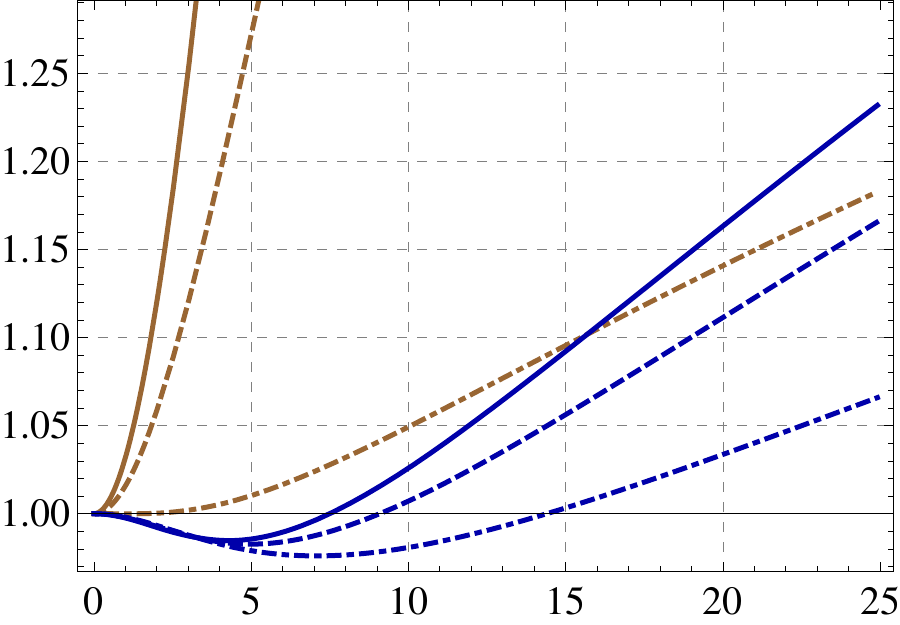}
\qquad
  \put(-510,25){\rotatebox{90}{$\chi_\mu^\mu/\chi^\mu_{\mu\mt{iso}}(T)$}}
         \put(-370,-10){$a/T$}
         \put(-333,25){\rotatebox{90}{$\chi_\mu^\mu/\chi^\mu_{\mu\mt{iso}}(T)$}}
         \put(-195,-10){$a/T$}
          \put(-160,25){\rotatebox{90}{$\chi_\mu^\mu/\chi^\mu_{\mu\mt{iso}}(T)$}}
         \put(-15,-10){$a/T$}
         \\
(g) & \qquad(h) & \qquad(i) \\
\end{tabular}
\end{center}
\caption{\small Plots of the trace of spectral density $\chi_\mu^\mu$ normalized with respect to the isotropic result at fixed temperature $\chi^\mu_{\mu\mt{iso}}(T)$. Curves of different colors denote different values of $\textswab{w}$ as follows  $\textswab{w}=$0.5 (black), 1 (brown), 1.5 (blue). The angles are $\vartheta=0$ (solid), $\pi/4$ (dashed), $\pi/2$ (dash-dotted). Columns correspond to different values of $\textswab{q}$: from left to right it is $\textswab{q}=0,0.5,1$. Rows correspond to different values of the quark mass: from top to bottom it is $\psi_\mt{H}=0,0.75,0.941$. Then, for instance, (f) corresponds to $\textswab{q}=1$, $\psi_\mt{H}=0.75$.}
\label{c4act}
\end{figure}

In Fig.~\ref{ctw} we plot the trace of the spectral density $\chi_\mu^\mu$ as a function of $\wn$, normalized with the corresponding trace $\chi_{\mu\mt{iso}}^\mu$ for an isotropic plasma at the same temperature. The same quantity as a function of $\qn$ is plotted in Fig.~\ref{ctq}, and in Figs.~\ref{c4bct} and \ref{c4act} as a function of the anisotropy parameter $a/T$.



\section{Discussion}
\label{sec5}

In this paper we have studied two important phenomenological probes of a strongly coupled anisotropic plasma, namely the in-medium production of photons and of dileptons. In order to model the plasma at strong coupling, we have used the dual anisotropic black brane solution found in \cite{Mateos:2011ix,Mateos:2011tv}. This geometry is static, regular on and outside the horizon, and asymptotically  AdS. The anisotropic equilibrium is due to a bulk axion field, corresponding on the gauge theory side to a marginally relevant deformation of the ${\cal N}=4$ SYM action. The insertion of flavor D7-branes in this background has allowed us to couple the ${\cal N}=4$ adjoint fields to fields in the fundamental representation, which we have called `quarks'.

First, we have completed the computation of the photon production rate initiated in \cite{photon}, where the plasma of the adjoint ${\cal N}=4$ fields was coupled to {\it massless} quarks. Here we have allowed these fields to have a non vanishing mass, thus bringing the analysis closer to the real world experiments performed at RHIC and LHC. Secondly, we have considered the possibility of the plasma emitting these massive fundamental fields in pairs, which with a slight abuse of notation we have called `dileptons', and we have computed their production rate. 

The main results of our analysis may be summarized as follows. As for the photon production rate, we have seen that, in general, an anisotropic plasma glows brighter than its isotropic counterpart at the same temperature, both in the case of massless fundamental fields, as found in \cite{photon}, and in the case of massive fundamental fields. Moreover, increasing the mass of these fields results in a decrease of the photon production rate and the DC conductivity of the plasma, as can be anticipated on general grounds. This also happens in the isotropic plasma considered in \cite{Mateos:2007yp}. 

As for the dilepton production rate, the analysis is made more complicated by the presence of an extra parameter, the magnitude of the spatial part of the momentum, $\qn$. We have studied the total production rate, normalized with respect to an isotropic plasma at the same temperature, first as a function of the frequency $\wn$ for fixed values of $\qn$ (see Fig.~\ref{ctw}), and then as a function of $\qn$ for fixed values of $\wn$ (see Fig.~\ref{ctq}). For fixed values of $\qn$, we have seen that the anisotropic rate is higher than the isotropic counterpart at small frequencies (the larger the anisotropy the larger the deviation) and tends to the isotropic one at large frequencies. Moreover, the production for momenta along the anisotropic direction $z$ is always larger than when the momenta are contained in the transverse $xy$-plane. Increasing the quark mass and/or $\qn$ increases the deviation from the corresponding isotropic results. For fixed values of $\wn$, we have found that for small spatial momentum the production rate does not depend on the quark mass nor on the angle, but only on the anisotropy and on $\wn$. Production along the anisotropic direction is always an increasing function of $\qn$ (in the regime of values we have explored), whereas the production along the transverse plane can be increasing or decreasing, depending on the value of $\wn$. 

The dependence of the dilepton production rate on the degree of anisotropy of the system is detailed in Figs.~\ref{c4bct} and \ref{c4act}. We find that in general the anisotropic rate is larger than the isotropic one, but, if the quark mass and $\wn$ are large enough, there is a range of anisotropies where it might be smaller. The dependence on the angle is always monotonic, either increasing or decreasing depending on the specific quantity which is studied.

It is interesting to compare our results with what happens at weak coupling. Dilepton production in weakly coupled anisotropic plasmas has been studied in \cite{Mauricio:2007vz,Martinez:2008di,Martinez:2008mc}, where an enhancement of the spectral functions with respect to the isotropic case has been reported, in agreement with the present analysis. In those works the dilepton production density per space-time volume and momentum space volume is computed using
\be
\dfrac{dR^{\ell^+\ell^-}}{d^4P}=\int \dfrac{d \vec{p}_1}{(2\pi)^3}\dfrac{d\vec{p}_2}{(2\pi)^3}f_q(\vec{p}_1)f_{\bar{q}}(\vec{p}_2)v_{q\bar{q}}{\sigma_{q\bar{q}}}^{\ell^+\ell^-}\delta^{(4)}(P-p_1-p_2), \label{mauridilrate}
\ee
where $f_{q,\bar q}$ are the phase space distribution functions of the quarks and antiquarks in the plasma, while $v_{q\bar{q}}$ and ${\sigma_{q\bar{q}}}^{\ell^+\ell^-}$ are their relative velocities and total cross section, respectively. The anisotropy of the plasma is modeled by including a parameter $\xi$ which encodes the type and strength of the anisotropy:
\be
f_{q,\bar{q}}(\vec{p},\xi,p_\mt{hard})={f_{q,\bar{q}}}^\mt{iso}(\sqrt{\vec{p}^2+\xi\, (\vec{p}\cdot \hat n)^2},p_\mt{hard}),
\ee
with ${f_{q,\bar{q}}}^\mt{iso}$ being the isotropic phase space distribution function, $p_\mt{hard}$ the hard momentum scale (which can be identified with the temperature $T$ when $\xi=0$), and $\hat n$ defining the anisotropic direction. What we provide in the present work is a non perturbative expression for the factors $f_{q,\bar{q}}(\vec{p},\xi,p_\mt{hard})$ used in the small coupling calculations. It is interesting to note that the enhancement of dilepton production appears to be a robust feature of anisotropic plasmas, which is present both at weak and strong coupling. Of course, the source of anisotropy in \cite{Mauricio:2007vz,Martinez:2008di,Martinez:2008mc} is different than the one we have used in this paper, so that comparisons have to be taken with a grain of salt. Nonetheless, these comparisons certainly motivate further analysis.

As for the generality of our results, we observe that while we have used a very specific source of anisotropy, namely a non-trivial axion, we expect our results to be quite general. We observe in fact that the equations of motion for the gauge fields (\ref{eomy1})-(\ref{eom3}) are solely dependent on the metric and dilaton, meaning that any source of anisotropy that gives rise to similar metric and dilaton (and no Kalb-Ramond field) will produce qualitatively similar results for the photon and dilepton production rates. It would be interesting, nonetheless, to compute these quantities in as many anisotropic backgrounds as possible, including e.g. the one of \cite{Janik:2008tc,Rebhan:2011ke}, to understand which features are really model-independent and therefore the most likely to be realized in the real-world~QGP. 

One thing we have left open is to determine what is the maximum value of $\psi_\mt{H}$ which results in a stable embedding of the flavor D7-branes. To address this question one needs to perform a careful analysis of the phase transition between the black hole and Minkowski embedding of the branes \cite{thermobrane}. This requires comparing the free energy of the system in the two phases. To do that one would presumably need to perform from scratch the holographic renormalization process done in \cite{yiannis} for the axion-dilaton gravity, including this time also the DBI action for the branes. The UV limit ($\epsilon \to 0$)  and the probe brane limit ($\nf/\nc \to 0$) do not in fact commute, in principle. This is because the dilaton of \cite{Mateos:2011ix,Mateos:2011tv} vanishes at the boundary and the introduction of additional D7-branes sources a dilaton that blows up before reaching the boundary; see e.g. \cite{Magana:2012kh}. This means that no matter how small $\nf/\nc$ is, it eventually overtakes the asymptotics and one cannot simply try to renormalize the D7-brane action in the fixed anisotropic background (as made evident by the fact that there do not seem to be enough counterterms to cancel all the divergences). 


\section*{Acknowledgements}
We thank Mariano Chernicoff, Daniel Fern\'andez, Mauricio Mart\'inez, and Yiannis Papadimitriou for useful discussions. We are supported in part by CNPq (VJ and DT), by FAPESP grant 2013/02775-0 (DT), and by DGAPA-UNAM grant IN117012 (AL and LP).


\appendix
\section{Solutions for ${\bf E}_{\bf 1}$ and ${\bf E}_{\bf 2}$}\label{concur}

In this appendix we describe how to construct the two linearly independent solutions $\textbf{E}_{\bf 1}$ and $\textbf{E}_{\bf 2}$ used in subsection \ref{dilep2}.

When solving the equation of motion for (\ref{Evector}) near the horizon, we assume that the fields $E_i$ behave like 
\begin{eqnarray}
E_i(u)=(\uh-u)^{\nu}e_i(u)\,,\qquad i=x,z\,,
\label{Ehorizon}
\end{eqnarray}
where $e_i(u)$ is some regular function at $\uh$. We obtain that the exponent $\nu$ for both components of this vector is the same as that for the $A_y$ mode, namely $\nu=\pm i \wn/2$. After imposing the infalling wave condition (by choosing the minus sign for $\nu$), the rest of the power series is linearly determined by the value of ${\bf E}$ at the horizon, ${\bf E}_\mt{H}$. Integrating from the horizon using any choice of such a vector ${\bf E}_\mt{H}$ would pick a particular solution to the equation of motion (\ref{matrixEOM}), which is linear. The general solution can then be written as a linear combination of any two linearly independent solutions ${\bf E}_a=(E_{x,a}\;\; E_{z,a})^\mt{T}$ and ${\bf E}_b=(E_{x,b}\;\; E_{z,b})^\mt{T}$. Using any two solutions ${\bf E}_a$ and ${\bf E}_b$ we can construct the matrix $\mathcal{E}$ that was needed in subsection \ref{dilep2}:
\begin{eqnarray}
\mathcal{E}\equiv(\textbf{E}_{\bf 1}\;\;\textbf{E}_{\bf 2})=(\textbf{E}_a\;\;\textbf{E}_b)\mathcal{I},
\label{Eab12IA}
\end{eqnarray}
where the matrix $\mathcal{I}$ is given by
\begin{eqnarray}
\mathcal{I}=(\textbf{E}_a\;\; \textbf{E}_b)^{-1}\big |_\mt{bdry}\,.
\label{IdefinitionA}
\end{eqnarray}
This makes $\mathcal{E}$ a matrix whose columns are solutions to (\ref{matrixEOM}) and that satisfies the desired property
\begin{eqnarray}
\mathcal{E}\big |_\mt{bdry}=\left(\begin{array}{cc}
1  &0  \\
0  &1  \\
\end{array}\right)\,.
\label{EbdryA}
\end{eqnarray}
Using (\ref{Eab12IA}) and (\ref{Egensolution}), we can write
\begin{eqnarray}
\textbf{E}_\mt{sol}=(\textbf{E}_a\;\; \textbf{E}_b)\mathcal{I}\left(\begin{array}{cc}
E_x^{(0)}\\
E_z^{(0)}\\
\end{array}\right)\,,\cr
\label{EgensolutionI}
\textbf{E}_\mt{sol}^{\prime}=(\textbf{E}_a^{\prime}\;\;\textbf{E}_{b}^{\prime})\mathcal{I}\left(\begin{array}{cc}
E_x^{(0)}\\
E_z^{(0)}\\
\end{array}\right),
\end{eqnarray}
and use it to write the boundary action (\ref{BdryActionMatrix}) as
\begin{eqnarray}
S_{\epsilon}=-2K_\mt{D7}\int dt\, d\vec x \left[(E_x^{(0)}\;\; E_z^{(0)})\mathcal{M}(\textbf{E}_a^{\prime}\;\;\textbf{E}_b^{\prime})\mathcal{I}\left(\begin{array}{cc}
E_x^{(0)}\\
E_z^{(0)}\\
\end{array}\right)\right]_{u=\epsilon}\,,
\label{BdryActionMatrix}
\end{eqnarray}
where (\ref{EbdryA}) was used.
If we define the matrix
\begin{eqnarray}
\mathcal{C}=\mathcal{M}(\textbf{E}_a^{\prime}\;\; \textbf{E}_b^{\prime})\big |_\mt{bdry}\mathcal{I}\,,
\label{definecalC}
\end{eqnarray}
we can see that
\bea
&& \frac{\delta^2S_{\epsilon}}{\delta E_x^{(0)2}}= -4K_\mt{D7} \mathcal{C}^{xx}\label{multiSECcalprima}\,, \cr
&& \frac{\delta^2S_{\epsilon}}{\delta E_z^{(0)2}}= -4K_\mt{D7} \mathcal{C}^{zz}\,, \cr
&& \frac{\delta^2S_{\epsilon}}{\delta E_z^{(0)}\delta E_x^{(0)}}= -2K_\mt{D7} (\mathcal{C}^{zx}+\mathcal{C}^{xz})\,.
\label{multiSECcal}
\eea

Using the technology developed in \cite{Kaminski:2009dh}, we will now see how to obtain the imaginary parts of the components of $\mathcal{C}$ from $u$-independent quantities that can be computed at the horizon, so that all of the information of the boundary is encoded exclusively in $\mathcal{I}$. The first step is to show that the matrix  $\tilde{\mathcal{C}}\equiv \mathcal{E}^{\dagger}\mathcal{M}\mathcal{E}^{\prime}-\mathcal{E}^{\dagger\,\prime}\mathcal{M}\mathcal{E}$ is $u$-independent. To prove this, we start by multiplying (\ref{matrixEOM}) on the left by $\mathcal{M}$ to obtain
\begin{eqnarray}
\left(\mathcal{M}\textbf{E}^{\prime}\right)^{\prime}+f(u)\mathcal{M}\textbf{E}=0,
\end{eqnarray}
which implies that the equation
\begin{eqnarray}
\left(\mathcal{M}\mathcal{E}^{\prime}\right)^{\prime}+f(u)\mathcal{M}\mathcal{E}=0
\label{matmateom}
\end{eqnarray}
holds for the matrix $\mathcal{E}$. If we multiply (\ref{matmateom}) on the left by $\mathcal{E}^{\dagger}$ and subtract from it its transpose conjugate multiplied on the right by $\mathcal{E}$, we are left with
\begin{eqnarray}
\mathcal{E}^{\dagger}\left(\mathcal{M}\mathcal{E}^{\prime}\right)^{\prime}-
\left(\mathcal{E}^{\dagger\prime}\mathcal{M}\right)^{\prime}\mathcal{E}=0\, . \label{indu}
\end{eqnarray}
Since
\be
\mathcal{E}^{\dagger}\left(\mathcal{M}\mathcal{E}^{\prime}\right)^{\prime}-
\left(\mathcal{E}^{\dagger\prime}\mathcal{M}\right)^{\prime}\mathcal{E}=\left(\mathcal{E}^{\dagger}\mathcal{M}\mathcal{E}^{\prime}-\mathcal{E}^{\dagger\,\prime}\mathcal{M}\mathcal{E}\right)^\prime ,
\ee
equation (\ref{indu}) proves that  $\tilde{\mathcal{C}}$ is indeed $u$-independent. Notice now that, since $\mathcal{E}$ reduces to the identity matrix at the boundary, we have 
\begin{eqnarray}
\tilde{\mathcal{C}}&=&\left(\mathcal{E}^{\dagger}\mathcal{M}\mathcal{E}^{\prime}-\mathcal{E}^{\dagger\,\prime}\mathcal{M}\mathcal{E}\right)\big |_\mt{H}=\left(\mathcal{E}^{\dagger}\mathcal{M}\mathcal{E}^{\prime}-\mathcal{E}^{\dagger\,\prime}\mathcal{M}\mathcal{E}\right)\big |_\mt{bdry}\cr &=&\left(\mathcal{M}\mathcal{E}^{\prime}-\mathcal{E}^{\dagger\,\prime}\mathcal{M}\right)\big |_\mt{bdry}\cr &=&\mathcal{C}-\mathcal{C}^{\dagger}\,.
\label{Ctildedefinition}
\end{eqnarray}
With (\ref{Ctildedefinition}) we can finally write
\begin{align}
\tilde{\mathcal{C}}^{xx}&= 2i\,\mbox{Im}\,\mathcal{C}^{xx}\,,\label{multiCtildeCprima}\\
\tilde{\mathcal{C}}^{zz}&= 2i\,\mbox{Im}\,\mathcal{C}^{zz}\,,\\
\tilde{\mathcal{C}}^{xz}+\tilde{\mathcal{C}}^{zx}&= 2i\,\mbox{Im}\,(\mathcal{C}^{xz}+\mathcal{C}^{zx})\, ,
\label{multiCtildeC}
\end{align}
which achieves the desired result of writing the imaginary parts of the correlators in terms of $\mathcal{M}$ and $\mathcal{E}$ evaluated at the horizon, leaving only $\mathcal{I}$ to be evaluated at the boundary.

The final expression for $\chi_{(2)}$ can then be obtained by inserting (\ref{multiSECcalprima}) into (\ref{chi1chi2}), and using (\ref{multiCtildeCprima})-(\ref{multiCtildeC}) to write
\begin{eqnarray}
&& \chi_{(2)}=-4K_\mt{D7}i
\left(
(k_0^2-q^2\sin^2\vartheta)\tilde{\mathcal{C}}^{xx}
+(k_0^2-q^2\cos^2\vartheta)\tilde{\mathcal{C}}^{zz}
-2q^2\cos\vartheta \sin\vartheta(\tilde{\mathcal{C}}^{xz}+\tilde{\mathcal{C}}^{zx})
\right).\cr &&
\label{Chi2Ctilde}
\end{eqnarray}


\section{Explicit near-boundary-expansion for the action (\ref{bActiondil})}
\label{app2}

We report here the explicit expression for the boundary action, eq. (\ref{bActiondil}). The action reads
\begin{equation}
S_{\epsilon}=-2K_\mt{D7}\int dt\, d\vec{x}\left[\mathcal{L}_{1}+\mathcal{L}_{2}+\mathcal{L}_{3}+\ldots
+O\left(u^{2}\right)\right]_{u=\epsilon}\,,
\end{equation}
where
\begin{align}
&\mathcal{L}_{1}=A_1\,E_x^{(0)2}+B_1\,E_z^{(0)2}+C_1\,E_x^{(0)}\,E_z^{(0)}\,,\cr
&\mathcal{L}_{2}=A_2\,E_x^{(0)}\,E_x^{(2)}+B_2\,E_x^{(0)}\,E_x^{(2)}+C_2\,E_x^{(0)}\,E_x^{(2)}+D_2\,E_x^{(0)}\,E_x^{(2)}\,,\cr
&\mathcal{L}_{3}=-\frac{\log u}{k_0^2}\left[(E_x^{(0)2}+E_z^{(0)2})\,k_0^2+(E_x^{(0)}\cos\vartheta - E_z^{(0)}\sin\vartheta)^2\,q^2 \right]\,,
\end{align}
and the various expansion coefficients are given by
\bea
A_1&=&\frac{(q^2-k_0^2)(2a^2-3(q^2-k_0^2+8\psi_1^2))}{2(q^2-k_0^2)(q^2-k_0^2+8 \psi_1^2)-6a^2\,q^2\,\mathrm{cos}^2\,\vartheta}+\cr
&&
+ \frac{6 a^2 q^4 \mathrm{cos}^4\vartheta - q^2 \mathrm{cos}^2\vartheta [6(q^2-k_0^2)(q^2-k_0^2+8\psi_1^2)-a^2 (4 q^2-10 k_0^2)+3 a^2 q^2 \mathrm{cos}^2\vartheta]}{4 k_0^2[(q^2-k_0^2)(2a^2-3(q^2-k_0^2+8\psi_1^2))+3 a^2 q^2 \mathrm{cos}^2\vartheta]}
\,,
\cr
B_1 &=& \frac{\left((5a^2+15q^2+48\psi_1^2-15k_0^2)(q^2-k_0^2)-3 a^2 q^2 \mathrm{cos}^2\vartheta\right)(q^2 \mathrm{sin}^2\vartheta-k_0^2)}{4 k_0^2[(q^2-k_0^2)(2a^2-3(q^2-k_0^2+8\psi_1^2))+3 a^2 q^2 \mathrm{cos}^2\vartheta]}\,,
\cr 
C_1 &=& \frac{q^2[3(7q^2-7k_0^2+32\psi_1^2)(q^2-k_0^2)-2a^2(q^2+2k_0^2)-6 a^2 q^2 \cos 2\vartheta]\sin 2 \vartheta}{8k_0^2((q^2-k_0^2)(2a^2-3(q^2-k_0^2+8\psi_1^2))+3a^2q^2 \mathrm{cos}^2\vartheta)}\,,
\cr 
A_2 &=& \frac{2\,\mathrm{cos}\,\vartheta [3(q^2-2k_0^2)(q^2-k_0^2+8\psi_1^2)-a^2(5q^2-4k_0^2)-q^2(5a^2-3(q^2-k_0^2+8\psi_1^2))\cos 2\vartheta]}{k_0^2[6(q^2-k_0^2)(q^2+8\psi_1^2-k_0^2)-a^2(7q^2-4k_0^2)-3 a^2 q^2 \cos 2\vartheta]}\,,
\cr 
B_2 &=&-\frac{48 q^2 \sin 2 \vartheta}{k_0^2[6(q^2-k_0^2)(q^2-k_0^2+8\psi_1^2)-a^2(7q^2-4k_0^2)-3 a^2 q^2 \cos 2\vartheta]}\,,
\cr 
C_2 &=&\frac{4 q^2 (5 a^2 - 3 (q^2-k_0^2+8\psi_1^2))\mathrm{cos}^2\vartheta \, \mathrm{sin}\,\vartheta}{k_0^2[6(q^2-k_0^2)(q^2-k_0^2+8\psi_1^2)-a^2(7q^2-4k_0^2)-3 a^2 q^2 \cos 2\vartheta]}\,,
\cr 
D_2 &=&\frac{96(q^2\sin^2\vartheta-k_0^2) }{k_0^2[6(q^2-k_0^2)(q^2-k_0^2+8\psi_1^2)-a^2(7q^2-4k_0^2)-3 a^2 q^2 \cos 2\vartheta]}\,.
\label{appformulas}
\eea


\end{document}